\documentclass[leqno]{siamltex}
\usepackage{geometry}
\usepackage{graphicx}
\usepackage{epstopdf}

\usepackage{amssymb,amsmath,amsfonts}
\usepackage{algpseudocode} \usepackage{algorithm} \usepackage{mathabx}
\usepackage[usenames,dvipsnames]{xcolor} \usepackage{url}
\usepackage{caption} \usepackage{subcaption} \usepackage{ulem}
\usepackage{morefloats} \usepackage{floatflt}

\graphicspath{ {.} }

\title{Numerical analysis of the vertex models for simulating grain boundary networks}

\author{C.~E.~Torres \thanks{Department of
    Mathematical Sciences, George Mason University, Fairfax VA 22015
    (Current address: Departamento de Inform\'atica, Universidad T\'ecnica Federico Santa Mar\'ia, Casilla 110-V, Valpara\'iso, Chile
    and 
Centro Cient\'ifico-Tecnol\'ogico de Valpara\'iso, Universidad T\'ecnica Federico Santa Mar\'ia, Casilla 110-V, Valpara\'iso, Chile ({\tt ctorres@inf.utfsm.cl}))} \and M.~Emelianenko \thanks{Department
    of Mathematical Sciences, George Mason University, Fairfax VA
    22015 ({\tt memelian@gmu.edu})} \and D.~Golovaty
  \thanks{Department of Mathematics,
    University of Akron, Akron OH 44325 ({\tt dmitry@uakron.edu})} \and D.~Kinderlehrer \thanks{Department of Mathematical Sciences 
Carnegie Mellon University 
Pittsburgh PA 15213 ({\tt davidk@andrew.cmu.edu})} \and S.~Ta'asan \thanks{Department of Mathematical Sciences 
Carnegie Mellon University 
Pittsburgh PA 15213} ({\tt shlomo@andrew.cmu.edu})}

\begin{document}
\maketitle
\slugger{siap}{xxxx}{xx}{x}{x--x}

\begin{abstract}
  Polycrystalline materials undergoing coarsening can be represented
  as evolving networks of grain boundaries, whose statistical
  characteristics describe macroscopic properties. The
  formation of various statistical distributions is
  extremely complex and is strongly influenced by topological changes
  in the network. This work is an attempt to elucidate the role of
  these changes by conducting a thorough numerical investigation of
  one of the simplest types of grain growth simulation models, the
  vertex model. While having obvious limitations in terms of its
  ability to represent realistic systems, the vertex model enables
  full control over topological transitions and retains essential
  geometric features of the network.

  We formulate a self-consistent vertex model and investigate the role
  of microscopic parameters on mesoscale network behavior. This
  study sheds light on several important questions, such as how
  statistics are affected by the choice of temporal and spatial
  resolution and rules governing topological changes.  Statistical
  analysis of the data produced by the simulation is performed for
  both isotropic and anisotropic grain boundary energies.
\end{abstract}



\pagestyle{myheadings} \thispagestyle{plain} \markboth{Torres, et.
  al. }{Numerical analysis of the vertex models}

\section{Introduction}
Polycrystalline materials  such as metals and ceramics are comprised of single crystallites, called grains, separated by their boundaries, called grain
boundaries.  The orientations, shapes, and arrangements of the grains have a direct relationship to macroscopic
materials properties. For example, the presence of grain boundaries
decreases thermal and electrical conductivity that affects the
performance of chips in microprocessors. Grain boundaries disrupt
motion of dislocations through a material, so reducing crystallite
size is a common way to improve strength and fracture toughness
in structures, as described by the Hall-Petch relationship \cite{smith2003foundations}. 

The grain and grain boundary configuration, or the microstructure of a material, is determined by a variety of factors, such as history of deformation, phase transitions, heat treatment, etc. In this paper, we are primarily interested in the process of microstructural relaxation known as coarsening. Evolution of the grain boundary network during coarsening is driven by the tendency of the system to reduce its total grain boundary surface energy spatially constrained that results in growth of some grains at the expense of others, as well as in disappearance and nucleation of both small grains and grain boundaries. In this process the average grain size increases, while the total surface area of the grain boundaries decreases. As it evolves, the grain boundary network begins to exhibit stable, self-similar statistical features that can be described by a finite number of time-dependent parameters. These parameters, in turn, can be used as continuum descriptors of the network. 

One of the principal goals of mathematical modeling of polycrystalline materials is to understand how the statistics of an evolving grain boundary network depend on the set of laws that govern the dynamics of the network at the microscopic scale. Here the laws in question describe the motion of grain boundaries and their junctions, as well as the criteria for nucleation and disappearance of grains and grain boundaries. Although these laws have been known for quite some time, their precise role in the development of macroscopic properties of the network is still not fully clear. A possible way to establish a connection between the statistical features of the network and the evolution of individual grain boundaries involves numerical experimentation using large-scale simulation models. As a starting point, it is therefore desirable to consider models that are as simple and as computationally inexpensive as possible, yet preserve the essential characteristics of the original grain boundary network. In what follows we concentrate on two-dimensional polycrystalline materials that can be thought of as, e.g., cross-sections of systems of columnar grains in aluminum films \cite{barmak12}.

Since the motion of grain boundaries is controlled by surface energy, it is normally modeled within the framework of curvature-driven growth. Under certain assumptions on relative mobilities of the boundaries and their junctions, it is possible to assume that the boundaries remain straight during the evolution so that the changes within the network can be described solely in terms of motion of the junctions, or vertices, of the graph formed by the boundaries. Note that for isotropic surface energies, only the junctions between three grain boundaries, as opposed to four or five, for instance, are stable. Vertex models that discount grain boundary motion
in favor of triple junction motion are used
extensively, both due
to their computational simplicity in handling extremely large scale
networks, and for the purpose of isolating properties local to triple junctions, for example triple
junction drag.  
Following the pioneering works of Fullman \cite{fullman} in the 50's and Frost {\it et al.} in the 80's
\cite{frost88}, a number of extensions of the original algorithm have
been proposed
\cite{Kawasaki}-\nocite{Otto03,kind04,weygand98, weygand2001, barrales10}\cite{syha10}. The vertex models have been able to reproduce many
characteristic features of the cellular pattern growth in foams
\cite{weaire83, weaire84} and to some extent in polycrystals \cite{zhang03}. They have
high flexibility, which motivates continued interest
in their use despite the existence of more sophisticated numerical
codes. The vertex model approach has been recently applied to the recrystallization of ferritic stainless steels \cite{sinclair04} and, more generally, to grain growth \cite{zhang03}, \cite{weygand04}. It has also been extensively used to validate topological theories of grain growth and Zener pinning \cite{Mullins86}, \cite{holm06}.
A comprehensive review of the relevant literature can be found in the chapter dedicated to vertex models in \cite{molodov2013}.

In this paper we develop a numerical algorithm for a version of a simple vertex model originally proposed by Kawasaki in \cite{Kawasaki}. Our main aims are to derive the set of rules for topological transitions within the network that are consistent with continuous evolution of vertices as well as to control the stability and accuracy of the code. This is done in order to eliminate numerical issues from the investigation of the role that various model parameters play in the development of statistical features. We demonstrate that, although simple, our model results in rich statistics that are reminiscent of what is observed in experiments and more sophisticated simulations. We emphasize, however, that our main motivation is not to replicate or explain experimental observations, but to ensure the correct characterization of the complexity of network behavior. In a subsequent publication we will investigate the formation of statistics using numerical experimentation with the vertex model developed here. This investigation will expand on our prior studies of a one-dimensional model \cite{begkt08,begkt08_1,beeekst11}.

The paper is organized as follows. In Section~\ref{extreme}, we start
by formulating a general energy-based model of an evolving grain boundary network. We formally demonstrate 
that this model reduces to a vertex model by assuming that the mobility of the vertices is much lower than 
the mobility of grain boundaries evolving via curvature-driven motion.  Next, in Sections  \ref{sec:analysis_flipping_rule} 
and \ref{sec:collision_time} we use semi-rigorous analysis of vertex dynamics to derive the set of neighbor switching rules 
as well as estimates of vertex collision times.  Stability analysis of the explicit numerical scheme used in the main algorithm 
is given in Section \ref{stability}. The full description of the algorithm appears in
Section~\ref{alg}, followed by the numerical results in
Section~\ref{num}.  We begin this section by testing the numerical procedure for accuracy and
convergence. The procedure is then employed to simulate coarsening of grain boundary networks containing a large number of grains. The geometry of configurations that develop in these simulations is described using the standard statistical measures for characterizing grain growth. These include distributions of relative areas of grains, dihedral angle, number of sides, among others. We are able to confirm spatiotemporal stability of the distributions that emerge in a network evolving via our numerical algorithm.  We find that the distributions are essentially independent of the level of numerical resolution as the network appears to pass through the sequence of similar states, possibly at different rates.  While mesoscopically the model is insensitive to various modifications, including the rules governing topological changes, the microscopic features of the network tend to differ with the scenario.
\section{Vertex model formalism}
\label{extreme}

Let us define the configuration and establish the law of evolution for our network. Suppose given a rectangular domain $R\subset\mathbb{R}^2$ that contains a
set $\mathbf\Gamma$ of $K>0$ smooth curves
$\Gamma_k:=\left\{ \mathbf{x} =\boldsymbol\xi_k(s),\ 0\leq s \leq
  L_k\right\},\ k=1,\ldots,K,$ that we will call {\it{grain
    boundaries}}, with $L_k>0$ the length of the $k$-th
boundary. On a grain boundary curve $\Gamma_k$, one can
define an orthogonal frame $\left\{\mathbf{b}_k,\mathbf{n}_k\right\}$,
where $$ \mathbf{b}_k=\frac{d \boldsymbol\xi_k}{ds}/\left|
  \frac{d\boldsymbol\xi_k}{ds}\right| \ \textrm{and} \
\mathbf{n}_k=\frac{d \mathbf b_k}{ds}/\left| \frac{d\mathbf
    b_k}{ds}\right| $$
Assuming periodic boundary conditions on $\partial R,$ all grain
boundaries can terminate only at junctions with other boundaries. We
denote the set of all junctions in $R$ by $\mathbf{X}:=\left\{
  \mathbf{x}_1^{n_1}, \mathbf{x}_2^{n_2}, \ldots,
  \mathbf{x}_M^{n_M}\right\},$ where the number of junctions,
$M\in\mathbb N$.  Here an {\it{$n-$tuple junction}} $\mathbf{x}_m^n$
is a terminal point of $n$ grain boundaries $\Gamma_{j_1}$, $
\Gamma_{j_2},\ldots,\Gamma_{j_n}$, for some $j_1,\
j_2,\ldots,j_n\in\left\{1,\ldots,K\right\}$. In the simplest and most
commonly studied type of a grain boundary network, the numbers
$n_1=n_2=\ldots=n_M=3,$ i.e., all elements of $\mathbf{X}$ are
{\it{triple junctions}}.

The grain boundaries contained in $\mathbf\Gamma$ subdivide the domain
$R$ into $N$ disjoint regions \sloppy
$\left\{\Sigma_1,\ldots,\Sigma_N\right\}=:\mathbf\Sigma$, called {\it
  grains}. With each grain $\Sigma_l\in\mathbf\Sigma$, $l=1,\ldots,N,$
we associate an {\it orientation} $\alpha_l\in[0,2\pi)$ and the set of
grain boundaries
$\partial\Sigma_l=\left\{\Gamma_{k_1},\ldots,\Gamma_{k_l}\right\}$
that enclose $\Sigma_l.$ Likewise, for each grain boundary $\Gamma_k,$
$k=1,\ldots,K,$ there are exactly two grains $\Sigma_{l_1(k)}$ and
$\Sigma_{l_2(k)}$ that are separated by $\Gamma_k.$

The {\it grain misorientation} parameter $\Delta \alpha_k$ is defined
as $\Delta\alpha_k:= \alpha_{l_2(k)}-\alpha_{l_1(k)},$ where
$k=1,\ldots,K.$ The {\it grain boundary energy,} $\gamma_k,$ will be
assumed to depend only on misorientation, i.e.
$\gamma_k=\gamma\left(\Delta \alpha_k\right)$ for every $k=1,\ldots,K$
and some given function $\gamma: \mathbb R\to\mathbb R$. The function
$\gamma$ is even and periodic with a period that depends on the
symmetries of crystalline lattices of neighboring grains.

We will assume that the grain boundary network evolves in time via
simultaneous motion of both the grain boundaries and their
junctions. In the course of this motion, some grains grow and some
shrink. Once the length of a grain boundary or the area of a grain
decreases to zero, we will say that a component of the network has
disappeared as a result of a {\it topological transition}. In order to
describe the evolution of the grain boundary network, both the laws of
continuous motion of the boundaries and junctions as well as the rules
governing the topological transitions must be specified. From now on
we will assume that the sets $\mathbf\Gamma,\mathbf X$ and
$\mathbf\Sigma$ depend on time $t>0.$

We begin the discussion of the network dynamics by considering a
period of time $[t_0,t_0+T]$ during which no topological transitions
occur. Introduce the total energy of the network
\begin{equation}
  \nonumber
  E(t)=\sum_{k=1}^K \int_0^{L_k} \gamma\left(\Delta \alpha_k\right) | \mathbf l_k(s,t) | ds, \quad \mathbf l_k=\frac{d\boldsymbol\xi_k}{ds},
\end{equation}
where all curves at $t_0$ are assumed to be parametrized with respect
to their arc-length and $L_k$ is the length of $\Gamma_k$ at the time
$t_0$. Denoting $\gamma_k=\gamma\left(\Delta\alpha_k\right)$, we
obtain
\begin{equation}
  \nonumber
  \frac{d}{dt}E(t)=\sum_{k=1}^K \int_0^{L_k} \gamma_k  \frac{\mathbf l_k}{| \mathbf l_k |} \cdot  \frac{\partial\mathbf l_k}{\partial t} ds
  =\sum_{k=1}^K \int_0^{L_k} \gamma_k  \mathbf b_k \cdot  \frac{\partial \mathbf v_k}{\partial s} ds = \sum_{k=1}^K \int_0^{L_k} \mathbf T_k \cdot  \frac{\partial \mathbf v_k}{\partial s} ds
\end{equation}
where $\mathbf T_k=\gamma_k \mathbf b_k$ denotes the capillary force,
also called the line tension. Further, $\mathbf v_k(s,t)$ denotes the
velocity of the material point $s$ on the curve $\Gamma_k$ at the time
$t$ so that $$\displaystyle{\frac{\partial \mathbf l_k}{\partial
    t}=\frac{\partial}{\partial t}\left(\frac{\partial
      \boldsymbol\xi_k}{\partial s}\right)=\frac{\partial}{\partial
    s}\left(\frac{\partial \boldsymbol\xi_k}{\partial
      t}\right)=\frac{\partial \mathbf v_k}{\partial s}}.$$ Integrating
by parts and using the Frenet formula $\displaystyle{\frac{\partial
    \mathbf b_k}{\partial s}=\kappa_k | \mathbf l_k| \mathbf n_k}$ we
obtain
\begin{equation}
  \frac{d}{dt}E(t)=-\sum_{k=1}^K \int_0^{L_k} \kappa_kV_k |\mathbf l_k|ds-\sum_{m=1}^M \mathbf v_m\cdot \sum_{l=1}^{n_m} \mathbf T_{m,l},
  \label{eq:dEdt_curvature_and_vertices}
\end{equation}
where $\mathbf T_{m,l}$ is the capillary force along the grain
boundary $\Gamma_{j_l}$ that ends at the triple junction
$\mathbf{x}_m^{n_m}.$ Further, $\kappa_k$ and $V_k=\mathbf
v_k\cdot\mathbf n_k$ are the curvature and the normal velocity of
$\Gamma_k,$ respectively.

The simplest framework to enforce energy dissipation is to assume that
the grain boundaries and their junctions follow a version of gradient flow
dynamics. Then the normal velocity of the boundary $\Gamma_k$ and the
velocity $\mathbf{v}_m:=\frac{d}{dt}\mathbf x_m^{n_m}$ of the triple
junction $\mathbf{x}_m^{n_m}$ can be written as
\begin{equation}
  \label{curvature}
  V_k=\mu_k\kappa_k
\end{equation}
and
\begin{equation}
  \label{motion_law}
  \displaystyle \mathbf v_m =\lambda_m \sum^{n_m}_{l=1}{\mathbf T_{m,l}},
\end{equation}
respectively. Here $\mu_k>0$ is the mobility of $\Gamma_k$ and
$\lambda_m>0$ is the mobility of $\mathbf x_m^{n_m}$. Then, using
\eqref{eq:dEdt_curvature_and_vertices}, we have
\begin{equation}
  \frac{d}{dt}E(t)=-\sum_{k=1}^K \mu_k\int_0^{L_k}  \kappa_k^2 |\mathbf l_k|ds - \sum_{m=1}^{M}\lambda_m{\left|\sum^{n_m}_{l=1}{\mathbf T_{m,l}}\right|^2} \leq 0
  \label{eq:dEdt_curvature}
\end{equation}
\eqref{curvature} and \eqref{motion_law} are known as the Mullins Equation \cite{Mullins:MS:1963,ISI:A1956WF82500011} and a variation of the Herring Condition \cite{Herring:SPSS:1952} respectively, cf. also \cite{ISI:000169336700007}.

When the grain boundary mobility is much higher than that of triple
junctions, the grain boundaries $\Gamma_k$ are essentially straight
lines throughout the evolution. The precise asymptotic reduction, which we will not treat here, requires boundary layer analysis near the junctions \cite{Otto_Selim13}. Hence the dynamics of the grain
boundary network is completely determined by the motion of the triple
junctions via the law \eqref{motion_law} which relates the velocity of
a triple junction to the sum of capillary forces acting on it. This
reduced model is known as a {\it vertex model} in the literature and
is a subject of study in this work.

In what follows, we will set $\lambda_m=1$ for all $m=1,\ldots,M$ and,
unless noted otherwise, assume that the grain boundary energy is only
weakly anisotropic, i.e., $\gamma(\Delta\alpha)=1+\varepsilon
f(\Delta\alpha),$ where $\varepsilon>0$ is small. As we will soon see,
this assumption ensures that all grain boundary junctions are, in
fact, triple junctions and $n_m=3$ for all $m=1,\ldots,M$. Thus we can
refer to triple junctions simply as $\mathbf{x}_m,\ m=1,\ldots,M$ by
dropping the superscript index $n_m$.

Suppose now that $\mathbf{x}_{m_1},\ \mathbf{x}_{m_2},\ \mathrm{and}\
\mathbf{x}_{m_3} $ denote three vertices connected to a vertex
$\mathbf{x}_m$ for every $m=1,\ldots,M$. Let $\gamma_{mm_1},\
\gamma_{mm_2},\ \mathrm{and}\ \gamma_{mm_3}$ be the grain boundary
energy of the straight edges connecting $\mathbf{x}_m$ with
$\mathbf{x}_{m_1},\ \mathbf{x}_{m_2},\ \mathrm{and}\
\mathbf{x}_{m_3},$ respectively. Then the law of the vertex motion
\eqref{motion_law} takes the form
\begin{equation}
  \label{law_motion_final}
  \dot{\mathbf x}_m=\sum_{i=1}^3\gamma_{mm_i}\frac{\mathbf{x}_{m_i}-\mathbf{x}_m}{\left\|\mathbf{x}_{m_i}-\mathbf{x}_m\right\|},\ m=1,\ldots,M.
\end{equation}

In order to fully describe the evolution of the grain boundary
network, it is still necessary to understand what happens during
topological transitions that alter the structure of the network. This
is the subject of the next section.

\section{Topological transitions}\label{sec:analysis_flipping_rule}
As discussed earlier, a topological
transition occurs when an element of a grain boundary network
disappears. This can happen when the length of a single edge, or area of a
single grain decreases to zero. In this section, we formulate a set of
rules that govern topological transitions. Our approach aims to
improve on existing literature by emphasizing the consistency between
continuous evolution of the network and discrete transitions.

\begin{figure}[!t]
  \centering
  \includegraphics[width=0.9\textwidth]{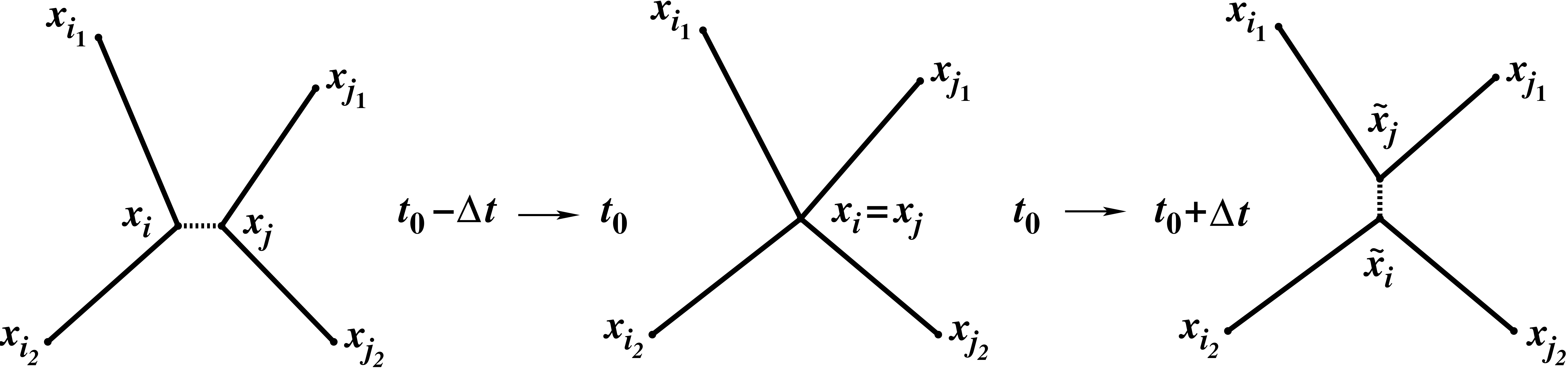}
  \caption{Neighbor switching.}
  \label{fig:sketch_flipping}
\end{figure}

Suppose first that an edge connecting two vertices $\mathbf{x}_i$ and
$\mathbf{x}_j$ disappears at some time $t_0>0$. We will then say that
$\mathbf{x}_i$ and $\mathbf{x}_j$ {\it collide} at the time $t_0$
forming a quadruple junction. When the grain boundary energy is weakly
anisotropic, this junction is unstable in the following sense: it is
possible to split the quadruple junction into two new triple junctions
$\mathbf{\tilde x}_i$ and $\mathbf{\tilde x}_j$ connected by an
infinitesimally short edge that will grow. Clearly, as indicated in
Fig.~\ref{fig:sketch_flipping}, the direction in which the splitting
occurs is not the same as the direction of the original collision. We
will refer to this event as a {\it neighbor switching}.

Next, we will use \eqref{law_motion_final} to determine the
orientation of the edge that forms as a result of neighbor switching. Fix a
sufficiently small $\Delta t>0$, then
\begin{align}
  \dot{\mathbf{x}}_i&=\gamma_{ii_1} \,
  \frac{\mathbf{x}_{i_1}-\mathbf{x}_i}{\|\mathbf{x}_{i_1}-\mathbf{x}_i\|}
  +\gamma_{ii_2} \,
  \frac{\mathbf{x}_{i_2}-\mathbf{x}_i}{\|\mathbf{x}_{i_2}-\mathbf{x}_i\|}
  +\gamma_{ij} \,
  \frac{\mathbf{x}_j-\mathbf{x}_i}{\|\mathbf{x}_j-\mathbf{x}_i\|},
  \label{eq:vel_1}\\
  \dot{\mathbf{x}}_j&=\gamma_{jj_1} \,
  \frac{\mathbf{x}_{j_1}-\mathbf{x}_j}{\|\mathbf{x}_{j_1}-\mathbf{x}_j\|}
  +\gamma_{jj_2} \,
  \frac{\mathbf{x}_{j_2}-\mathbf{x}_j}{\|\mathbf{x}_{j_2}-\mathbf{x}_j\|}
  +\gamma_{ij} \,
  \frac{\mathbf{x}_i-\mathbf{x}_j}{\|\mathbf{x}_i-\mathbf{x}_j\|},
  \label{eq:vel_2}
\end{align}
when $t\in(t_0-\Delta t,t_0).$ Note that the first two terms in both
equations are continuous functions of $t$ on $[t_0-\Delta t, t_0]$ if
we assume that $\mathbf{x}_i$ and $\mathbf{x}_j$ are continuous on
$[t_0-\Delta t, t_0].$ Subtracting \eqref{eq:vel_1} from
\eqref{eq:vel_2} we have
\begin{equation}
  \label{eq:diffvel}
  ({\mathbf{x}}_j-{\mathbf{x}}_i)\dot\, =\mathbf{p_-}(t)-2\gamma_{ij}\,\frac{\mathbf{x}_j-\mathbf{x}_i}{\|\mathbf{x}_j-\mathbf{x}_i\|},
\end{equation}
where
\begin{equation} {\bf p}_-=\gamma_{jj_1} \,
  \frac{\mathbf{x}_{j_1}-\mathbf{x}_j}{\|\mathbf{x}_{j_1}-\mathbf{x}_j\|}
  +\gamma_{jj_2} \,
  \frac{\mathbf{x}_{j_2}-\mathbf{x}_j}{\|\mathbf{x}_{j_2}-\mathbf{x}_j\|}
  -\gamma_{ii_1} \,
  \frac{\mathbf{x}_{i_1}-\mathbf{x}_i}{\|\mathbf{x}_{i_1}-\mathbf{x}_i\|}
  -\gamma_{ii_2} \,
  \frac{\mathbf{x}_{i_2}-\mathbf{x}_i}{\|\mathbf{x}_{i_2}-\mathbf{x}_i\|}
\end{equation}
satisfies $\mathbf{p_-}\in C([t_0-\Delta t,t_0]).$ Let
$\mathbf{x}_j-\mathbf{x}_i=\rho\mathbf{n}$, where
$\rho=\left\|\mathbf{x}_j-\mathbf{x}_i\right\|$ and
$\mathbf{n}=(\cos{\theta},\sin{\theta})$ and set
$\boldsymbol\tau=(-\sin{\theta},\cos{\theta})$. Rewriting \eqref{eq:diffvel}, we have
\begin{equation}
  \dot\rho\mathbf{n}+\rho\dot\theta\boldsymbol\tau=\mathbf{p_-}-2\gamma_{ij}\mathbf{n},
  \label{eq:kdot0}
\end{equation}
or by orthogonal decomposition,
\begin{subequations}
  \begin{equation}
    \dot{\rho}=\mathbf{p_-}\cdot\mathbf{n}-2\,\gamma_{ij},
    \label{eq:right_before_flipping_1}
  \end{equation}
  \begin{equation}
    \dot{\theta}=\frac{\mathbf{p_-}\cdot\boldsymbol\tau}{\rho}.
    \label{eq:right_before_flipping_2}
  \end{equation}
  \label{eq:right_before_flipping}
\end{subequations}

In Appendix 1, we use \eqref{eq:right_before_flipping} to show that
$$\lim_{t\to t_0^-}\mathbf{p_-}\cdot\boldsymbol\tau=0,$$ since $\lim_{t\to t_0^-}\rho=0.$ Thus $\mathbf{n}_-:=\lim_{t\to t_0^-}\mathbf{n}$  must be parallel to $\mathbf{p_-}(t_0)$. The analogous arguments on $(t_0,t_0+\Delta t)$ demonstrate that $\mathbf{n}_+:=\lim_{t\to t_0^+}\mathbf{n}$ must be parallel to $\mathbf{p_+}(t_0)$, where
\begin{equation} {\bf p}_+=\gamma_{\tilde{j}j_1} \,
  \frac{\mathbf{x}_{j_1}-\mathbf{\tilde
      x}_j}{\|\mathbf{x}_{j_1}-\mathbf{\tilde x}_j\|} +\gamma_{\tilde
    j j_1} \, \frac{\mathbf{x}_{i_1}-\mathbf{\tilde
      x}_j}{\|\mathbf{x}_{i_1}-\mathbf{\tilde x}_j\|} -\gamma_{\tilde
    ij_2} \, \frac{\mathbf{x}_{j_2}-\mathbf{\tilde
      x}_i}{\|\mathbf{x}_{j_2}-\mathbf{\tilde x}_i\|} -\gamma_{\tilde
    i i_2} \, \frac{\mathbf{x}_{i_2}-\mathbf{\tilde
      x}_i}{\|\mathbf{x}_{i_2}-\mathbf{\tilde x}_i\|},
\end{equation}
per Fig.~\ref{fig:sketch_flipping}.  Note that the edge that had
existed before the transition should disappear if the condition
\begin{equation}
  \|{\bf p}_-(t_0)\|-2\gamma_{ij}<0.
  \label{condition2}
\end{equation}
is satisfied. Further, an analog of \eqref{eq:right_before_flipping_1}
shows that
\begin{equation}
  \|{\bf p}_+(t_0)\|-2\gamma_{\tilde i\tilde j}>0,
\end{equation}
guarantees that the newly formed edge will grow.

Both of these inequalities simultaneously hold for collisions in grain
boundary networks with isotropic grain boundary energy, and we expect
them to hold in the case of weak anisotropy. Indeed, in grain growth
simulations described below, we numerically observed that the
condition \eqref{condition2} is always satisfied, as long as the
anisotropy is not too strong.  For large anisotropy quadruple
junctions may become stable, as shown in Section \ref{quad_junction}.

\section{Collision time estimate}\label{sec:collision_time}

We can use the evolution equations \eqref{eq:right_before_flipping} to
estimate whether a pair of the adjacent vertices of the grain
boundary network will collide during a given time step $\Delta
t$. This estimate is essential to detect topological transitions
within the numerical procedure that will be discussed in the
subsequent sections.

Given the current time $t=t_c$, suppose that the edge connecting the
vertices ${\mathbf x}_i$ and ${\mathbf x}_j$ becomes extinct at the
time $t_c+t_{ext}$, where $t_{ext}<\Delta t$. Assuming that $\Delta
t>0$ is sufficiently small and using the continuity of ${\mathbf p}_-$
on the interval $[t_c,t_c+t_{ext}]$, we have that ${\mathbf
  p}_-(t)={\mathbf p}_-(t_c)+o(1)$. Now consider the system of
equations
\begin{subequations}
  \begin{equation}
    \dot{\bar\rho}={\mathbf p}\cdot\bar{\mathbf n}-2\,\gamma_{ij},
    \label{eq:ct1}
  \end{equation}
  \begin{equation}
    \dot{\bar\theta}=\frac{\mathbf{p}\cdot\bar{\boldsymbol\tau}}{\bar\rho}.
    \label{eq:ct2}
  \end{equation}
  \label{eq:ct}
\end{subequations}
on $[t_c,t_c+\bar t_{ext}]$ satisfying $\bar\rho(t_c)=\rho(t_c)$ and
$\bar\theta(t_c)=\theta(t_c)$. Here ${\mathbf p}:={\mathbf
  p}_-(t_{c})$ and $\bar\rho\left(\bar t_{ext}\right)=0$. Taking the
derivative of (\ref{eq:ct1}), multiplying the resulting equation by
$\bar\rho$, and using \eqref{eq:ct} we obtain
\begin{align*}
  \bar\rho\ddot{\bar\rho} &= \bar\rho\,{\mathbf p}\cdot\dot{\bar{\mathbf n}}= \left({\mathbf p}\cdot\bar{\boldsymbol\tau} \right)\bar\rho\dot{\bar\theta}={\left({\mathbf p}\cdot\bar{\boldsymbol\tau}\right)}^2= {\|{\mathbf p}\|}^2-{\left({\mathbf p}\cdot\bar{\mathbf n}\right)}^2 \\
  & ={\|{\mathbf
      p}\|}^2-{\left(\dot{\bar\rho}+2\gamma_{ij}\right)}^2={\|{\mathbf
      p}\|}^2-{\dot{\bar\rho}}^2-4\gamma_{ij}\dot{\bar\rho}-4\gamma_{ij}^2.
\end{align*}
Rearranging terms then gives
\begin{equation}
  \label{eq:s1}
  \left(\bar\rho\,\left(\dot{\bar\rho}+ 4\,\gamma_{ij}\right)\right)\dot\,=\|\mathbf{p}\|^2-4\,\gamma_{ij}^2.
\end{equation}
Note that this equation no longer involves the angular coordinate
$\bar\theta$.

Integrating \eqref{eq:s1} once, leads to
\begin{equation}
  \label{eq:s2}
  \bar\rho\,\left(\dot{\bar\rho}+ 4\,\gamma_{ij}\right)=\left(\|\mathbf{p}\|^2-4\,\gamma_{ij}^2\right)\left(t-t_c\right)+\rho\left(t_c\right)\,\left(\dot\rho\left(t_c\right)+ 4\,\gamma_{ij}\right)
\end{equation}
on $(t_c,t_c+\bar t_{ext})$. Suppose that
$\|\mathbf{p}\|^2-4\,\gamma_{ij}^2<0$, then the right hand side of the
equation \eqref{eq:s2} vanishes when
$t-t_c=\rho\left(t_c\right)\,\left(\dot\rho\left(t_c\right)+
  4\,\gamma_{ij}\right)/\left(4\,\gamma_{ij}^2-\|\mathbf{p}\|^2\right).$
We claim that $\bar\rho$ becomes zero at the same time, i.e,
\[\bar
t_{ext}=\frac{\rho\left(t_c\right)\,\left(\dot\rho\left(t_c\right)+
    4\,\gamma_{ij}\right)}{4\,\gamma_{ij}^2-\|\mathbf{p}\|^2}=\frac{\rho\left(t_c\right)\,\left({\mathbf
      p}_-\left(t_c\right)\cdot{\mathbf n}\left(t_c\right)+
    2\,\gamma_{ij}\right)}{4\,\gamma_{ij}^2-\|\mathbf{p}\|^2},\] where
the final expression follows from the definition of ${\mathbf p}$ and
\eqref{eq:right_before_flipping_1}. Indeed, by our assumption that
$\|\mathbf{p}\|^2-4\,\gamma_{ij}^2<0$ and from \eqref{eq:ct1}, the
expression $\dot{\bar\rho}+ 4\,\gamma_{ij}={\mathbf
  p}\cdot\bar{\mathbf n}+2\,\gamma_{ij}$ is strictly positive and
bounded on $(t_c,t_c+\bar t_{ext})$. It then follows that $\bar\rho$
vanishes along with the right hand side of the equation \eqref{eq:s2}.

Finally, since ${\mathbf p}_-(t)={\mathbf p}_-(t_c)+o(1)$ on
$[t_c,t_c+t_{ext}]$, we have that $\bar t_{ext}$ is the leading order
approximation to $t_{ext}$, i.e.,
\begin{equation}
  \label{eq:t_lambda_in_use}
  t_{ext}=\frac{\rho\left(t_c\right)\,\left({\mathbf p}_-\left(t_c\right)\cdot{\mathbf n}\left(t_c\right)+ 2\,\gamma_{ij}\right)}{4\,\gamma_{ij}^2-\|\mathbf{p}_-\left(t_c\right)\|^2}\left(1+o(1)\right).
\end{equation}

\section{Stability Analysis} \label{stability}

Here we present arguments to show that the explicit numerical scheme proposed
in this paper is stable. For simplicity, consider a discretization of
the system \eqref{eq:right_before_flipping} in the isotropic case when
the grain boundary energy is identically equal to one
\begin{equation}
  \label{eq:d0}
  \left\{
    \begin{split}
      & \rho_{i+1}=\rho_i+\left({\mathbf p}_-(t_i)\cdot {\mathbf
          n}(t_i)-2\right)\Delta t, \\ &
      \theta_{i+1}=\theta_i+\frac{\Delta t}{\rho_i}{\mathbf
        p}_-(t_i)\cdot {\boldsymbol\tau}(t_i).
    \end{split}
  \right.
\end{equation}
To simplify this system further, suppose that ${\mathbf
  p}_-(t_i)={\mathbf
  p}=p(\cos{\theta_p},\sin{\theta_p})=\mathrm{const}$ for all
$i=1,2,3,\ldots$ and $\left|\theta_0-\theta_p\right|\ll 1$. Then the
system \eqref{eq:d0} takes the form
\begin{equation}
  \label{eq:d12}
  \left\{
    \begin{split}
      &
      \rho_{i+1}=\rho_i+\left(p\cos(\theta_p-\theta_i)-2\right)\Delta
      t, \\ & \theta_{i+1}=\theta_i+\frac{p\Delta
        t}{\rho_i}\sin{\left(\theta_p-\theta_i\right)}.
    \end{split}
  \right.
\end{equation}
Linearization of the system \eqref{eq:d12} in $\theta_i$ around
$\theta_p$ gives
\begin{equation}
  \label{eq:d1}
  \left\{
    \begin{split}
      & \rho_{i+1}=\rho_i+(p-2)\Delta t, \\ &
      \theta_{i+1}-\theta_p=\left(1-\frac{p}{\rho_i}\Delta
        t\right)(\theta_i-\theta_p).
    \end{split}
  \right.
\end{equation}
If $p<2$, then $\rho$ should vanish after
$N:=\left\lfloor\rho_0{(2-p)}^{-1}{\Delta t}^{-1}\right\rfloor$ time
steps. Possible problems with stability may, therefore, arise when $N$
is large, that is when $p$ is close to $2$. This situation corresponds
to a local equilibrium of the grain boundary network when all angles
between adjacent edges are close to $120^\circ$ in the vicinity of the
disappearing edge. Consider the worst case scenario when $p$ remains
close to $2$ for a long time (this is unlikely in real simulations as
the motion of other vortices will likely cause $p$ to change). Suppose
that $\Delta t$ satisfies the condition $0<p\,\Delta t /\rho_0<1$,
then $$\Delta t=\frac{\alpha\rho_0}{p},$$ where
$0<\alpha<1$. Substituting this expression into \eqref{eq:d1}, the
second equation in \eqref{eq:d1} takes the form
\begin{equation}
  \label{eq:d2}
  \theta_{i+1}-\theta_p=\left(1-\frac{\alpha}{1-\frac{i}{x}}\right)\left(\theta_i-\theta_p\right),
\end{equation}
where $x=\frac{p}{\alpha(2-p)}$. Using the same notation, we have
$N=\lfloor{x}\rfloor$. It now follows that
\begin{equation}
  \label{eq:d3}
  \theta_i-\theta_p=\left(\theta_0-\theta_p\right)\prod_{j=1}^i\left(1-\frac{\alpha}{1-\frac{j}{x}}\right),
\end{equation}
where $i=1,2,3,\ldots,\lfloor{x}\rfloor-1$ since we do not need to
determine $\theta$ when $\rho=0$. Suppose that $p\uparrow 2$, then
$x\to\infty$ and the magnitude of the factors in the product in
\eqref{eq:d3} is close to $1-\alpha$ when $j$ is small. On the other
hand, when $j$ is close to $\lfloor x\rfloor$,
\[\left|1-\frac{\alpha}{1-\frac{j}{x}}\right|\gg 1,\]
and the value of the product is largest when $i=\lfloor
x\rfloor-1$. Thus there are no issues with numerical stability if the
product
\begin{equation}
  \label{eq:d4}
  \Phi(\alpha,x):=\prod_{j=1}^{\lfloor x\rfloor-1}\left|1-\frac{\alpha}{1-\frac{j}{x}}\right|
\end{equation}
remains finite for large $x$. If $\Gamma(x)$ is the $\Gamma$-function,
in Appendix 2 we show that
\begin{equation}
  \label{eq:d5}
  \Phi(\alpha,x)=\frac{1}{\pi}\frac{\Gamma(1+x-\lfloor x \rfloor)\Gamma((1-\alpha)x)\Gamma(\lfloor x \rfloor-x+\alpha x)}{\Gamma(x)}\sin{\pi\left((1-\alpha)x-\lfloor(1-\alpha)x\rfloor\right)},
\end{equation}
as long as $\alpha x>1$. Since
$\max_{1\leq\lambda\leq2}\Gamma(\lambda)=1$, we have that
\[0\leq\Phi(\alpha,x)\leq\frac{1}{\pi}\frac{\Gamma((1-\alpha)x)\Gamma(\lfloor
  x \rfloor-x+\alpha x)}{\Gamma(x)},\] for all $x>0$ as long as
$\alpha x>1$. Further, if $0<\alpha<1$ is fixed and $x\gg1$, then the
monotonicity and asymptotics of $\Gamma$ for large values of its
argument imply that
\[0\leq\Phi(\alpha,x)\leq\frac{1}{\pi}\frac{\Gamma((1-\alpha)x)\Gamma(\alpha
  x)}{\Gamma(x)}\sim{\left(\alpha^\alpha{\left(1-\alpha\right)}^{1-\alpha}\right)}^x\to0\
\mathrm{as}\ x\to0.\] Then $\theta_{N-1}\to 0$ as $p\to2$ and the
numerical scheme is stable.

\section{Vertex code algorithm description}
\label{alg}

The main algorithm can be decomposed into two parts: discrete and
continuous, describing topological transitions and motion of triple
junctions, respectively. Both of these processes depend on the time
resolution $\Delta t$.  The discrete component of the algorithm
detects and carries out topological transitions within
$\left[t,t+\Delta t\right]$, while the continuous component evolves
the triple junctions from time $t$ to time $t+\Delta t$. The procedure
is described in details in Algorithm 1 below.

\begin{algorithm}
  \caption{Main algorithm.}
  \label{alg:Main}
  \begin{algorithmic}[1]
    \State GRAINS(0,:) $\gets$ Initial configuration at time $t=0$.
    \State $\Delta t_0 \gets $ Upper bound for $\Delta t$.  \State
    $\Delta t \gets \Delta t_0$ \While{Stopping Criteria Not
      Satisfied} \State $t_{ext} \gets $ Compute extinction times for
    all grain boundaries.  \State $\mathcal{L}_1 \gets $ Select grain
    boundaries such that $0<t_{ext} < \Delta t$ \If{$\mathcal{L}_1$ is
      empty} \State GRAINS($t+\Delta t$,:) $\gets$ Evolve grain data
    structure GRAINS($t$,:) to time $t+\Delta t$.  \Else \State
    $\mathcal{L}_1 \gets $\text{Sort} $\mathcal{L}_1$ \text{in
      increasing order of} $t_{ext}$ \State $\text{tmp} \gets \{\}$,
    $\mathcal{L}_2 \gets \{\}$, $l_2 \gets \{\}$ \For{$l \in
      \mathcal{L}_1$} \If{$\text{triple junctions}(l) \cap
      \text{tmp}=\emptyset$} \State $\mathcal{L}_2 \gets$ Add grain
    boundary $l$.  \State $\text{tmp} \gets$ Add $\text{triple
      junctions}(l)$.  \State $l_2 \gets l$.  \Else \State $\Delta t
    \gets \displaystyle{\frac{t_{ext}(l_2)+ t_{ext}(l)}{2}}$.  \State
    \textbf{Break-loop}.
    \EndIf
    \EndFor
    \For{$m \in \mathcal{L}_2$} \State $nl$, $nr$ $\gets$ Numbers of
    sides of the grains adjacent to the grain boundary $m$.
    \If{$nl=3$ {\bf and} $nr=3$} \State \Return ERROR \ElsIf{$nr=3$
      {\bf or} $nl=3$} \State Remove the adjacent 3-sided
    grain. Replace it with an edge, as shown in
    Fig.~\ref{fig:grain_removal} \Else \State Flip the grain
    boundary $m$, as shown in Fig.~\ref{fig:flipping_full}.
    \EndIf
    \State GRAINS($t+\Delta t,m$) $\gets$ Evolve GRAINS($t,m$) to time
    $t+\Delta t$.
    \EndFor
    \State GRAINS($t+\Delta t,\neg \mathcal{L}_2$) $\gets$ Evolve
    GRAINS($t,\neg \mathcal{L}_2$) to time $t+\Delta t$.  \State
    $\Delta t \gets \Delta t_0$.
    \EndIf
    \EndWhile
  \end{algorithmic}
\end{algorithm}

The continuous part of the evolution is relatively straightforward and can
be achieved by solving a system of ODEs by means of any available
numerical scheme, e.g. MATLAB {\tt ode45} routine. The principal aim
of this implementation is to isolate and fully resolve topological
transitions. This is done by dynamically adapting the time step
$\Delta t$ using the formula \eqref{eq:t_lambda_in_use} within the
following procedure.

\begin{figure}[!t]
  \centering
  \begin{subfigure}[b]{0.3\textwidth}
    \centering
    \includegraphics[height=0.6\textwidth]{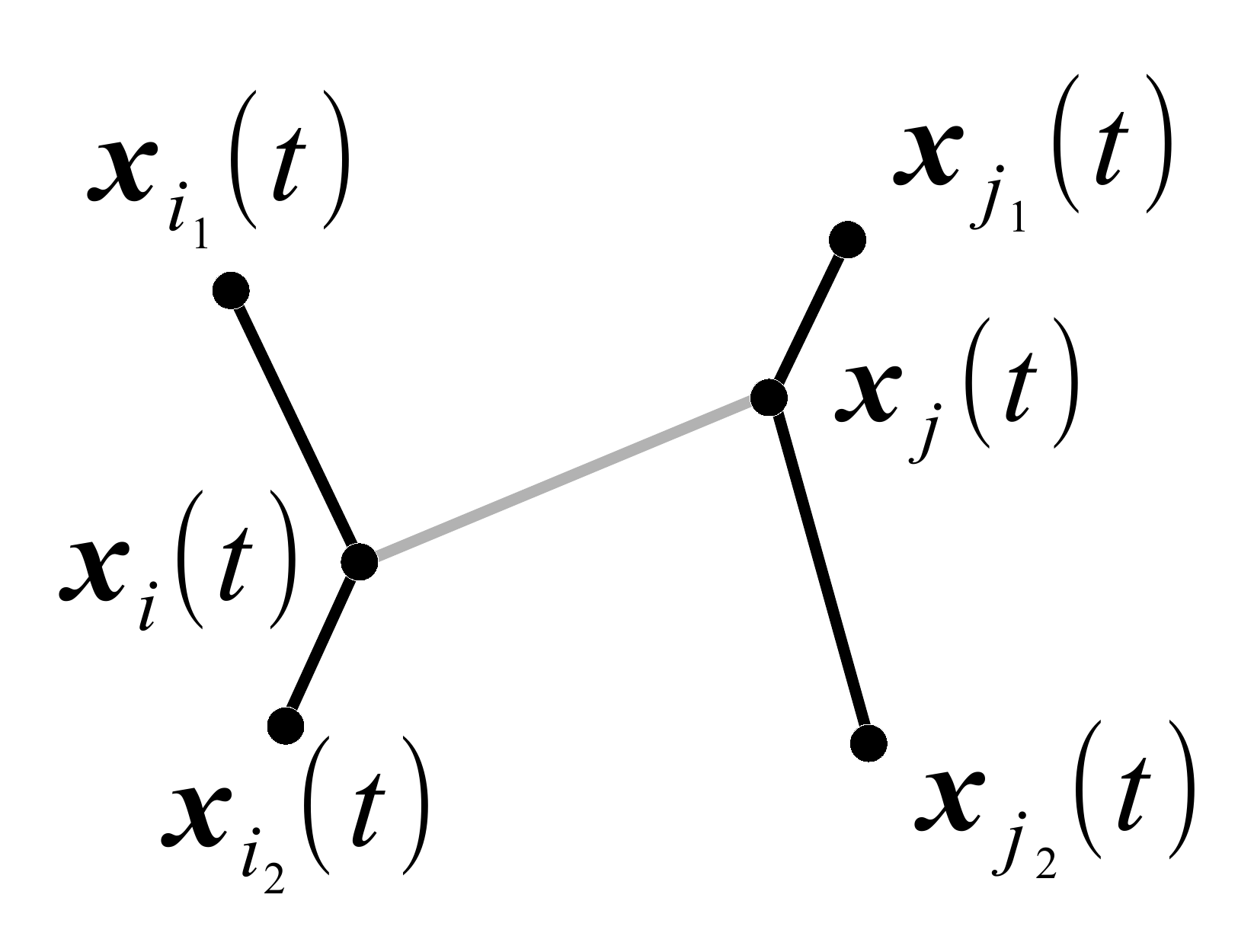}
    \caption{Step 1: Detect a neighbor switching event.}
    \nonumber
  \end{subfigure}
  \hspace{3em}
  \begin{subfigure}[b]{0.3\textwidth}
    \centering
    \includegraphics[height=0.6\textwidth]{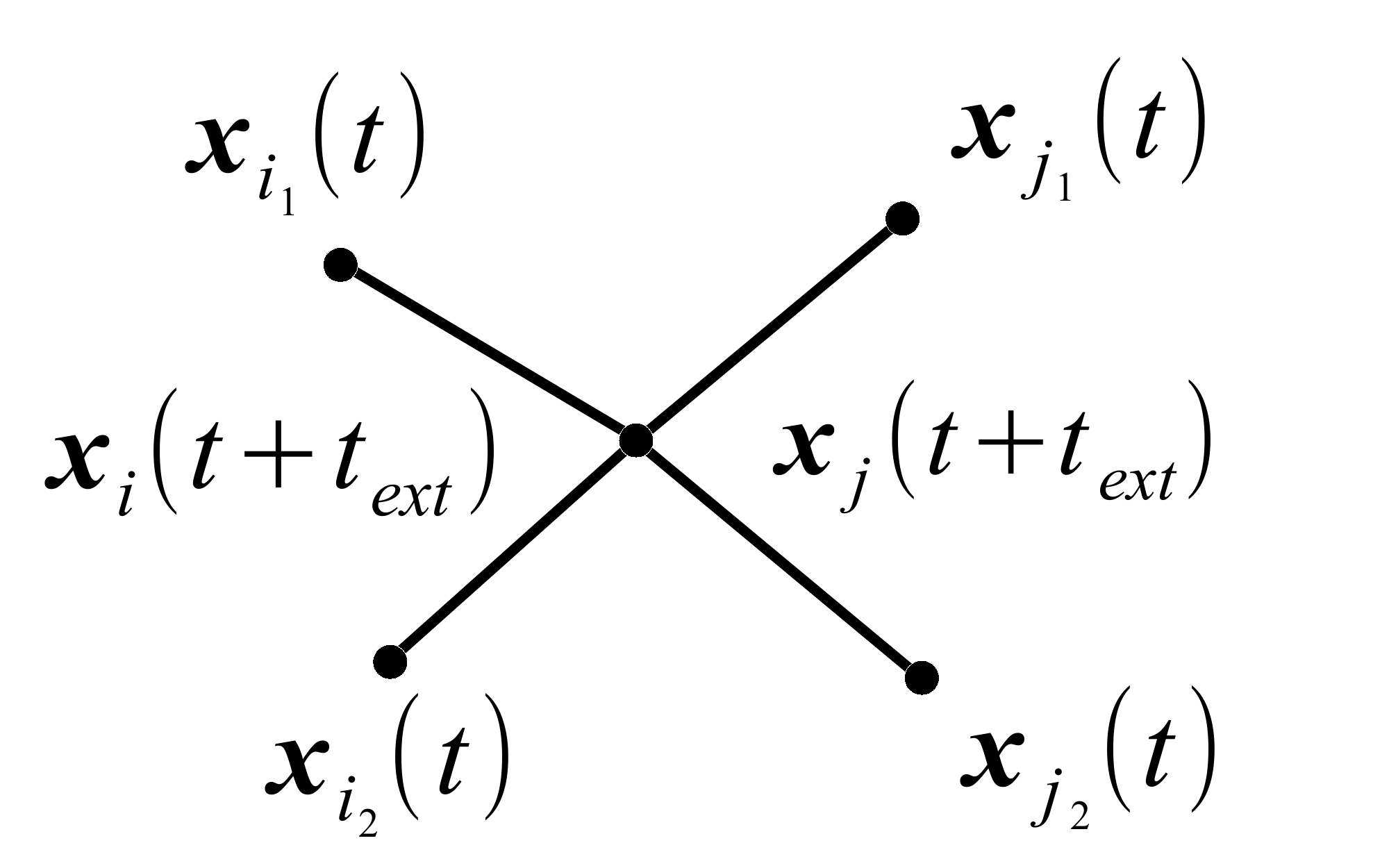}
    \caption{Step 2: Advance the colliding triple junctions $i$ and
      $j$ to their positions at time $t+t_{ext}$.}
  \end{subfigure}
  
  \begin{subfigure}[b]{0.3\textwidth}
    \centering
    \includegraphics[height=0.6\textwidth]{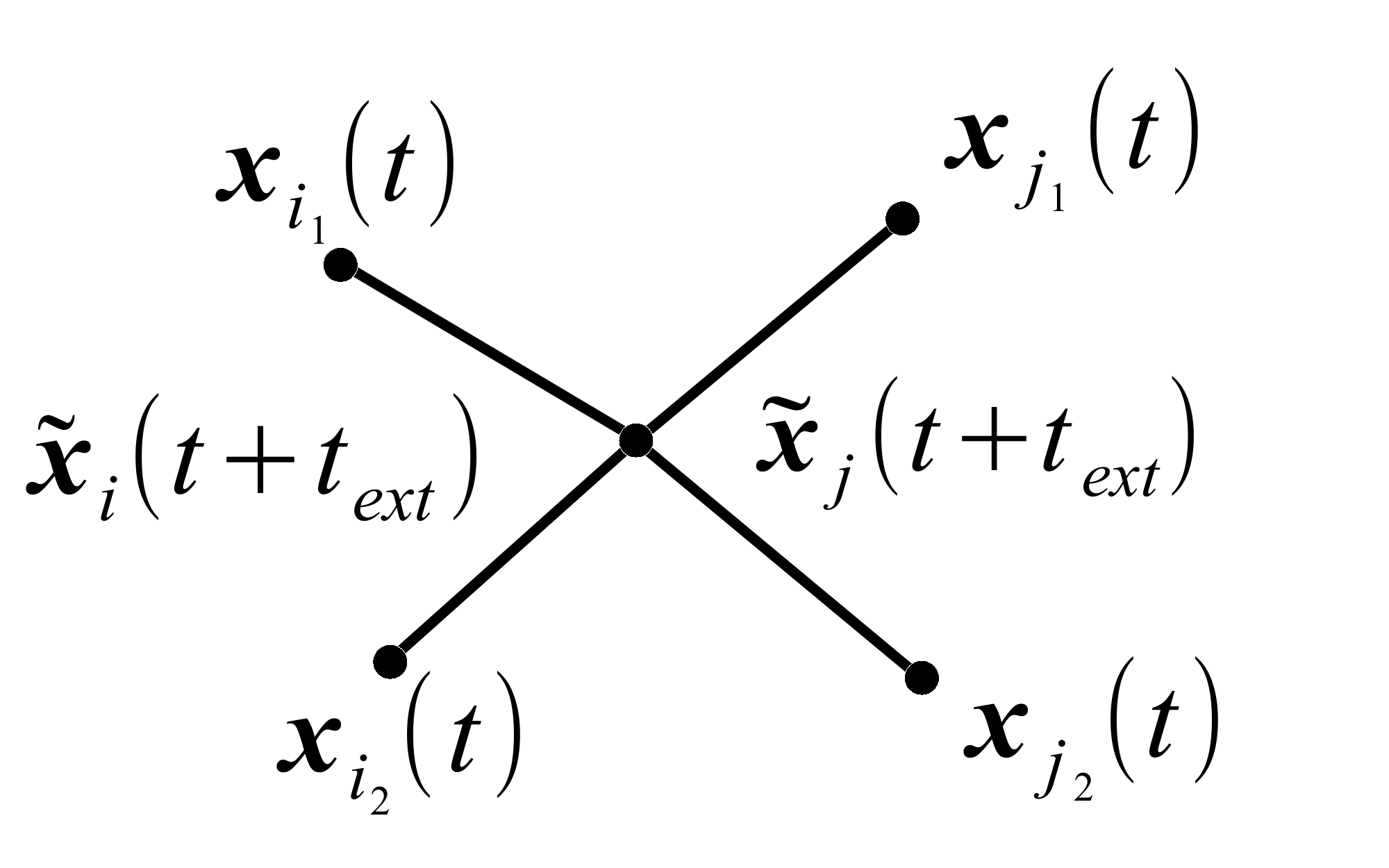}
    \caption{Step 3: Exchange the neighbors of colliding triple
      junctions $i$ and $j$.}
  \end{subfigure}
  \hspace{3em}
  \begin{subfigure}[b]{0.3\textwidth}
    \centering
    \includegraphics[height=0.6\textwidth]{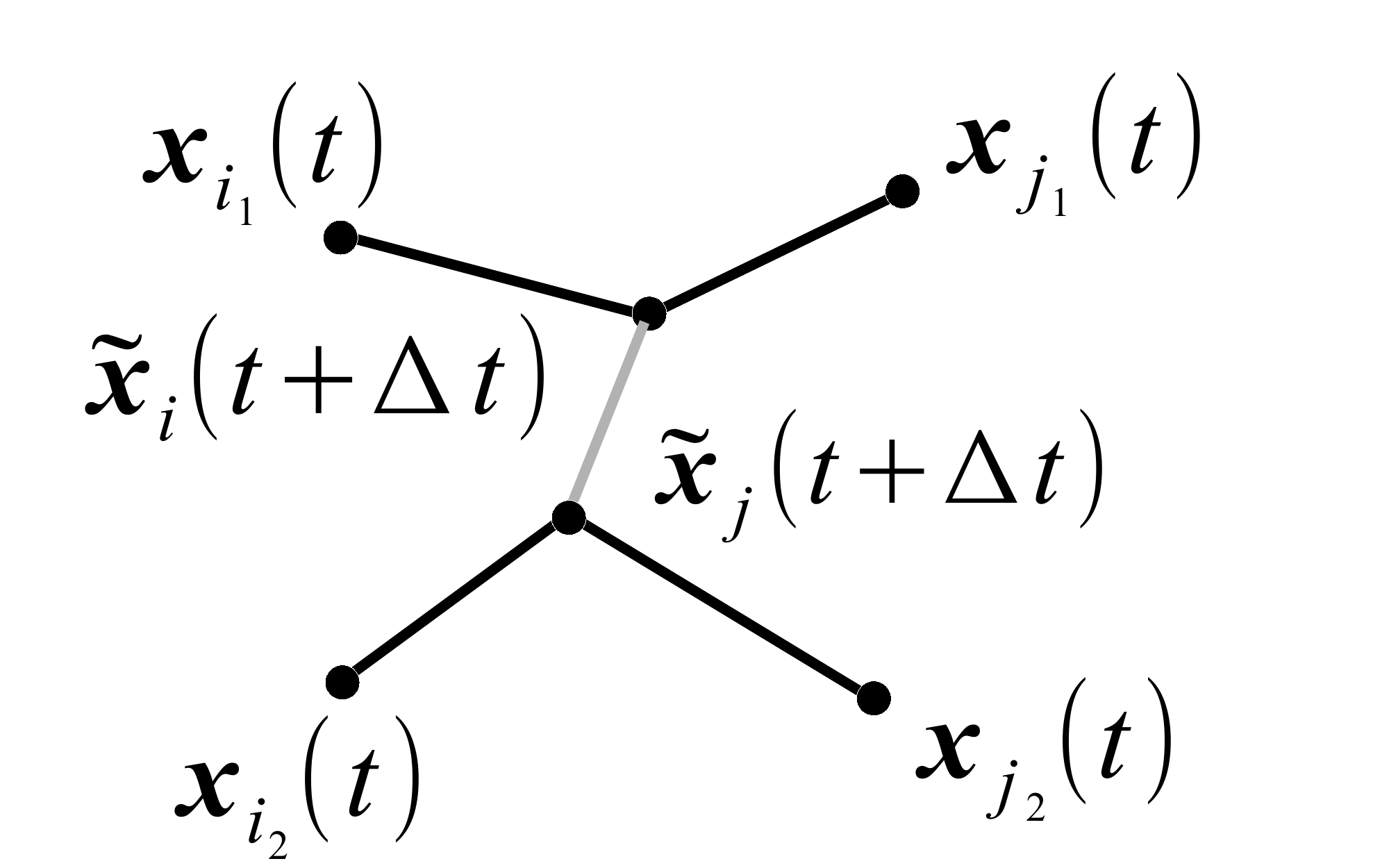}
    \caption{Step 4: Advance positions of the triple junctions $i$ and
      $j$ from time $t+t_{ext}$ to time $t+\Delta t$.}
  \end{subfigure}
  
  \begin{subfigure}[b]{0.3\textwidth}
    \centering
    \includegraphics[height=0.55\textwidth]{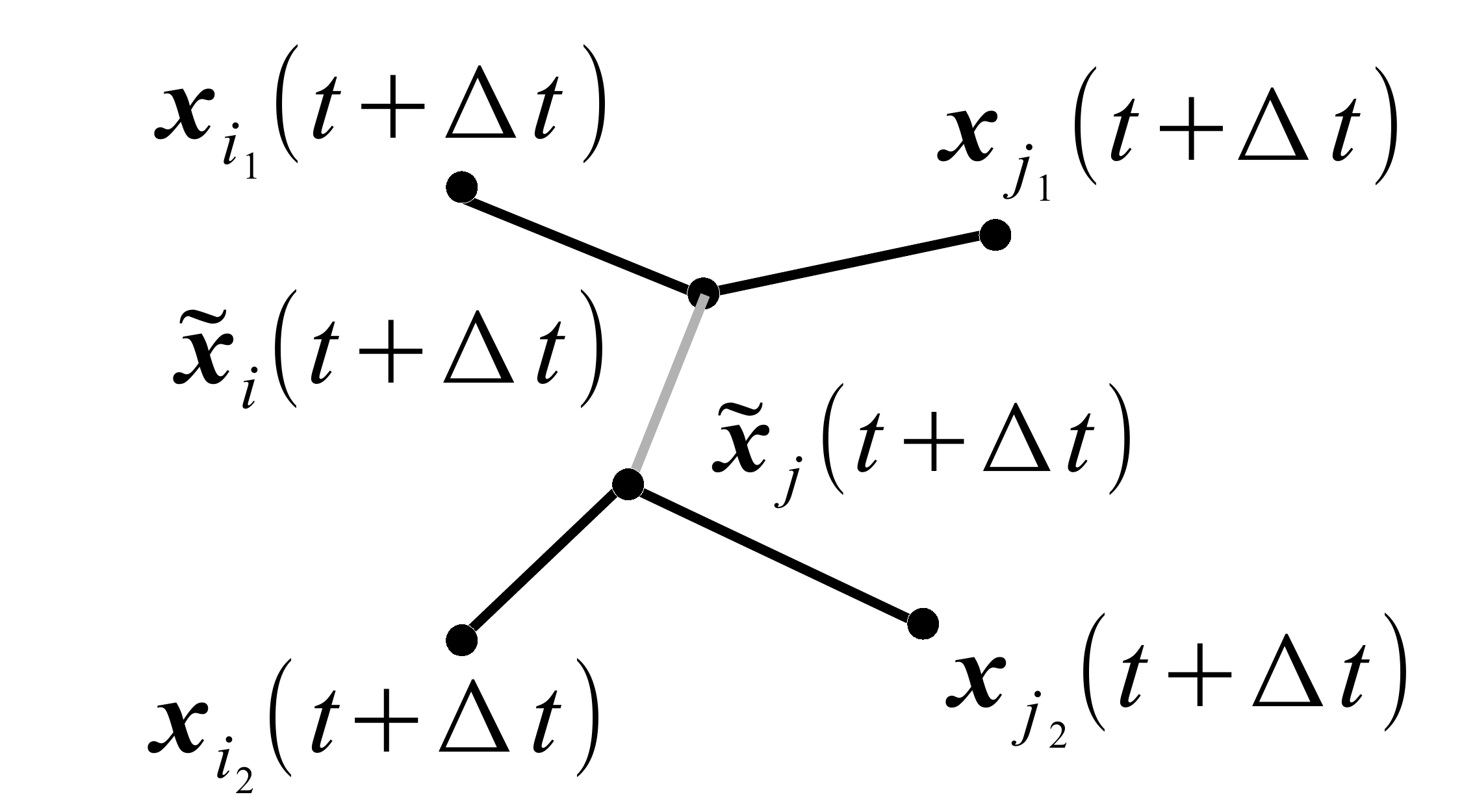}
    \caption{Step 5: Fix positions of junctions $i$ and $j$ and move
      the remaining junctions to new positions at time $t+\Delta t$.}
  \end{subfigure}
  \caption{Neighbor switching algorithm.}
  \label{fig:flipping_full}
\end{figure}

Initially, $\Delta t$ is set equal to a prescribed value of $\Delta
t_0$. On each time step, the algorithm estimates extinction times for
all grain boundaries and selects only those which fall in the interval
$[t,t+\Delta t]$. Then the corresponding grain boundaries are sorted
according to their extinction times. Next we move along this list and
record those boundaries whose vertices have not yet been
encountered. This process continues until either all vertices are
exhausted, or a boundary with an already recorded vertex has been
detected. In the latter case, the process stops and the time step is
adjusted to only allow extinction of the boundaries that have been
recorded. This ensures that there is spatial separation between
topological transitions. Once the time step has been adapted,
transitions that involve all recorded boundaries are implemented and
the rest of the network is allowed to evolve in a continuous
fashion. Then the time step is reset to the original value of $\Delta
t_0$ and the procedure is repeated until 80\% of the grains have been
eliminated.

\begin{figure}[!t]
  \centering
    \includegraphics[height=0.2\textwidth]{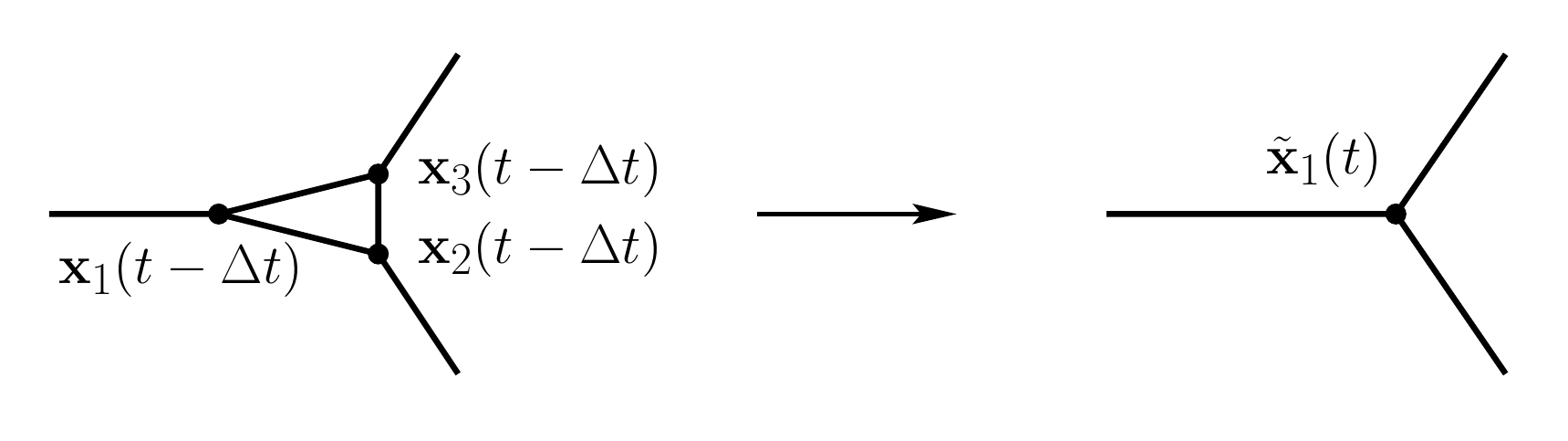}
  \caption{Configurations resulting in grain removal.  The location of
    the new triple junction $\widetilde{\mathbf{x}}_1(t)$ is set to be
    the collision point for the two triple junctions that collide
    first.}
  \label{fig:grain_removal}
\end{figure}

There are two types of transition events: (1) neighbor switching, and (2) 
grain removal. From the detection point of view
outlined above, these events are indistinguishable. On the other hand,
their resolution is completely different. At first, all detected
events are considered to be a neighbor switching event, as shown
in Fig.~\ref{fig:flipping_full}.  However, when one of the two
grains adjacent to the disappearing grain boundary has only three
sides, we proceed to remove that grain, as depicted in
Fig.~\ref{fig:grain_removal}. The neighbor switching is
done according to the rules discussed in
Section~\ref{sec:analysis_flipping_rule}. We assume that the
anisotropy is small enough so that a shrinking grain boundary cannot
be simulatneously adjacent to two 3-sided grains, and return an error
if this happens.

Note that only 3-sided grains are allowed to be removed. Thus, for a
5-sided grain to be able to disappear, it needs first to become a
4-sided grain and then a 3-sided grain. The number of sides of a given
grain can change either through a neighbor switching event, or as a consequence
of a neighboring grain disappearance. Finally, when a switching of neighbors is performed, 
the length of the new grain boundary is
computed to be proportional to $\Delta t-t_{ext}$, where $t_{ext}$ is
the extinction time of the grain boundary in question.

\section{Numerical convergence study}

\begin{figure}[!t]
    \centering
    \includegraphics[height=6cm]{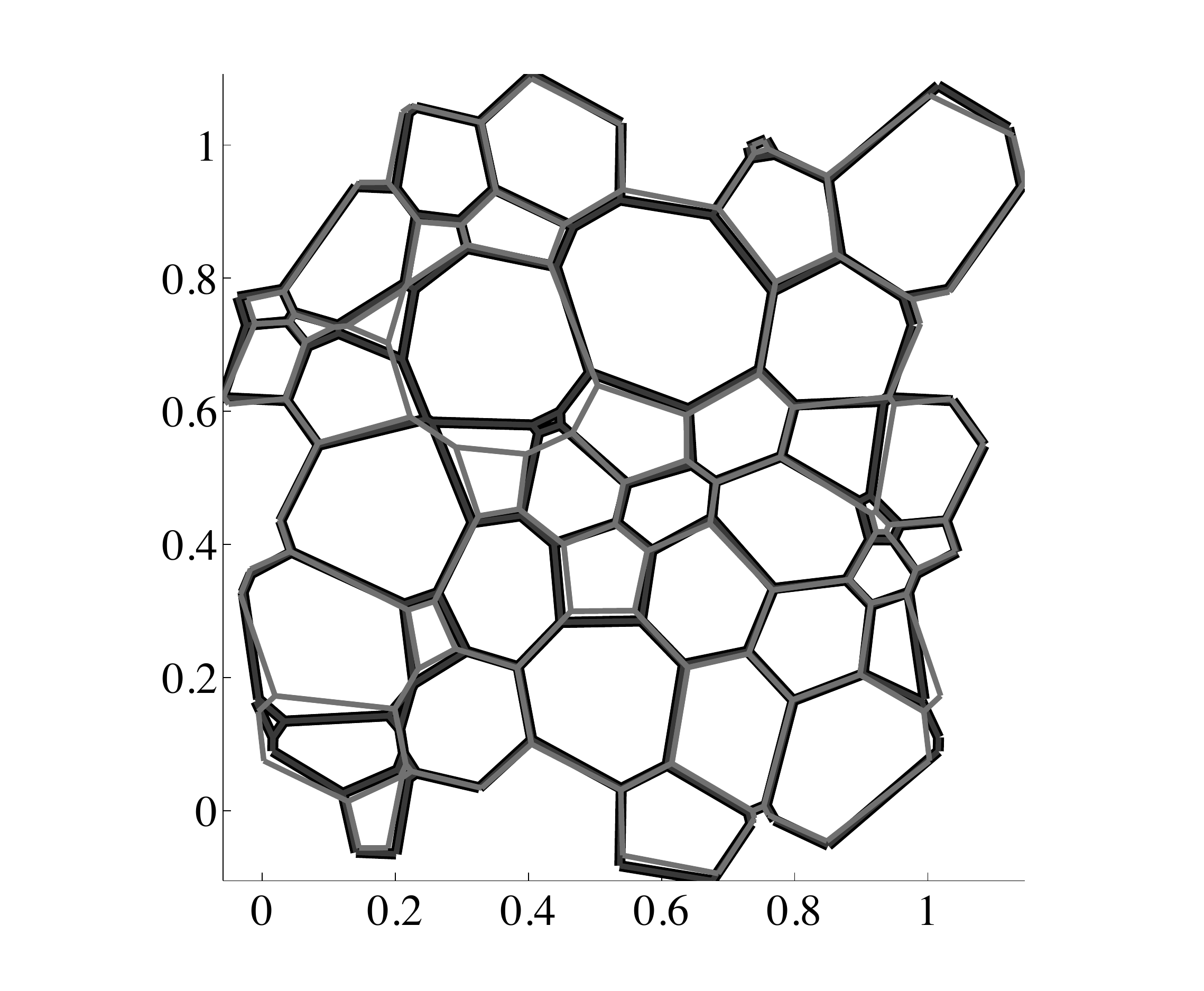}
    \caption{Grain boundary networks evolved from the same configuration with 200 grains up to t=0.45 when approximately 80\% of the grains were removed. The simulations were run with three different maximum time steps: $\Delta t_0=10^{-2}$ (light grey), $10^{-5}$ (dark grey), and $10^{-6}$ (black).}
    \label{fig:c1}
  \end{figure}

\begin{figure}[!t]
    \centering
    \includegraphics[height=5cm]{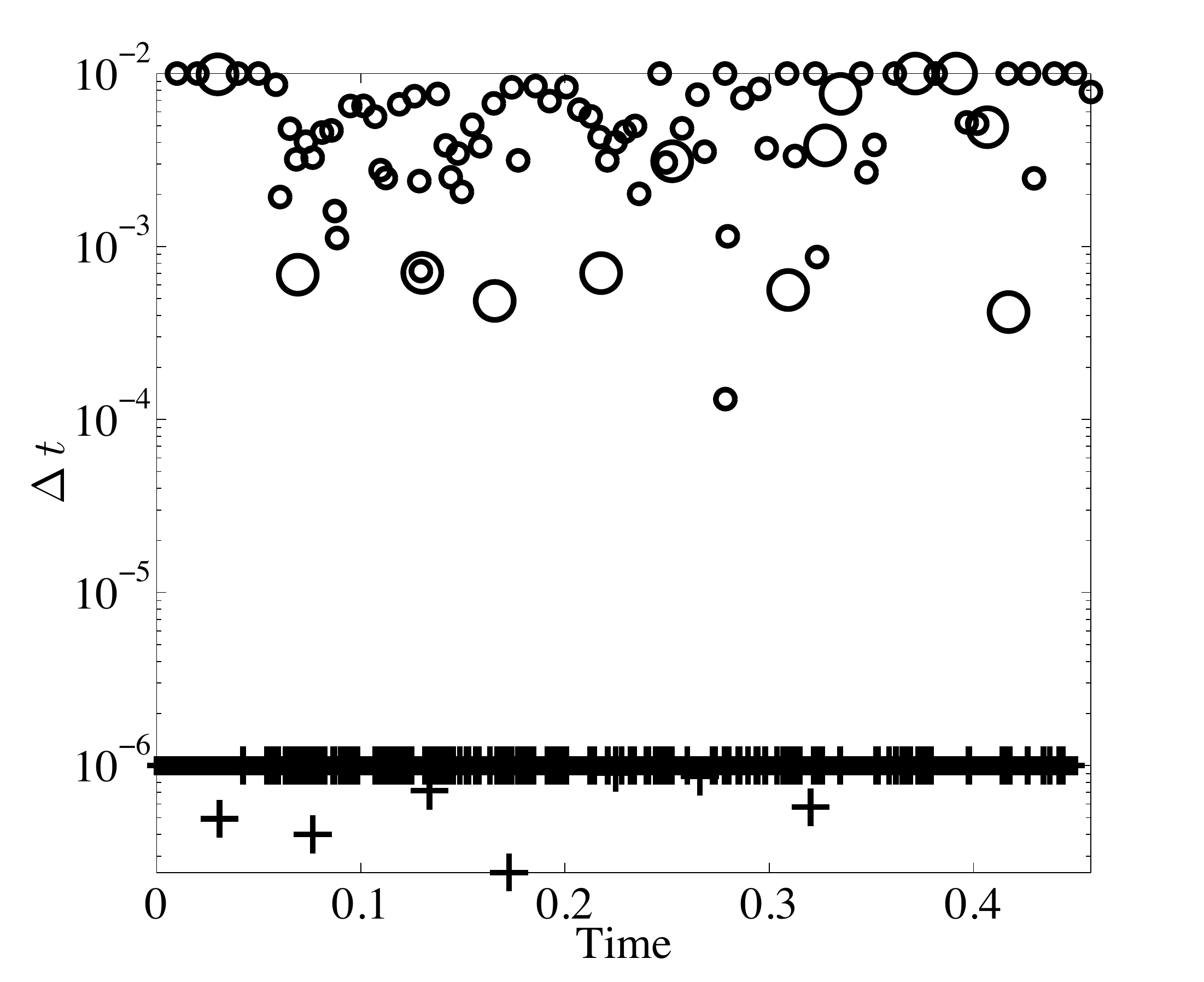}
    \caption{Evolution of the timestep for $\Delta t_0=10^{-2}$ ($\circ$) and $\Delta t_0=10^{-7}$ ($+$). Larger markers represent the instances when the decrease in time step was caused by grain removal in the absence of neighbor switching.}
    \label{fig:c2}
\end{figure}

Although a rigorous analytical investigation is beyond the scope of
this paper, in this section we present a numerical study subjecting
the proposed algorithm to several tests for accuracy and
convergence. The first test is designed to check how well the numerical
procedure is able to handle topological transitions via time-step
adaptation. For this purpose, we compare the results of several
simulations using different values of the maximum time step $\Delta
t_0$ in Figs.~\ref{fig:c1}-\ref{fig:c2}. Fig. \ref{fig:c1} shows grain boundary
structures obtained from the same initial configuration with $200$ grains 
at $t=0.45$ when approximately $80\%$ of the grains 
were removed. Three different simulations with maximum time step sizes 
$\Delta t_0=10^{-2},\ 10^{-5},\mbox{ and }10^{-6}$ were performed. 
It is evident that, with the exception of the
grain structure obtained when $\Delta t_0=10^{-2}$---the coarsest
maximum time step---all grain boundaries lie directly on top of each
other.  In Fig.~\ref{fig:c2} we plot the dependence between the
simulation time and the size of time step for $\Delta t_0=10^{-2}\mbox{ and }10^{-6}$. 
The figure shows that fewer time step refinements are performed for smaller $\Delta t_0$ and that
for a very small $\Delta t_0$, refinements are only needed when a
grain removal occurs.

\begin{figure}[!t]
  \centering
    \includegraphics[height=5cm]{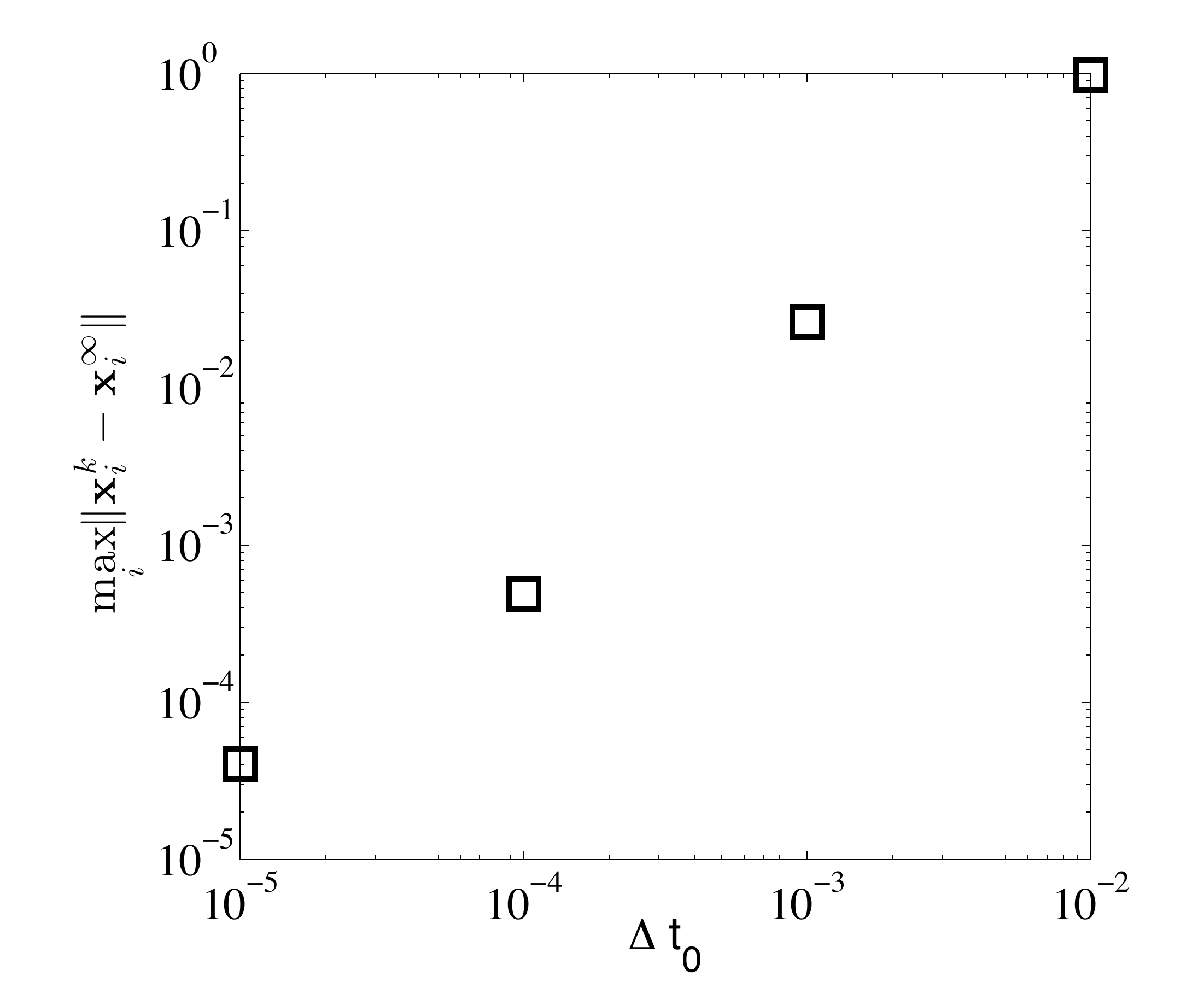}
    \caption{Deviation between positions of triple junctions corresponding to grain boundary networks for various $\Delta t_0$ compared to a well-resolved simulation with $\Delta t_0=10^{-7}$. All simulations started from the same initial data and were compared at the same absolute time $t=0.45$.}
    \label{fig:c3}
  \end{figure}

Next we test our algorithm for convergence using the following
measures. First we run a set of simulations corresponding to several
$\Delta t_0$. We consider the simulation with the smallest $\Delta
t_0=10^{-6}$ as ``well-resolved'' and benchmark the results of other
simulations against it. In Fig.~\ref{fig:c3} we consider the
difference between the positions of triple junctions produced by a
given and the well-resolved simulations at to the time $t=0.45$. Only the remaining 
common junctions are considered in this calculation. It can be seen that the error
decreases linearly toward zero as $\Delta t_0\to 0$. In fact, the
number of non-common triple junctions remaining at the time $t$
decreases to zero as well and it becomes exactly zero when $\Delta
t_0=10^{-4}$.

These tests allow the conclusion that the proposed algorithm successfully
handles both topological transitions and the grain boundary motion and
is numerically stable.


  \begin{figure}[!t]
    \centering
    \includegraphics[height=5cm]{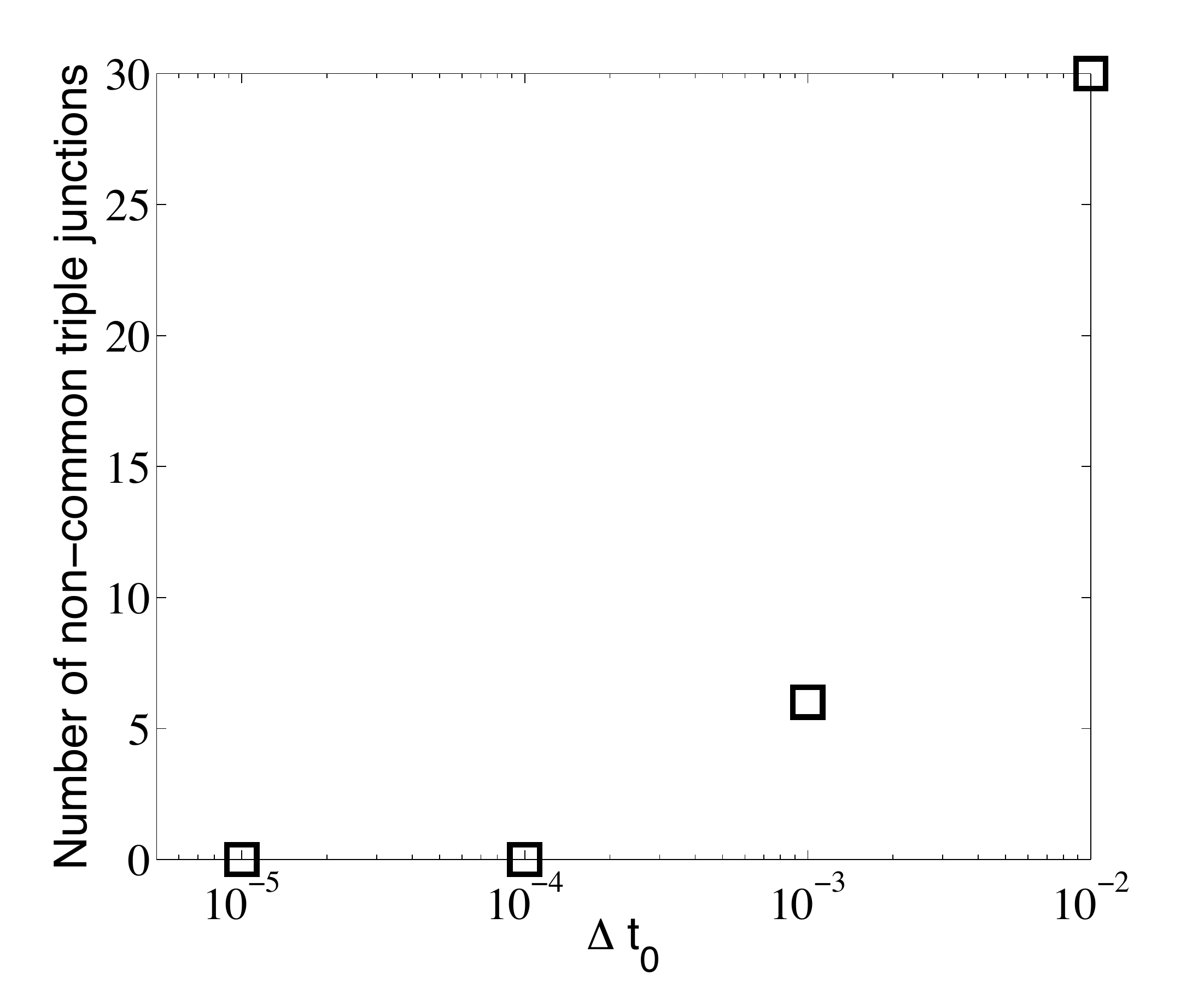}
    \caption{Number of non-common triple junctions between grain boundary networks for various $\Delta t_0$ and a well-resolved simulation with $\Delta t_0=10^{-6}$.  All simulations started from the same initial data and were compared at the same absolute time $t=0.45$.}
    \label{fig:c4}
  \end{figure}
  
\section{Numerical results}
\label{num}
  
      \subsection{Statistics}

      In this section we analyze the statistics for networks with isotropic and weakly anisotropic grain boundary energy. In all cases the network is initialized via a Voronoi
      tessellation using points uniformly distributed in the computational domain and assuming periodic boundary conditions. All networks initially contained $100,000$ grains and were evolved until $80\%$ of the grains were removed.

\begin{figure}[!t]
        \begin{subfigure}[b]{0.45\textwidth}
          \centering
          \includegraphics[height=5cm]{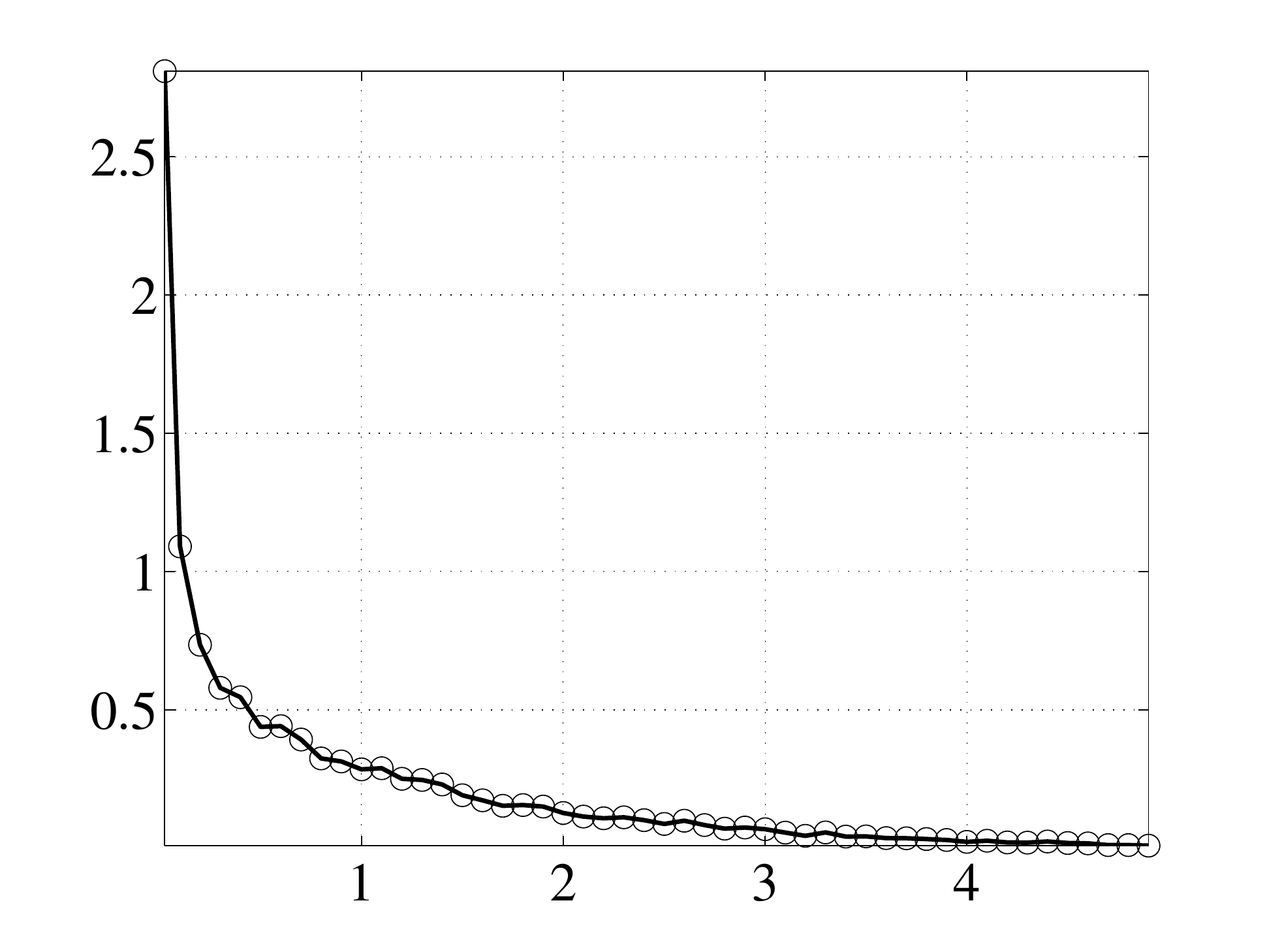}
          \caption{linear scale.}
          \label{fig:iso_stat_rel_area_lin}
        \end{subfigure}
        \begin{subfigure}[b]{0.45\textwidth}
          \centering
          \includegraphics[height=5cm]{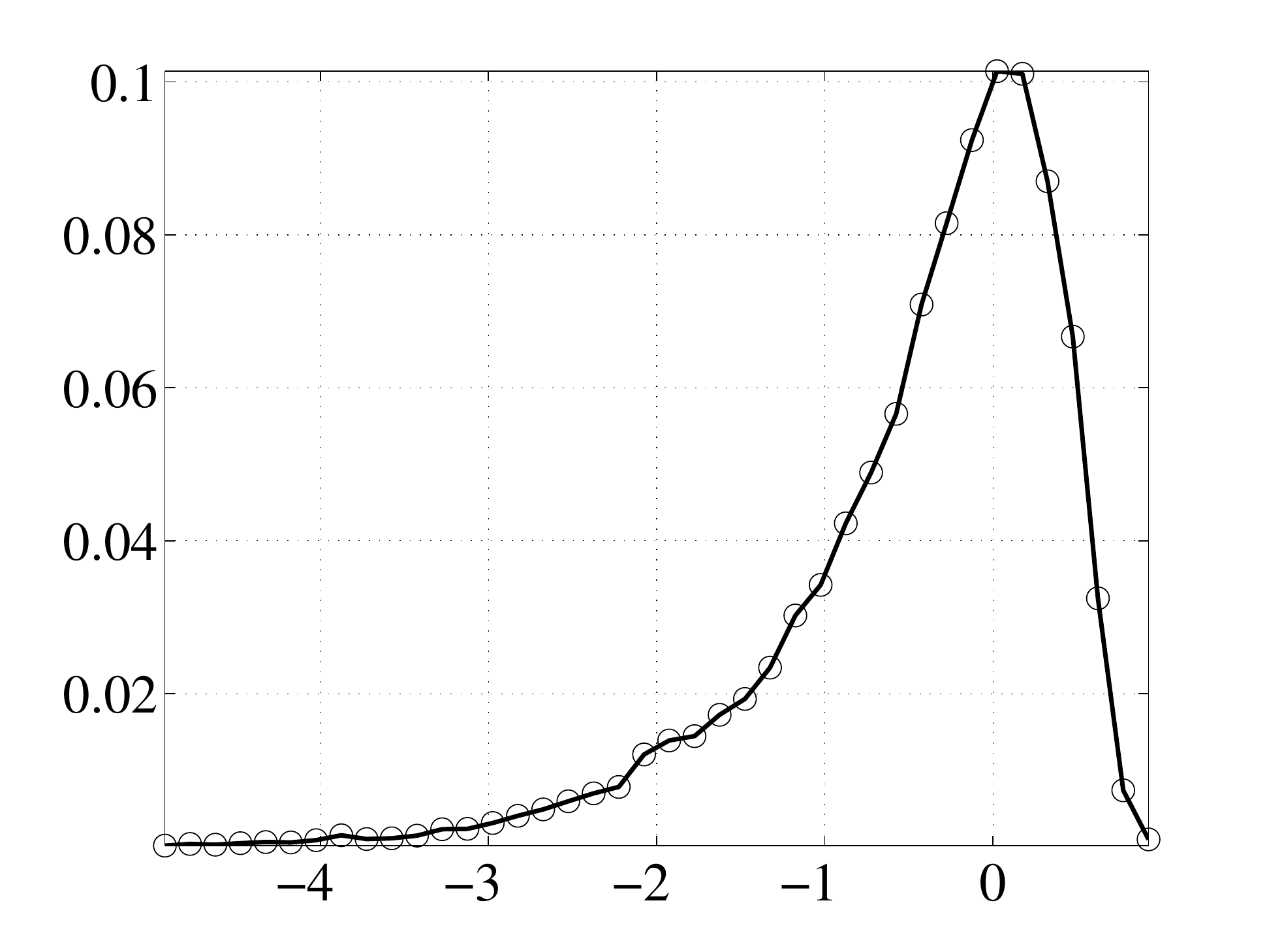}
          \caption{log scale.}
          \label{fig:iso_stat_rel_area_log}
        \end{subfigure}
        \caption{Relative grain area distribution in the isotropic network consisting of 20,000 grains that evolved from an initial configuration of 100,000 grains.}
        \label{fig:iso_stat}
      \end{figure}

      Fig.~\ref{fig:iso_stat} depicts the relative area distribution using linear and log scales. We observe that the
      distribution is skewed toward grains with smaller areas (this
      is emphasized in the log-scale plot). There is a notable difference
from statistics produced by a curvature-driven simulation; in fact,
       grains with smaller areas tend to have a smaller rate of area change  (Fig.~\ref{fig:dAdt_ave}). As the result 
       a grain boundary network evolved via triple 
       junctions motion tends to have more small grains.

\begin{figure}[!t]
        \begin{subfigure}[b]{0.45\textwidth}
          \centering
          \includegraphics[height=5cm]{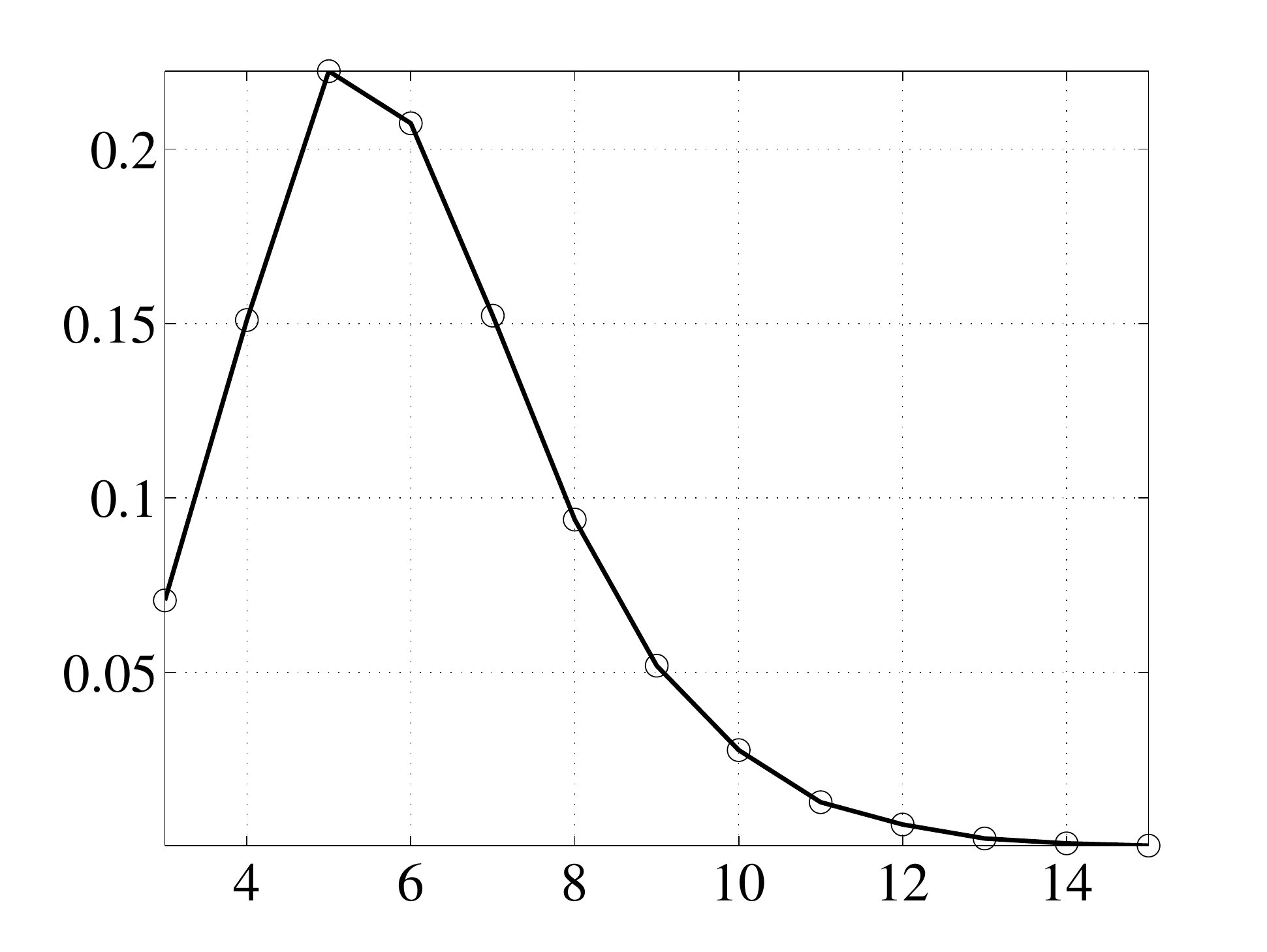}
          \caption{}
          \label{fig:iso_stat_n-sided}
        \end{subfigure}
        \begin{subfigure}[b]{0.45\textwidth}
          \centering
          \includegraphics[height=5cm]{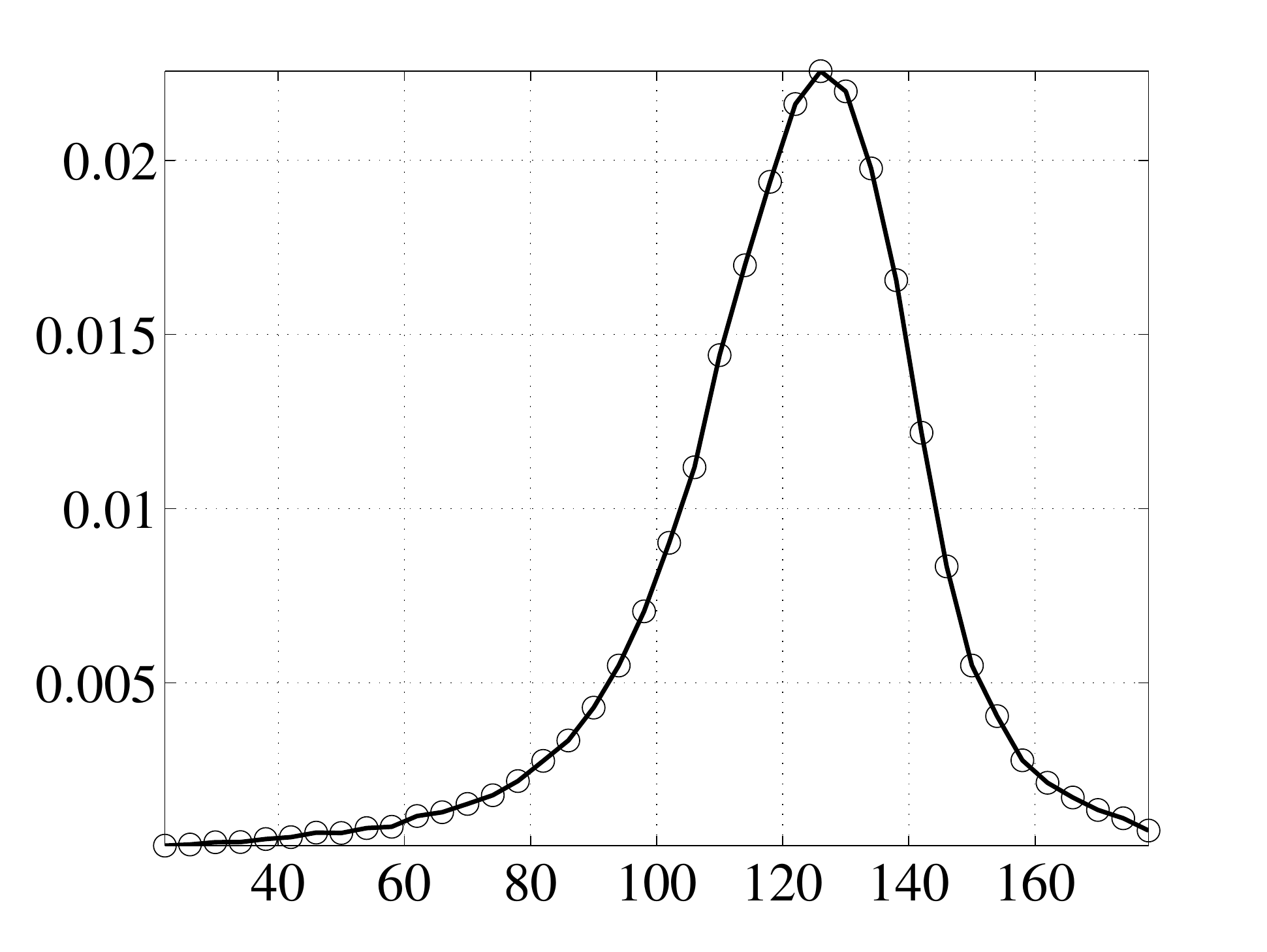}
          \caption{}
           \label{fig:iso_stat_dihedral}
        \end{subfigure}
                \caption{Distributions for the: (a) number of sides of a grain; and (b) dihedral angle.  The data is collected using the isotropic network of 20,000 grains that evolved from an initial configuration consisting of 100,000 grains.}
      \label{fig:iso_stat_1}
      \end{figure}

      The distribution of the number of sides per grain is shown in Fig.~\ref{fig:iso_stat_1}\subref{fig:iso_stat_n-sided} and demonstrates 
      a bias toward three-sided grains. We attribute this to the fact that grains need to go through a cascade of decreasing number of sides 
      before they can be removed,  i.e., a grain needs to become a three-sided grain before 
      it is allowed to disappear.  The largest proportion of grains are five-sided.  Another 
      important feature is that when the number of sides exceeds $n=12$, the value 
      of the probability density function is very small compared to the values when $n\leq 12$.  This is relevant to understanding statistics presented
       below because the sample size for large $n$ is small.

      The third statistic (Fig.~\ref{fig:iso_stat_1}\subref{fig:iso_stat_dihedral}) is the dihedral angle distribution.  This
      distribution does not seem to be centered at $120^\circ$ but instead
      shows a small shift toward angles larger than $120^\circ$.

\begin{figure}[!t]
        \begin{subfigure}[b]{0.45\textwidth}
          \centering
          \includegraphics[height=5cm]{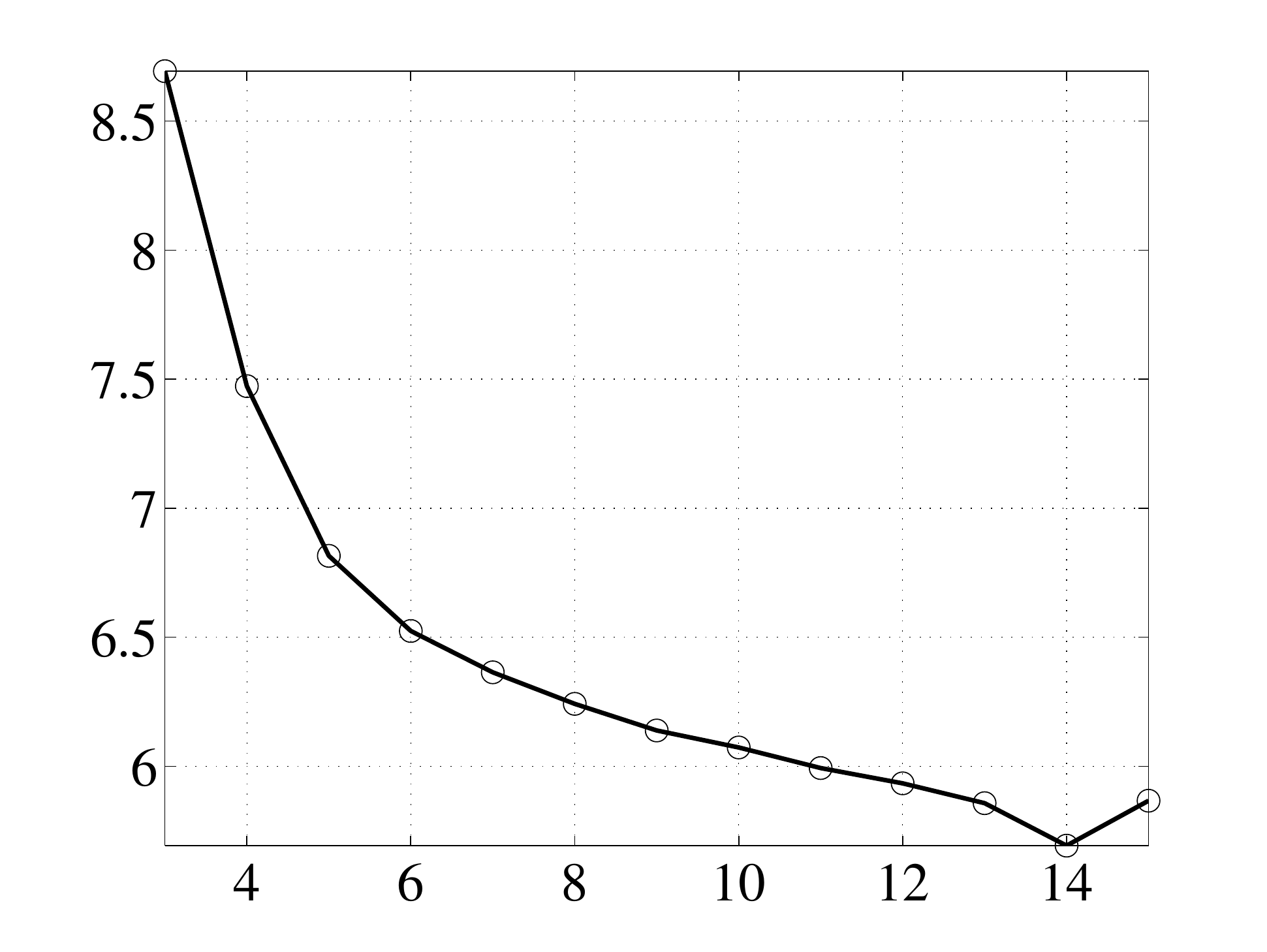}
          \caption{}
          \label{fig:iso_stat_ave_side_class_neigh}
        \end{subfigure}
        \begin{subfigure}[b]{0.45\textwidth}
          \centering
          \includegraphics[height=5cm]{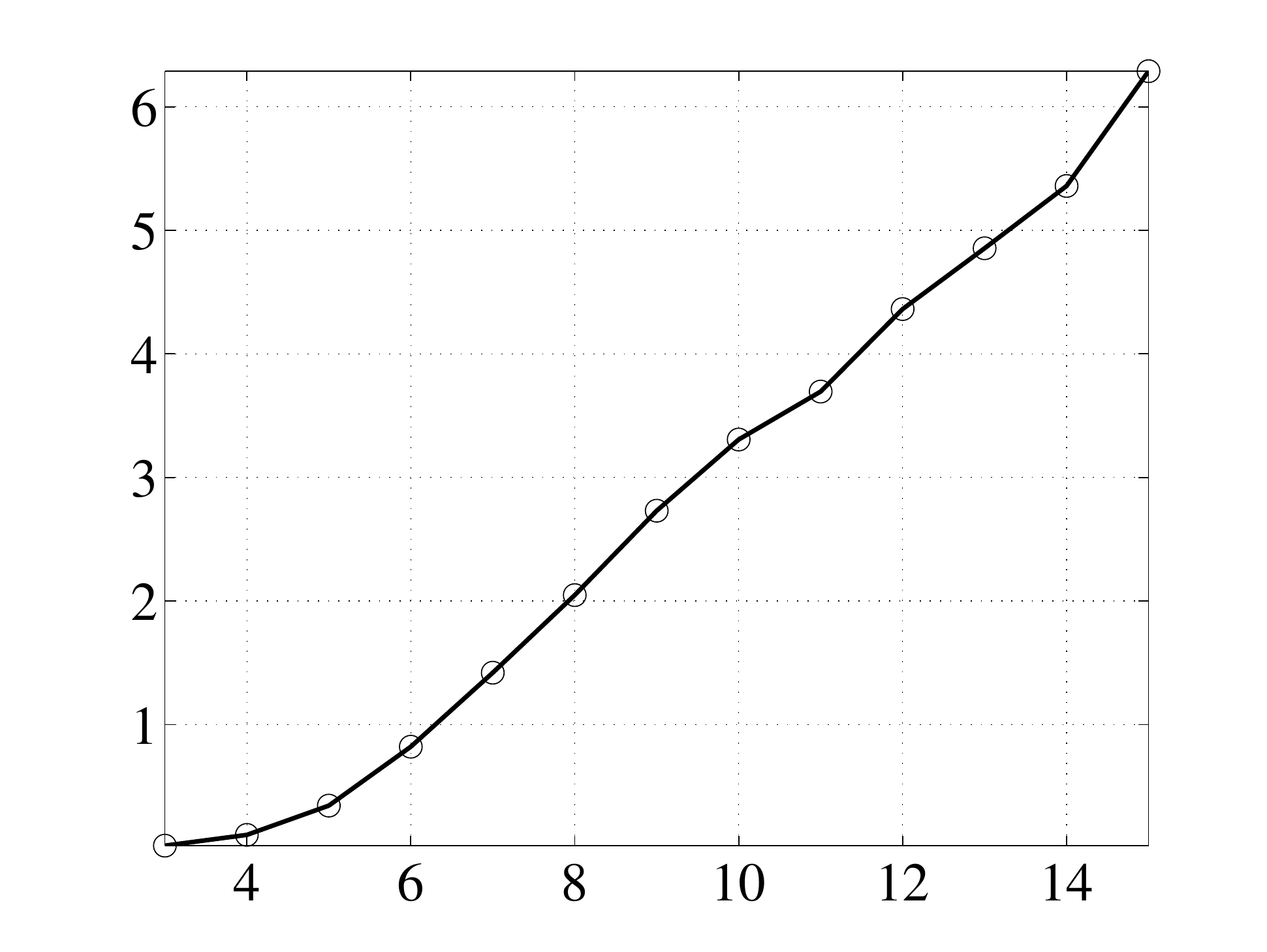}
          \caption{}
          \label{fig:iso_stat_red_ave_side_class}
        \end{subfigure}
        \caption{For grains with a given number of sides: (a) Average number of sides of neighboring grains; (b) Average relative area of neighboring grains. The data is collected using the isotropic network of 20,000 grains that evolved from an initial configuration consisting of 100,000 grains.}
        \label{fig:side_class}
      \end{figure}

\begin{figure}[!t]
        \begin{subfigure}[b]{0.45\textwidth}
          \centering
          \includegraphics[height=5cm]{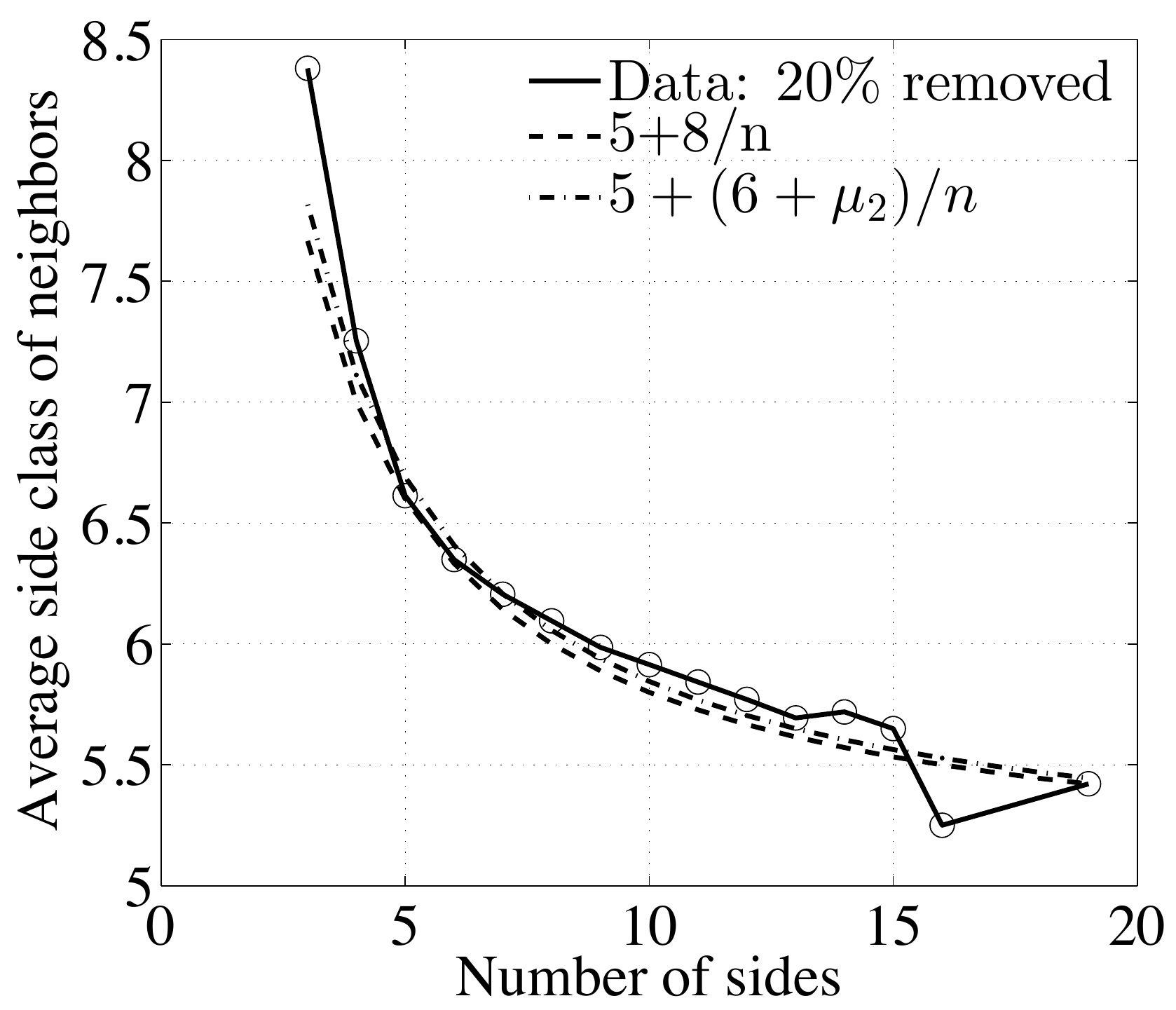}
          \caption{}
          \label{fig:iso_stat_aboav_20}
        \end{subfigure}
        \begin{subfigure}[b]{0.45\textwidth}
          \centering
          \includegraphics[height=5cm]{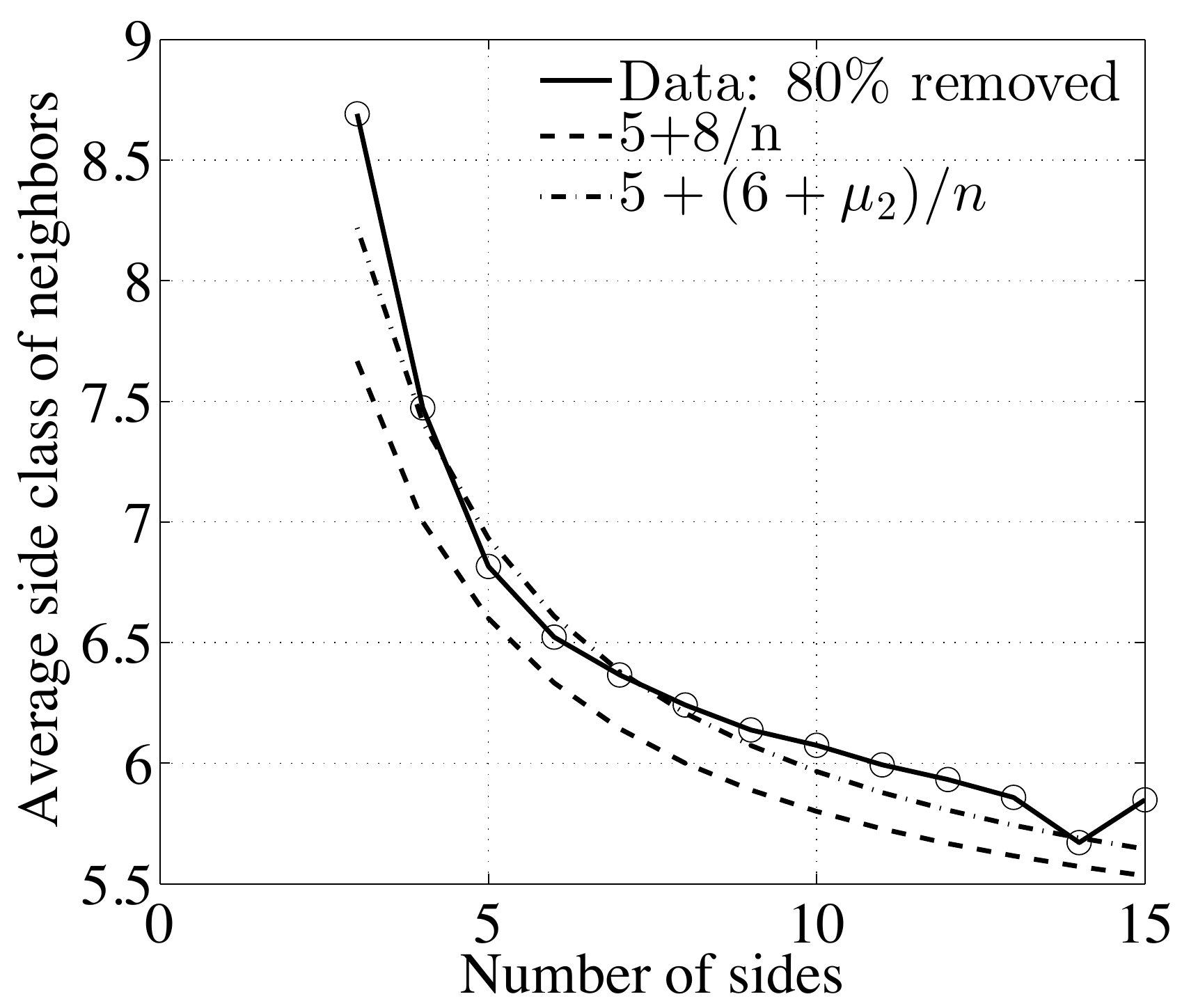}
          \caption{}
          \label{fig:iso_stat_aboav_80}
        \end{subfigure}
        \caption{Comparison with Aboav and Aboav-Weaire laws for an isotropic network. (a) $20\%$ of the grains have been removed; (b) $80\%$ of the grains have been removed.}
        \label{fig:aboav}
      \end{figure}

      Finally, we computed the average number of sides of neighbors for grains with a given number of sides (Fig.~\ref{fig:side_class}\subref{fig:iso_stat_ave_side_class_neigh}) as well as 
      the reduced average area of the neighbors (Fig.~\ref{fig:side_class}\subref{fig:iso_stat_red_ave_side_class}). The comparison to the empirical Aboav and Aboav-Weaire laws \cite{thompson} is shown in Fig.~\ref{fig:aboav} for the systems from which the 20\% and 80\% of the grains were removed, respectively. The Aboav-Weaire law appears to provide a better fit in both cases.

\begin{figure}[!t]
        \begin{subfigure}[b]{0.45\textwidth}
          \centering
          \includegraphics[height=5cm]{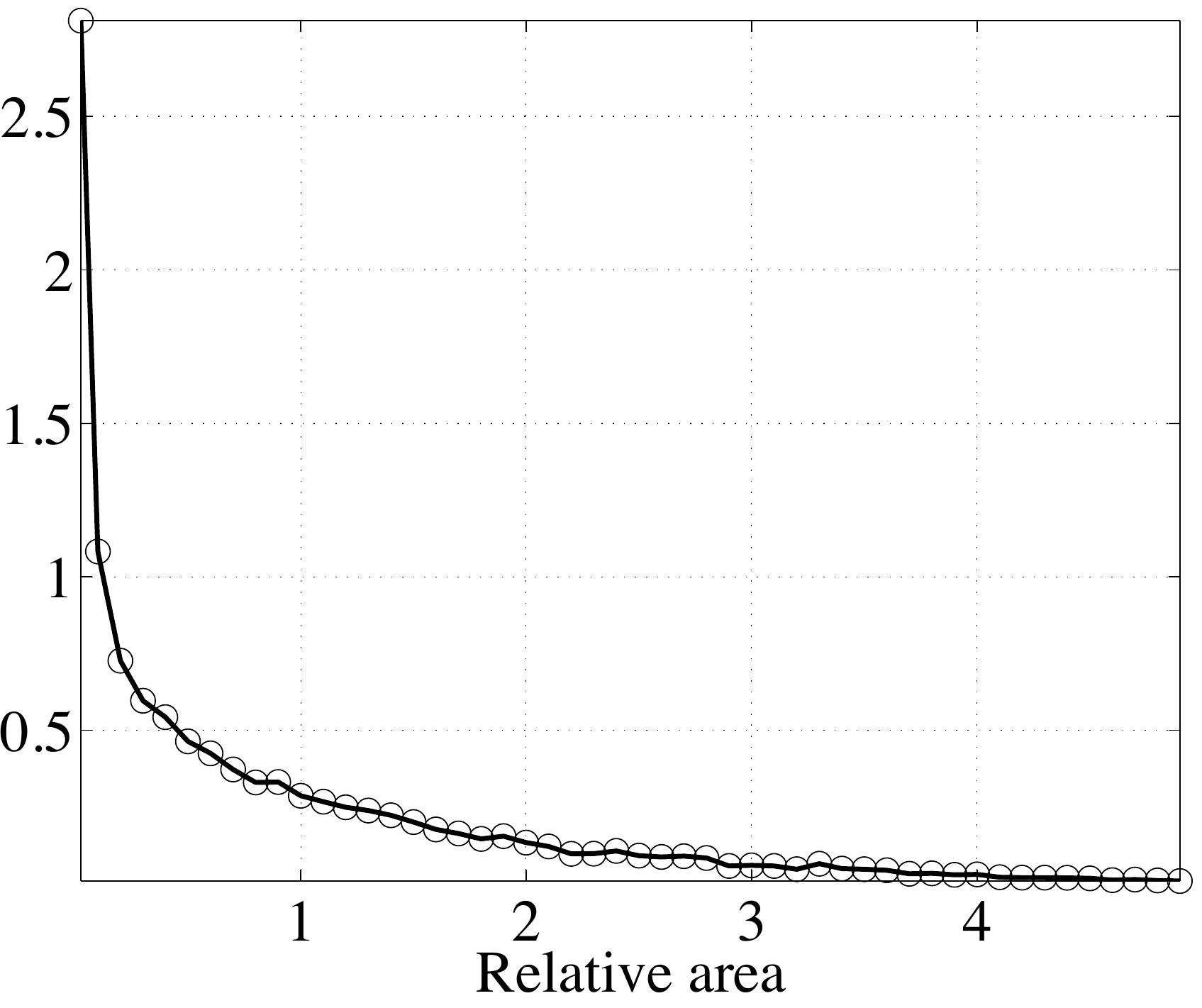}
          \caption{linear scale.}
          \label{fig:iso_stat_rel_area_lin_ani}
        \end{subfigure}
        \begin{subfigure}[b]{0.45\textwidth}
          \centering
          \includegraphics[height=5cm]{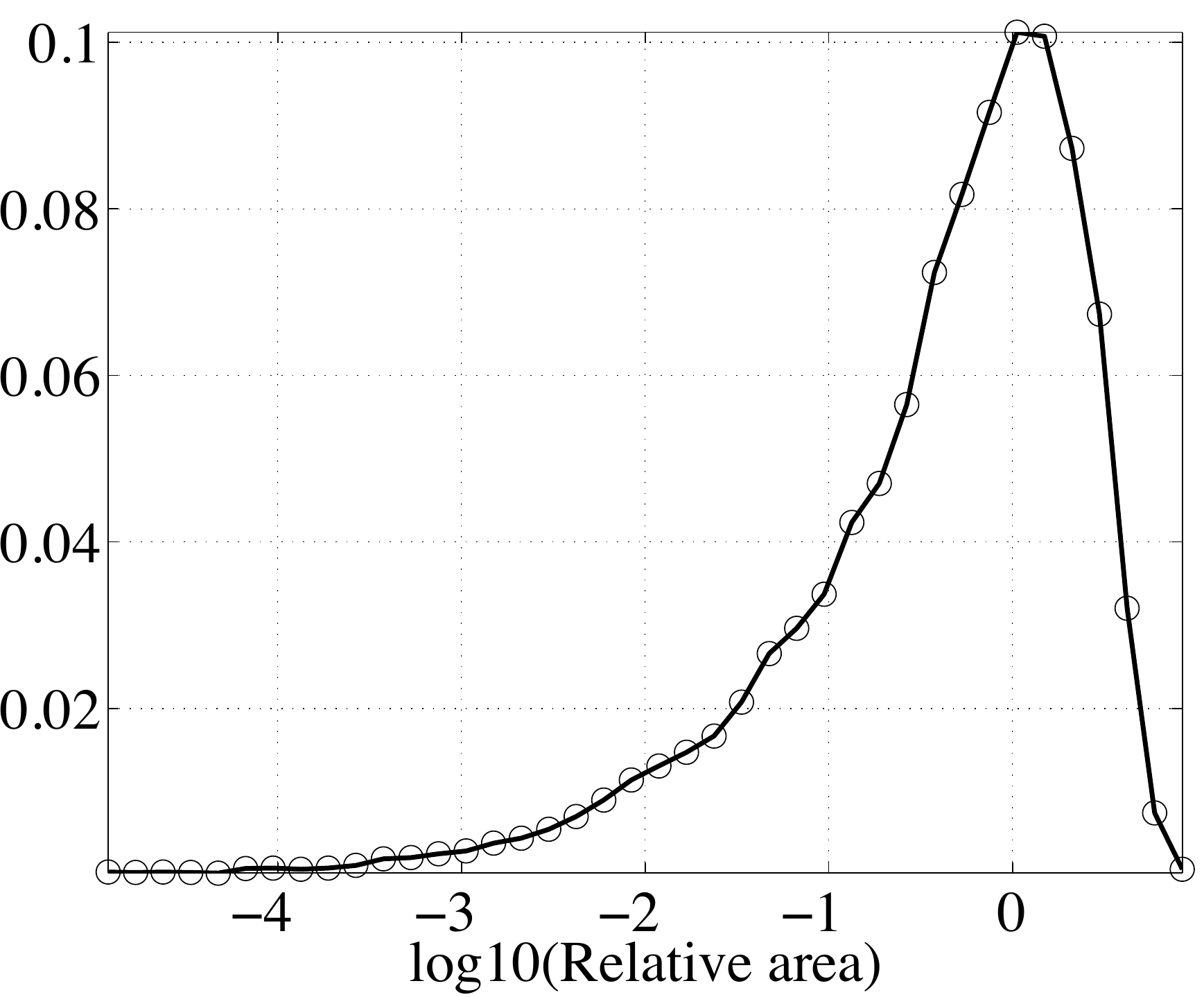}
          \caption{log scale.}
          \label{fig:iso_stat_rel_area_log_ani}
        \end{subfigure}
        \caption{Relative grain area distribution in the anisotropic network consisting of 20,000 grains that evolved from an initial configuration of 100,000 grains.}
        \label{fig:iso_stat_ani}
      \end{figure}

\begin{figure}[!t]
        \begin{subfigure}[b]{0.45\textwidth}
          \centering
          \includegraphics[height=5cm]{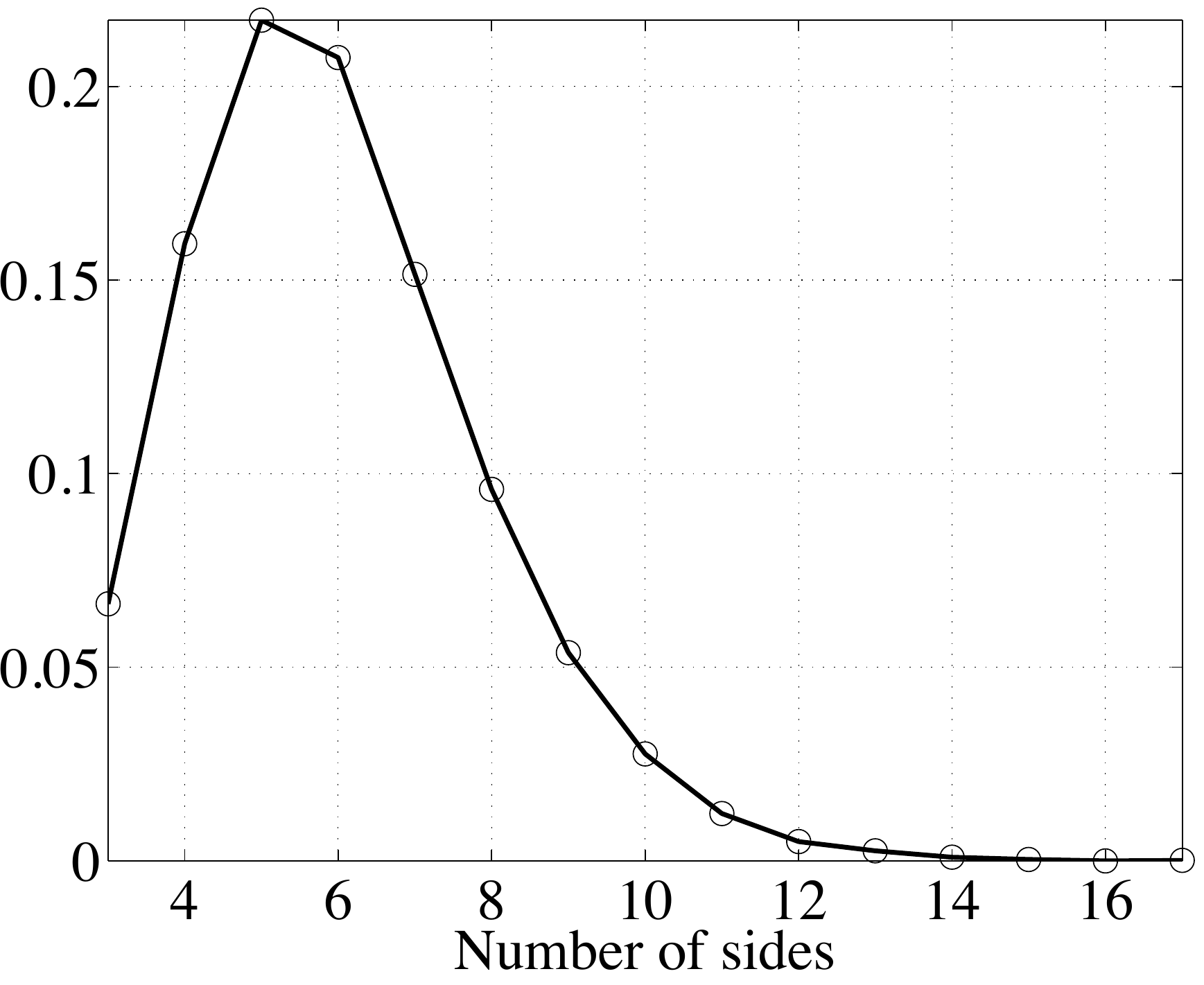}
          \caption{}
          \label{fig:iso_stat_n-sided_ani}
        \end{subfigure}
        \begin{subfigure}[b]{0.45\textwidth}
          \centering
          \includegraphics[height=5cm]{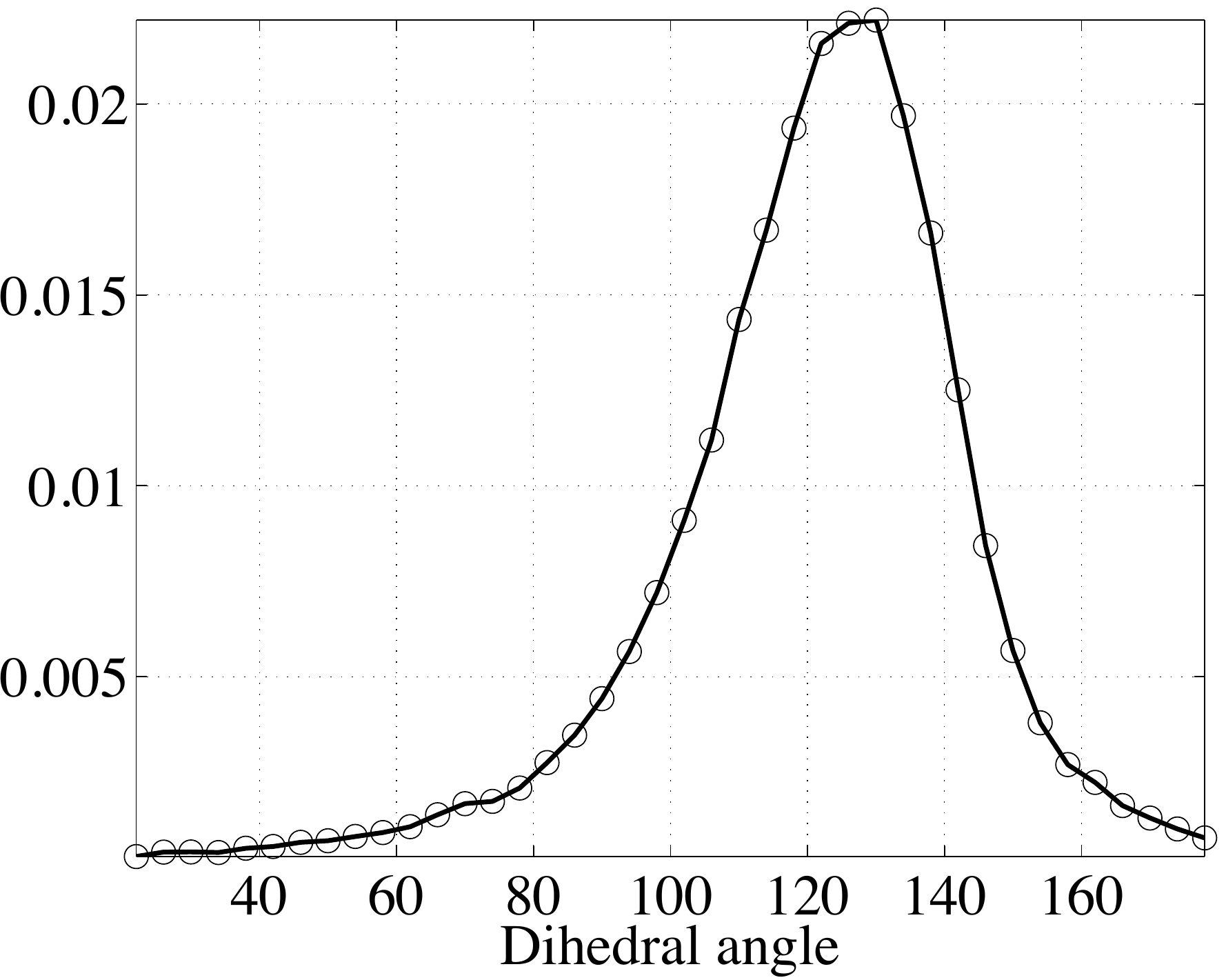}
          \caption{}
           \label{fig:iso_stat_dihedral_ani}
        \end{subfigure}
        \caption{Distributions for the: (a) number of sides of a grain; and (b) dihedral angle.  The data is collected using the anisotropic network of 20,000 grains that evolved from an initial configuration consisting of 100,000 grains.}
      \label{fig:iso_stat_1_ani}
      \end{figure}

\begin{figure}[!t]
        \begin{subfigure}[b]{0.45\textwidth}
          \centering
          \includegraphics[height=5cm]{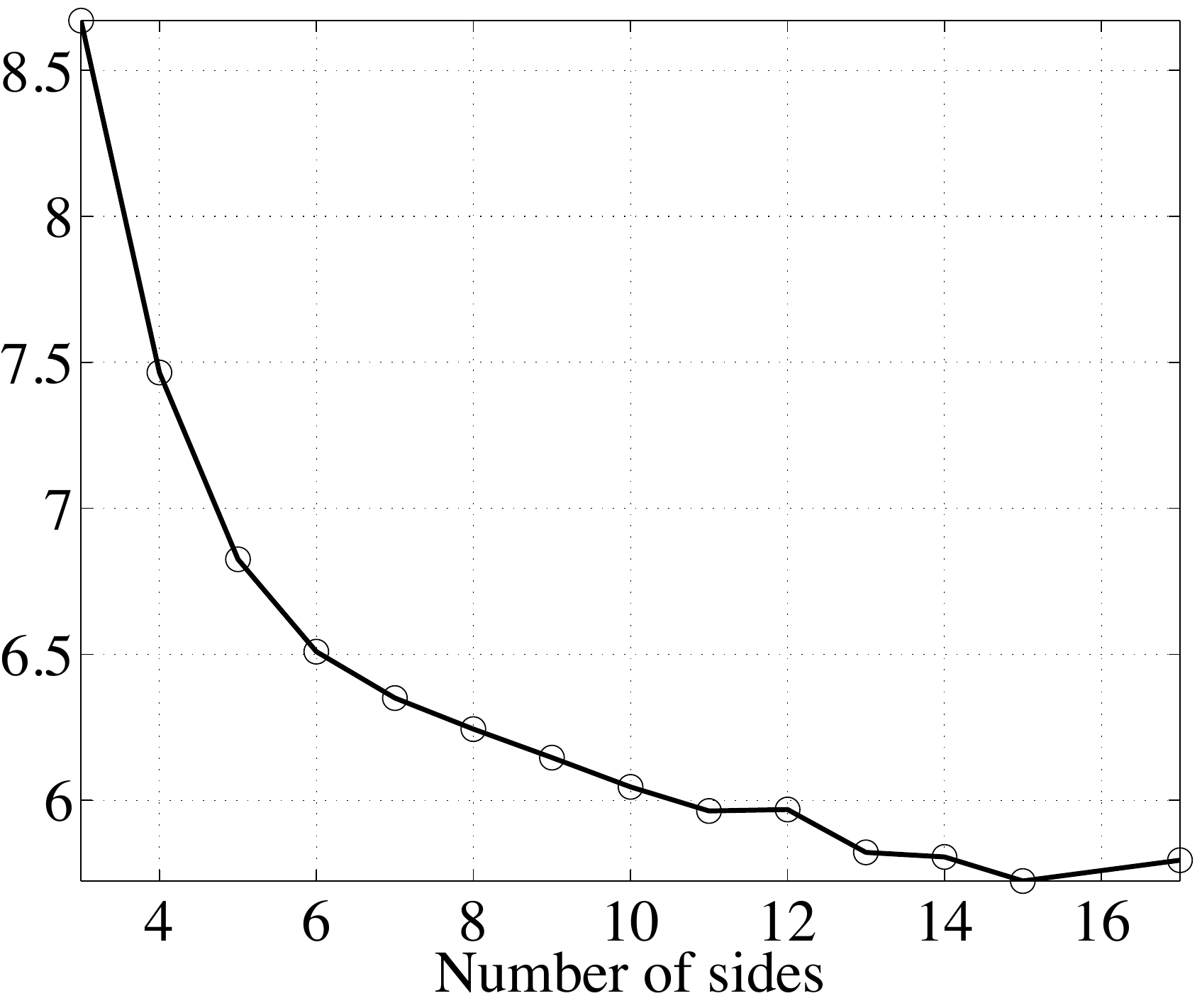}
          \caption{}
          \label{fig:iso_stat_ave_side_class_neigh_ani}
        \end{subfigure}
        \begin{subfigure}[b]{0.45\textwidth}
          \centering
          \includegraphics[height=5cm]{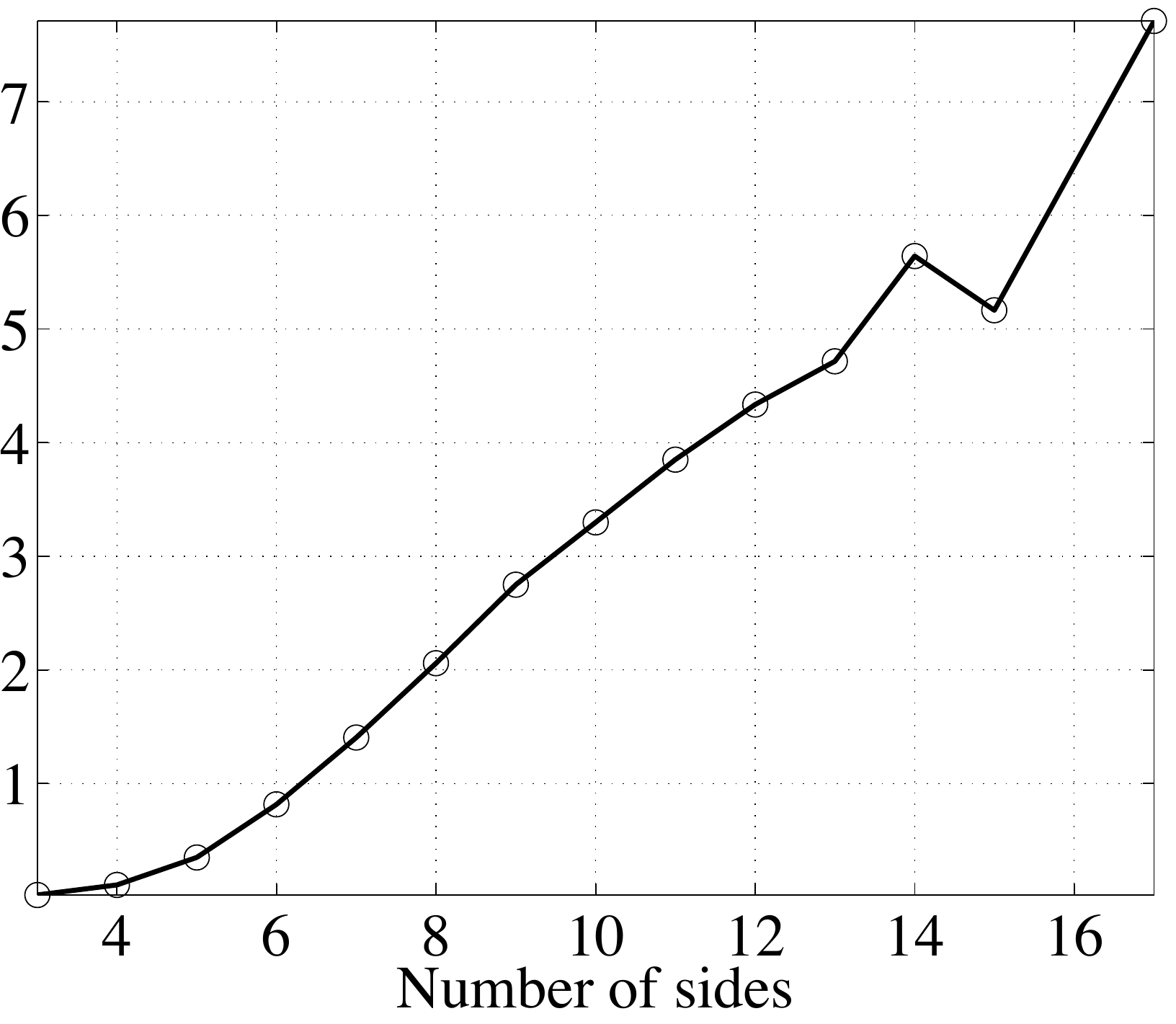}
          \caption{}
          \label{fig:iso_stat_red_ave_side_class_ani}
        \end{subfigure}
        \caption{For grains with a given number of sides: (a) Average number of sides of neighboring grains; (b) Average relative area of neighboring grains. The data is collected using the anisotropic network of 20,000 grains that evolved from an initial configuration consisting of 100,000 grains.}
        \label{fig:side_class_ani}
      \end{figure}

The distributions obtained for grain boundary networks with weakly anisotropic grain boundary energy $\gamma(\Delta \alpha)=0.95-0.05\,\cos^3(4\,\Delta \alpha)$ are shown in Figs. \ref{fig:iso_stat_ani}-\ref{fig:side_class_ani}. Note that the statistical features that develop in this case are essentially identical to those observed when the grain boundary energy is isotropic. 

      \subsection{Self-similarity}

      We have tested the distributions we have observed for
      self-similarity.  We found that self-similarity does indeed develop and
      is very consistent for all distributions of interest.  Here a critical issue is that a sufficiently large 
      number of grains is needed initially for a clear trend to develop. In our experience, the statistics 
      do not depend on the details of the initial configuration; these can only affect the time needed to 
      achieve stable statistics. Here the initial set of grains was generated using the Voronoi 
      construction for a uniformly distributed random collection of points in a rectangular domain.

\begin{figure}[!t]
        \begin{subfigure}[b]{0.45\textwidth}
          \centering
          \includegraphics[height=5cm]{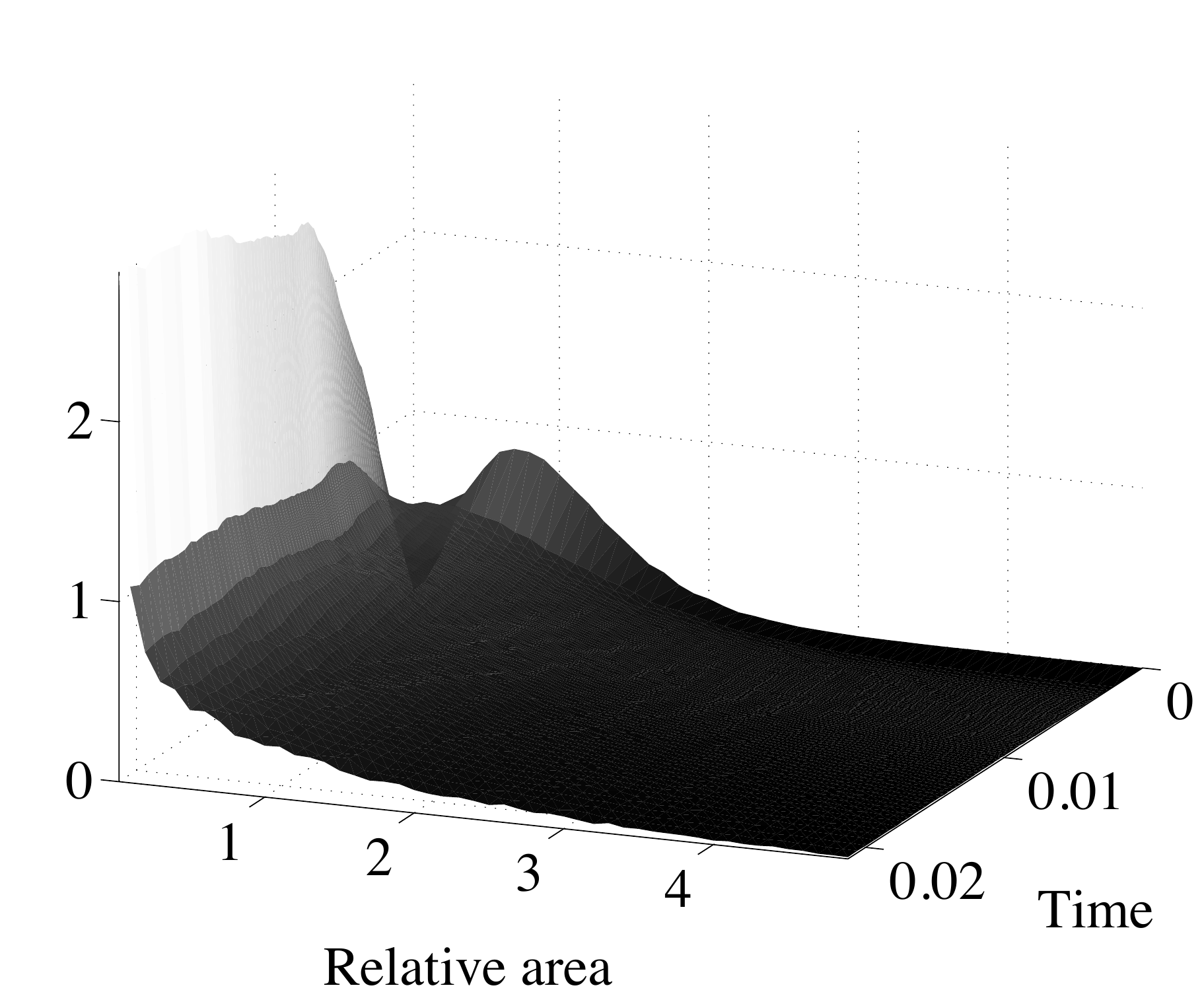}
          \caption{}
          \label{fig:self_iso_01}
        \end{subfigure}
        \begin{subfigure}[b]{0.45\textwidth}
          \centering
          \includegraphics[height=5cm]{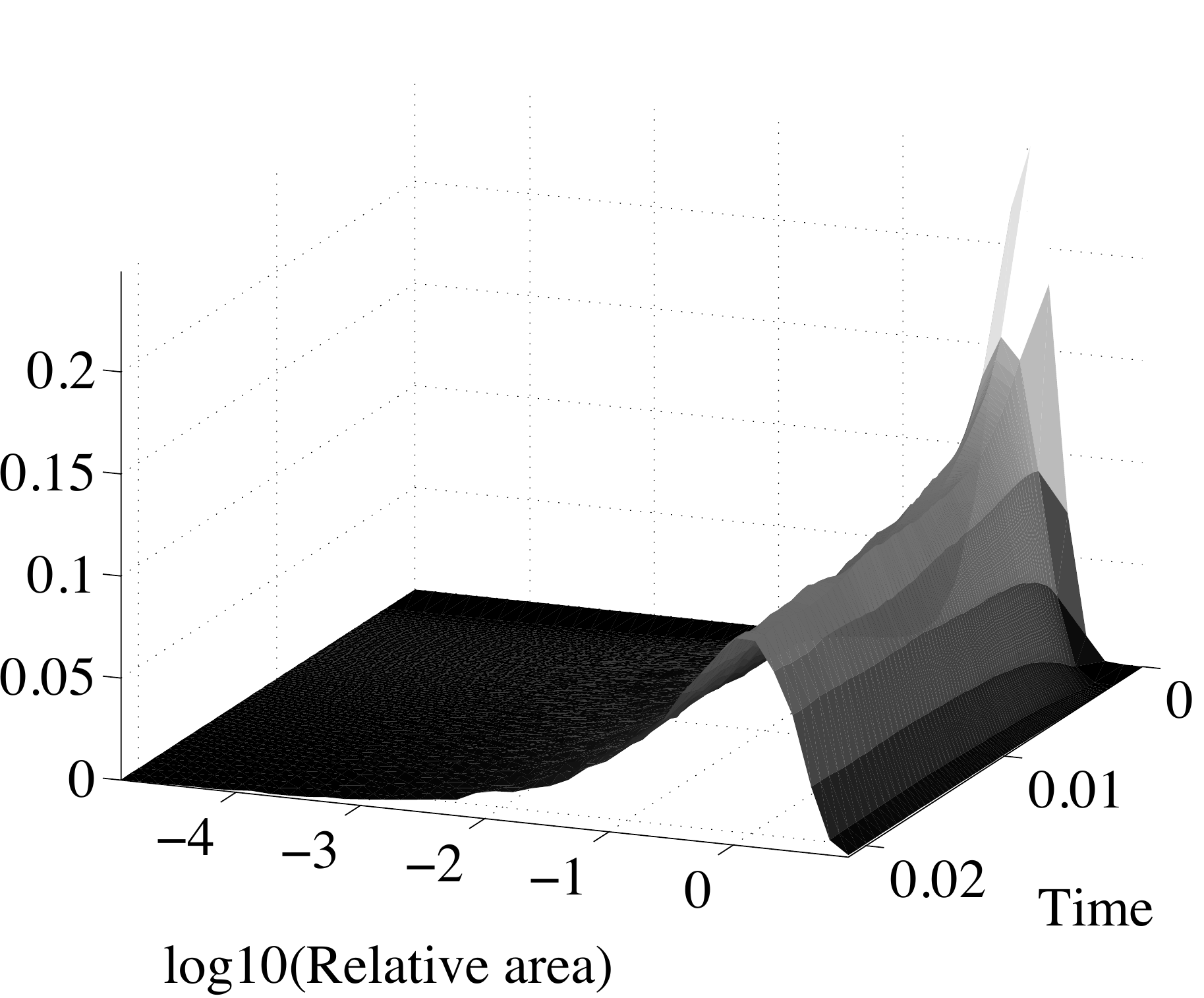}
          \caption{}
          \label{fig:self_iso_02}
        \end{subfigure}
        \caption{Evolution in time of the relative grain area distribution. The statistics was collected for every time step while evolving grain boundary networks that 
      initially contained $100,000$ grains up until $20,000$ grains remained.}
        \label{fig:self_iso_1}
      \end{figure}

\begin{figure}[!t]
        \begin{subfigure}[b]{0.45\textwidth}
          \centering
          \includegraphics[height=5cm]{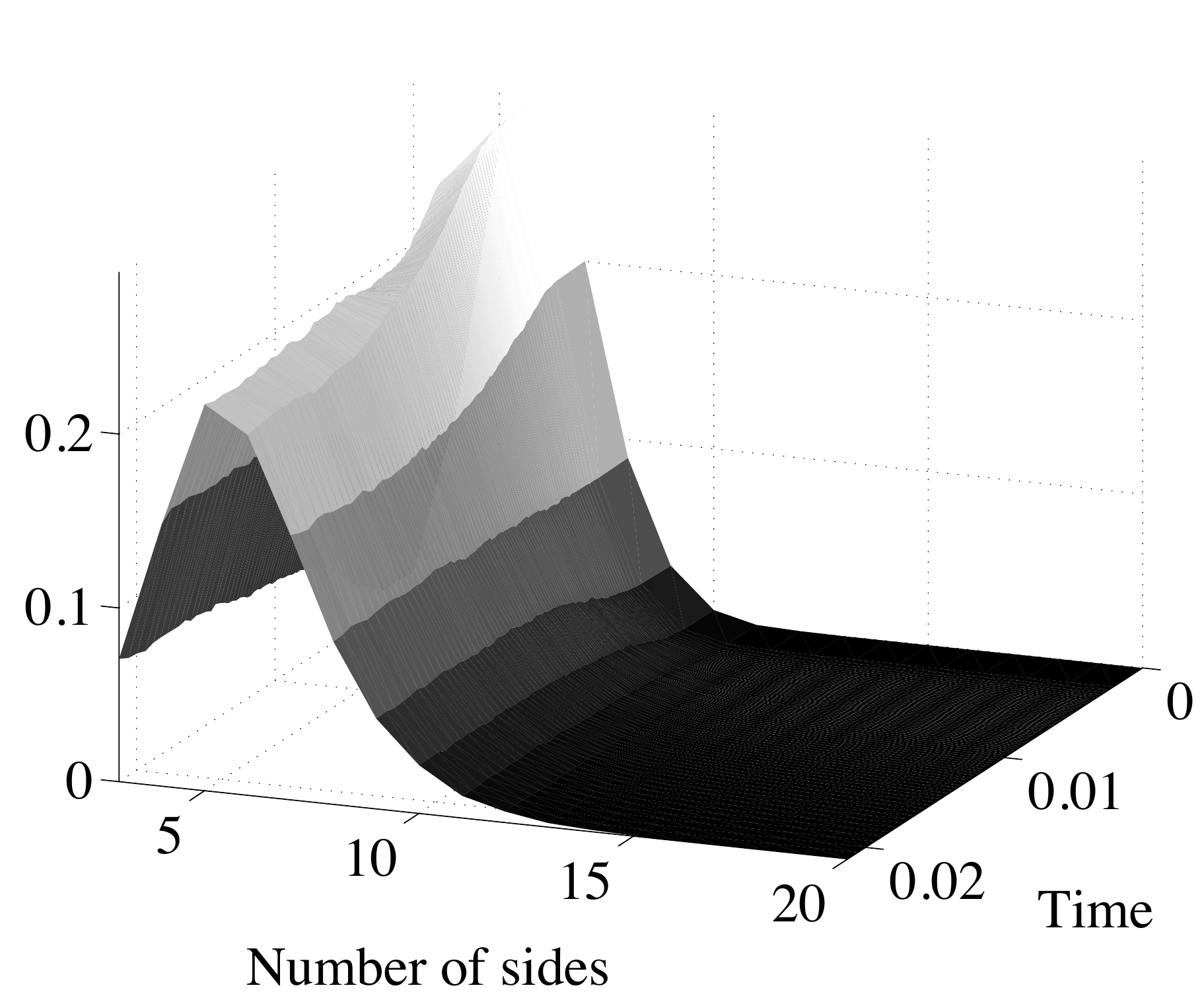}
          \caption{}
          \label{fig:self_iso_03}
        \end{subfigure}
        \begin{subfigure}[b]{0.45\textwidth}
          \centering
          \includegraphics[height=5cm]{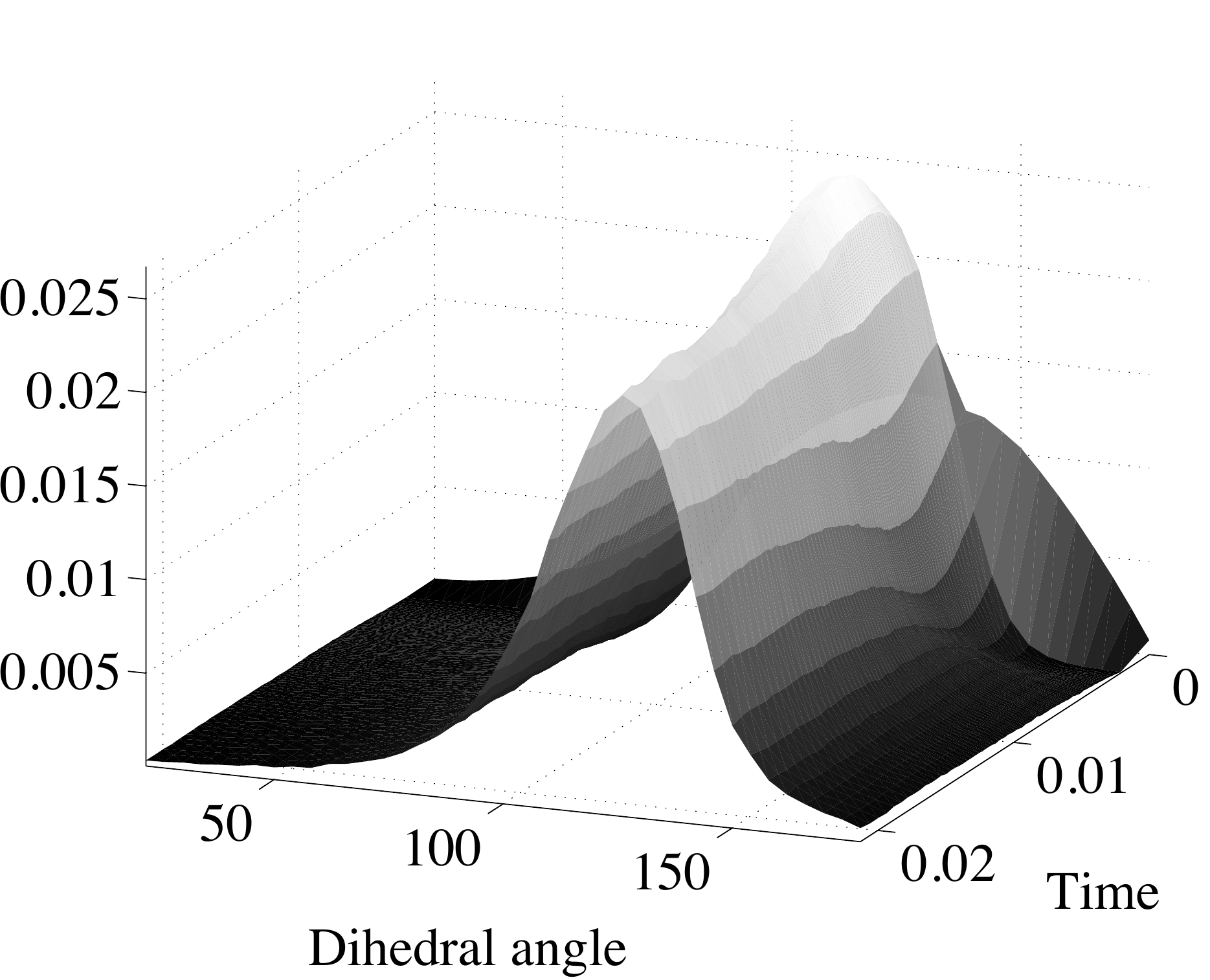}
          \caption{}
          \label{fig:self_iso_04}
        \end{subfigure}
       \caption{Evolution in time of distributions for the: (a) number of sides of a grain; and (b) dihedral angle. The statistics was collected for every time step while evolving grain boundary networks that initially contained $100,000$ grains up until $20,000$ grains remained.}
      \label{fig:self_iso_2}
      \end{figure}

      Figs.~\ref{fig:self_iso_1} and \ref{fig:self_iso_2} show the development 
      of various distributions over time.  The plots were produced by evolving grain boundary networks that 
      initially contained $100,000$ grains up until $20,000$ grains remained.

      \subsection{Comparison of different neighbor switching rules}

\begin{figure}[!t]
        \begin{subfigure}[b]{0.45\textwidth}
          \centering
          \includegraphics[height=5cm]{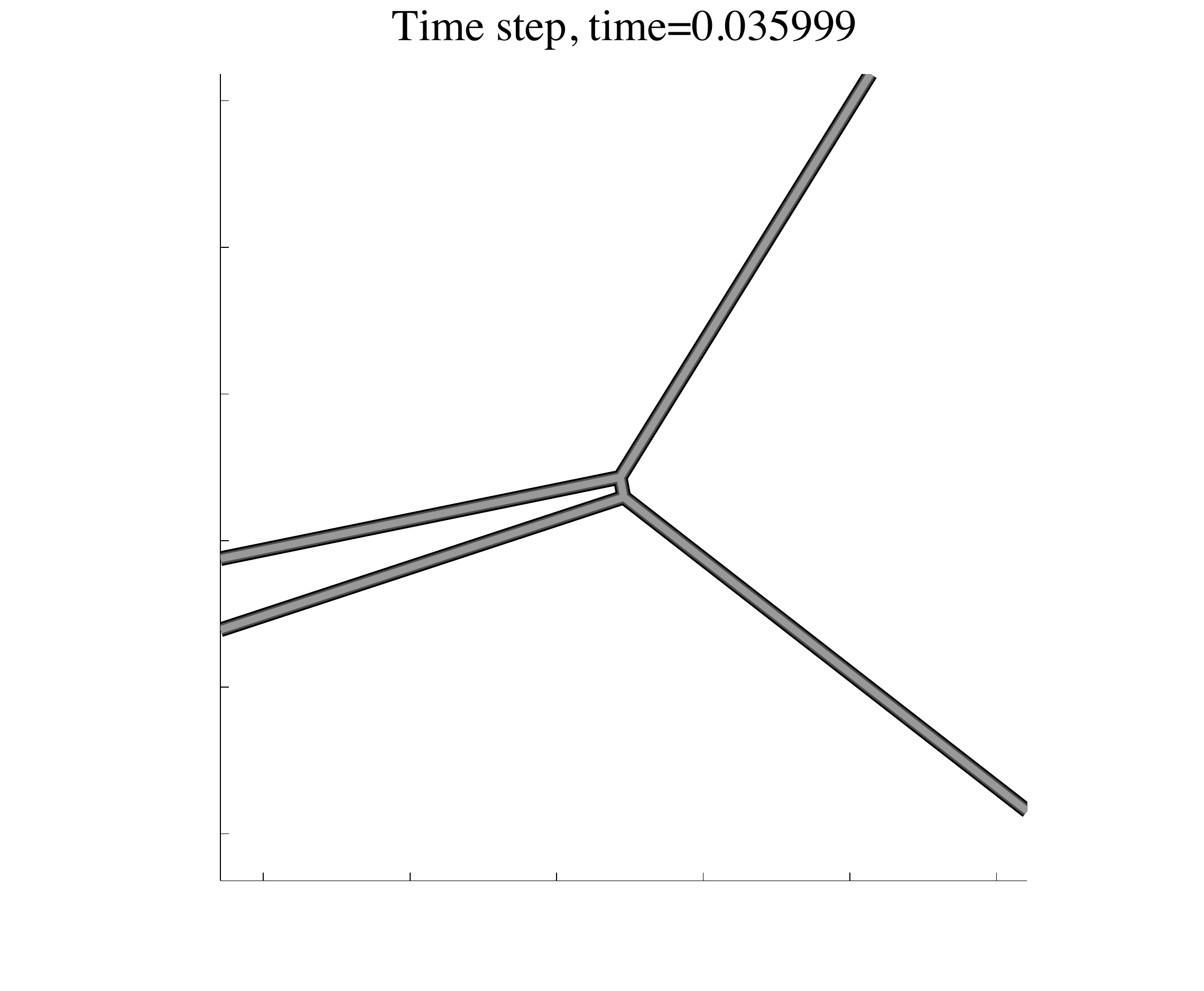}
          \caption{}
          \label{fig:comp-flip-rule_01}
        \end{subfigure}
        \begin{subfigure}[b]{0.45\textwidth}
          \centering
          \includegraphics[height=5cm]{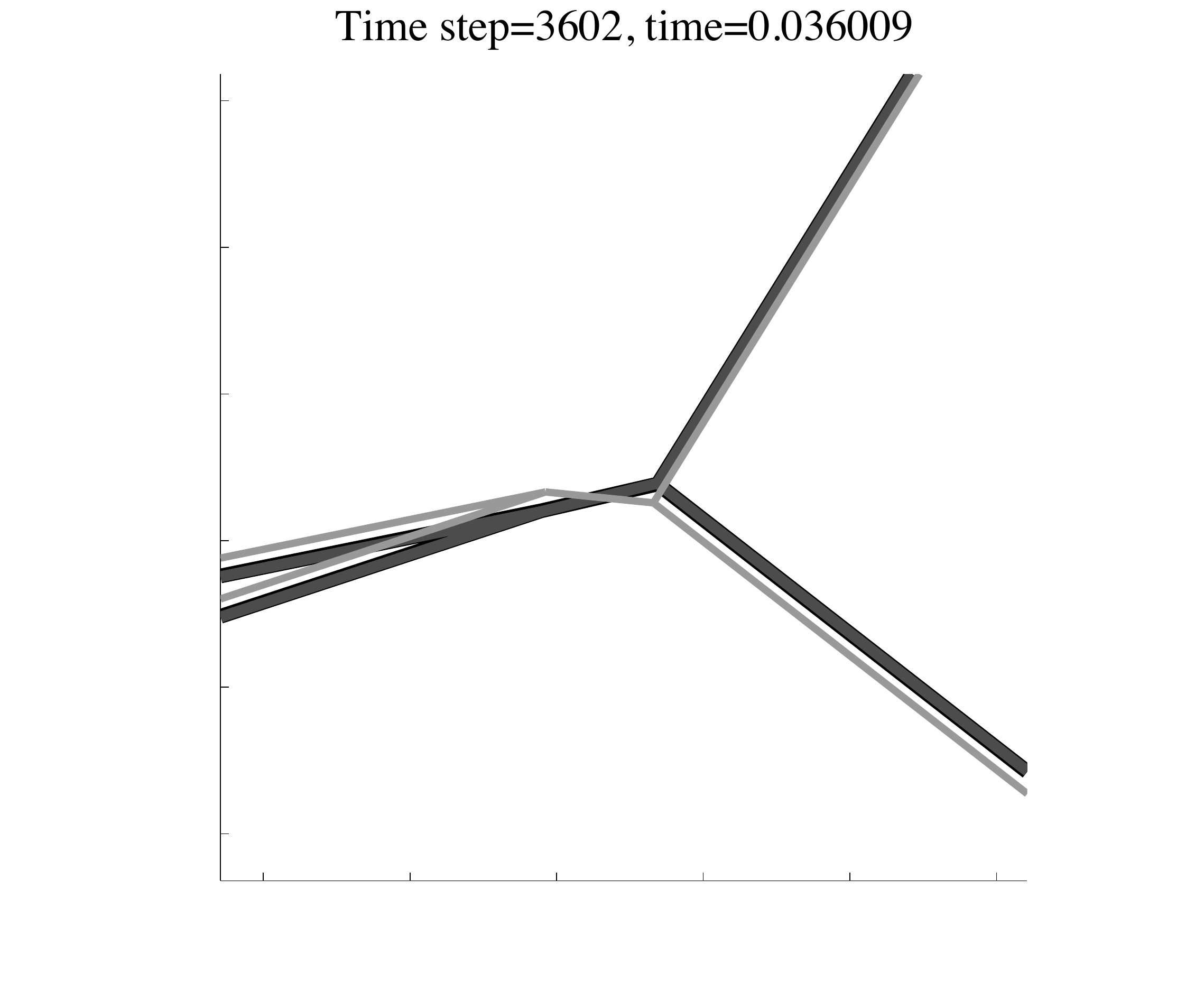}
          \caption{}
          \label{fig:comp-flip-rule_02}
        \end{subfigure}
        \\
       \begin{subfigure}[b]{0.45\textwidth}
          \centering
          \includegraphics[height=5cm]{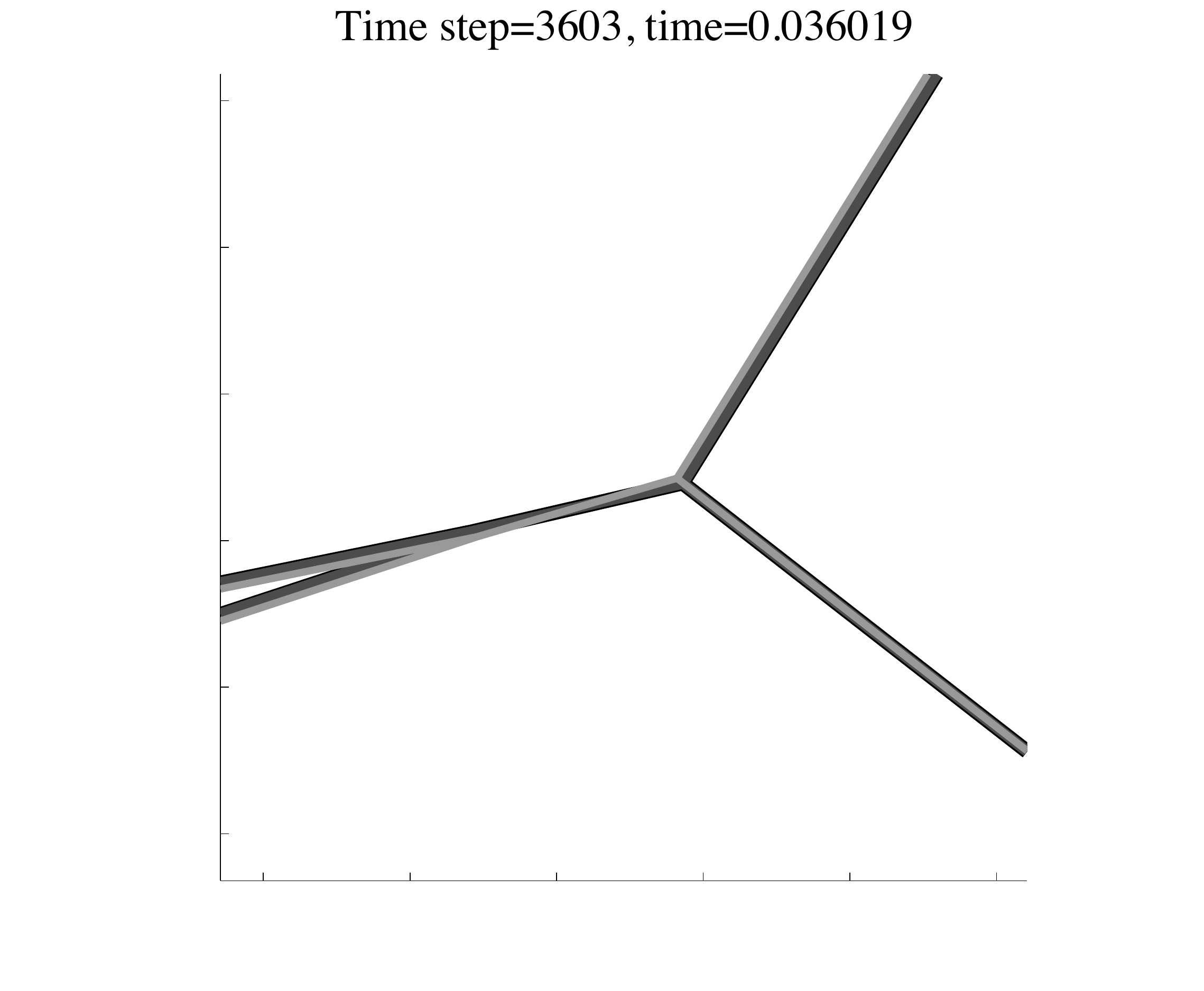}
          \caption{}
          \label{fig:comp-flip-rule_03}
        \end{subfigure}
        \begin{subfigure}[b]{0.45\textwidth}
          \centering
          \includegraphics[height=5cm]{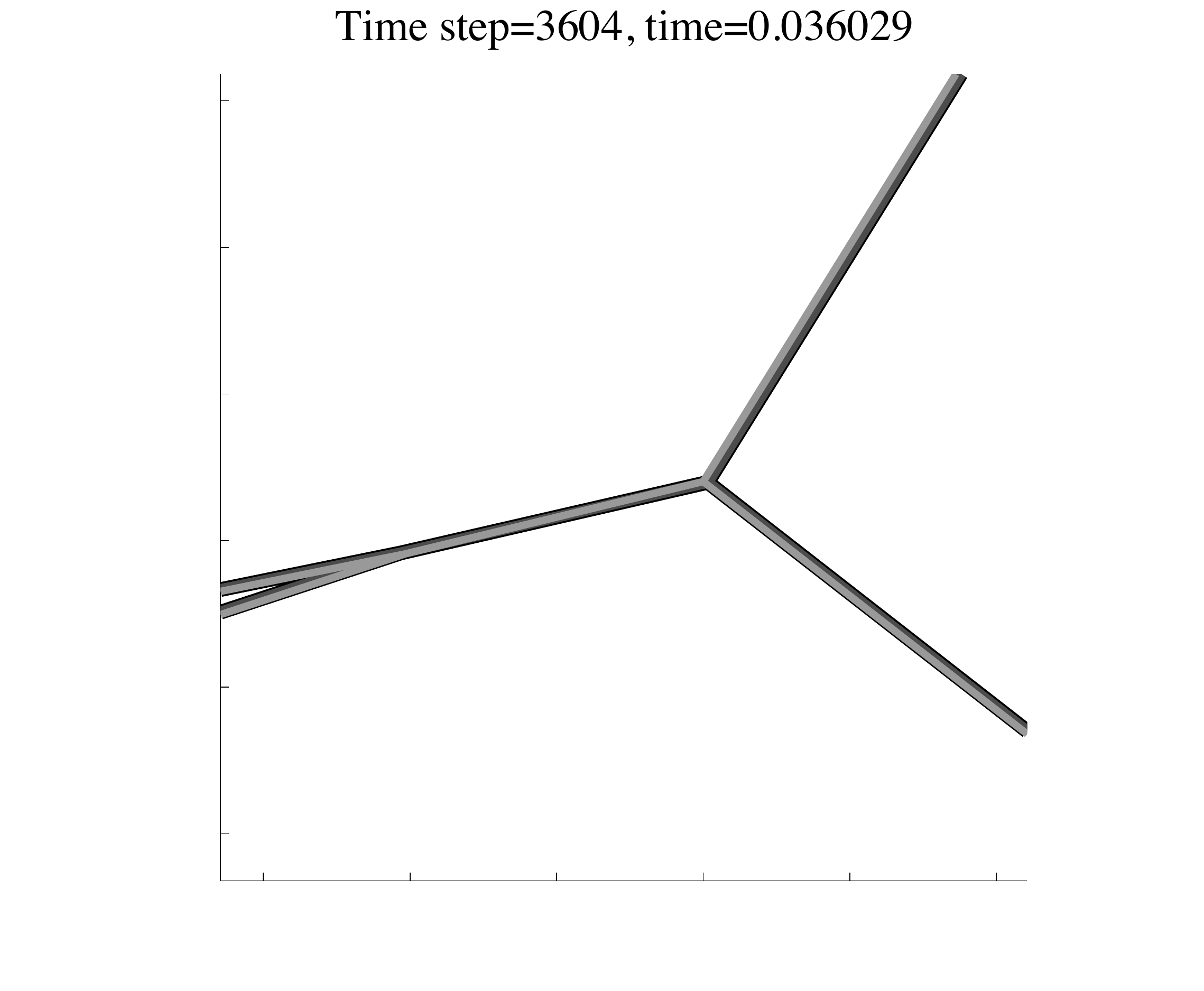}
          \caption{}
          \label{fig:comp-flip-rule_04}
        \end{subfigure}
        \caption{Comparison of three neighbor switching rules: Maximum dissipation
          rate (light gray) \cite{kind04}, maximum force (gray) \cite{barrales10}, the approach described in this paper (black). After a short period of time, configurations are essentially indistinguishable when using a small time step.}        \label{fig:comp-flip-rule}
      \end{figure}

      We have compared three types of neighbor switching rules used in the
      literature: maximum dissipation rate \cite{kind04}, maximum force
      \cite{barrales10}, and the proposed approach.  Fig.~\ref{fig:comp-flip-rule} shows a part of the grain boundary network immediately preceding 
      the first neighbor switching event and for several time steps afterward. The three rules result in a different initial orientation of the newly formed 
      grain boundary that has the length proportional to $\Delta t-t_{ext}$.  In all cases, subsequent evolution of the network corrects the angle to the
      one enforced by the continuous part of the algorithm. Overall, it appears that a ``suboptimal'' neighbor switching rule leads to accumulating errors 
      that result in a grain boundary network that differs significantly from that produced using the ``optimal'' rule, starting from 
      the same initial conditions. However, singular behavior of the continuous part of the dynamical system at the time of the neighbor switch, combined with
       a smaller maximum time step $\Delta t_0$, makes this the error less significant, per Fig.~\ref{fig:comp-flip-rule}d. In all cases, the statistical features of the
       network seem to be unaffected by the type of the rule used. 
      
      From our simulations, it also appears that the networks evolving via different neighbor switching rules (or via the same rule, but with a different maximum 
      time step), move through a very similar sequence of configurations in the state space, albeit at different times.

\subsection{Rate of area change for an $N$-sided grain}

      \begin{figure}[!t]
        \begin{subfigure}[b]{0.45\textwidth}
          \centering
          \includegraphics[height=5cm]{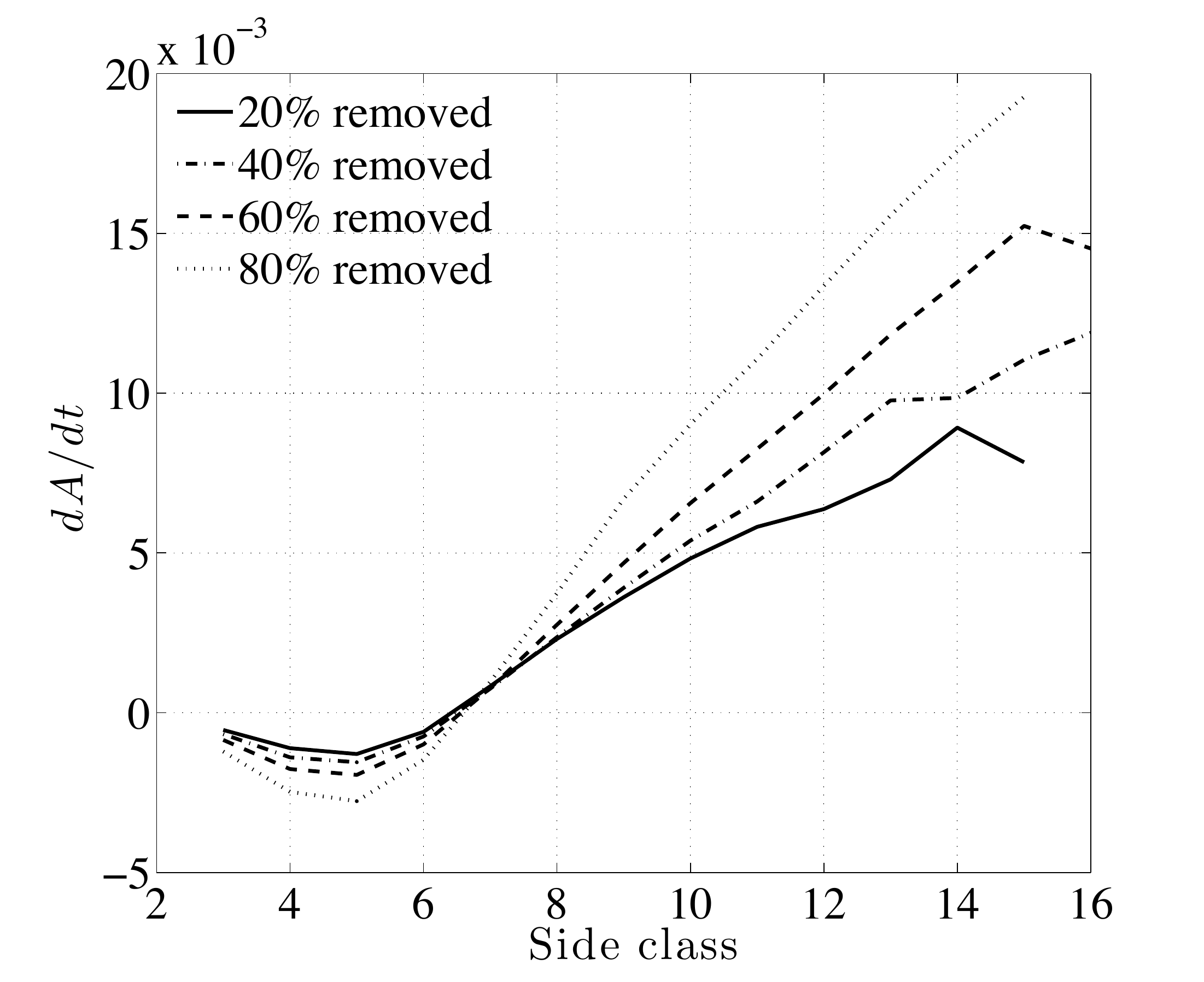}
         \caption{}
          \label{fig:dAdt_ave}
        \end{subfigure}
        \begin{subfigure}[b]{0.45\textwidth}
          \centering
          \includegraphics[height=5cm]{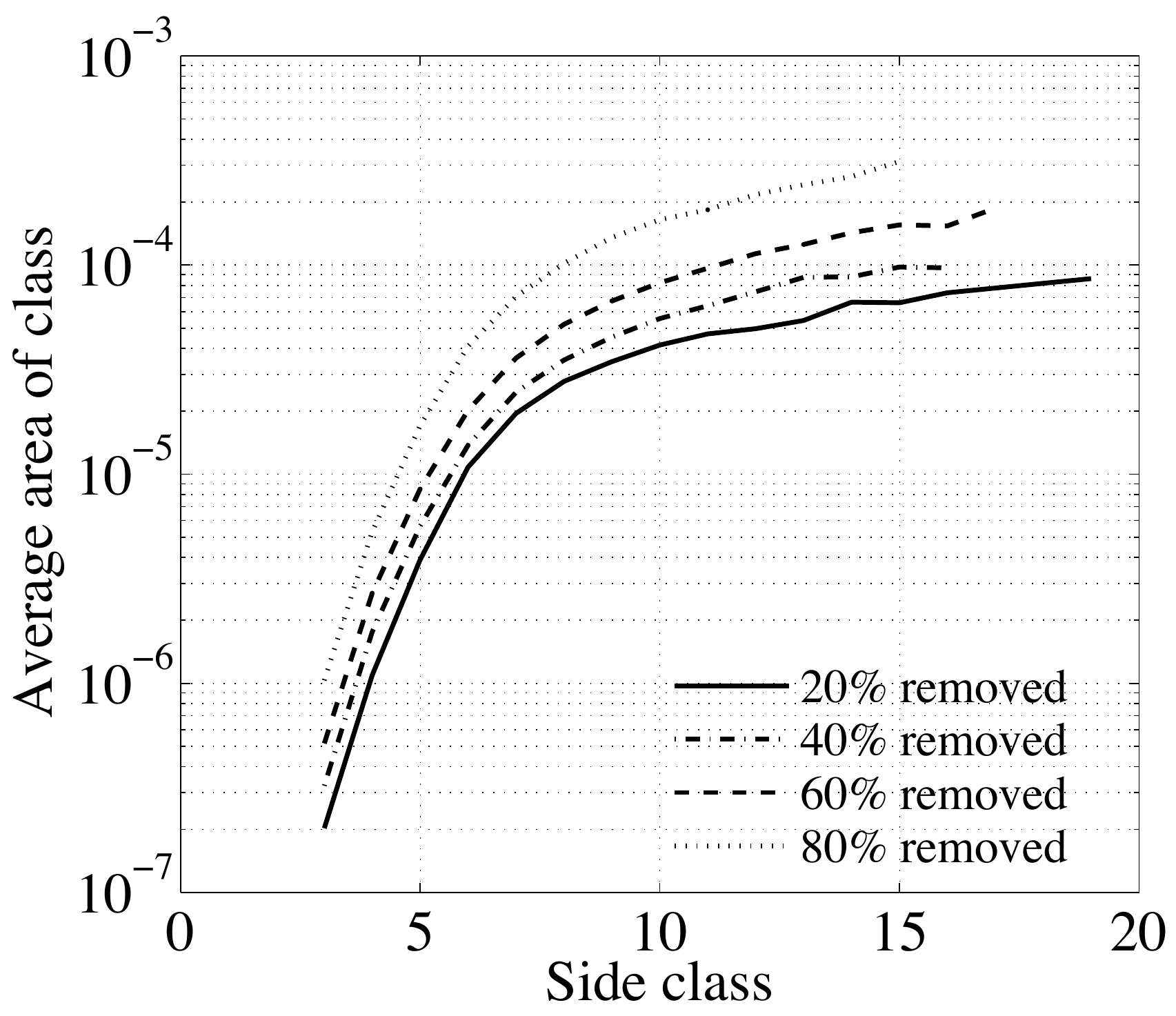}
         \caption{}
          \label{fig:Ave_area_time}
        \end{subfigure}
        \caption{(a) Average rate of change of grain area vs. the number of sides of the grain. (b) Average grain area vs. the number of sides of the grain.}
      \label{fig:dAdt_A}
      \end{figure}
      
A well-known result for curvature-driven grain growth is the von Neumann-Mullins
$(n-6)$-rule.  The rule states that, given constant mobility and constant anisotropy, and assuming that 
the network satisfies the Herring condition  (angles between the boundaries meeting at a triple junction 
are all equal to $120^\circ$ in an isotropic case),
we have
\begin{equation*}
  \frac{dA}{dt}=c (n-6),
\end{equation*}
where $c>0$ is a known constant. In a vertex algorithm the $n-6$ rule does not hold. Indeed, in Fig.~\ref{fig:dAdt_ave} 
we observe that, although the relation for $n$ between $5$ and $15$ is close to being linear, it is far from that for grains with a smaller number of sides. The distribution depends on time and has a self-similar shape, however the nature of the observed dependence is still an open problem that will be addressed in a future publication. Fig.~\ref{fig:Ave_area_time} shows that the average area for each class of grains with a given number of sides grows over time. This same behavior is observed in curvature codes \cite{kind04}.

\subsection{Stability}

We have tested for stability distributions that develop for networks evolving via our algorithm in both isotropic 
and anisotropic cases. Here we only present the results for the isotropic case due to space
constraints and because the conclusions are qualitatively similar.

The test has been performed on a sample with $100,000$
grains initially, where the simulation was run until $50\%$ of grains where
removed.  We decomposed the resulting sample into 12 spatially smaller subsets of equal area 
and collected various statistics for each sample. These were compared to the output of 10
simulations with initially $20,000$ grains that were run until only
$4,000$ were left. The statistics were computed for these
smaller samples. We also analyzed the statistics for a single $50,000$
sample.

\begin{figure}[!t]
  \begin{subfigure}[b]{0.45\textwidth}
    \centering
    \includegraphics[height=5cm]{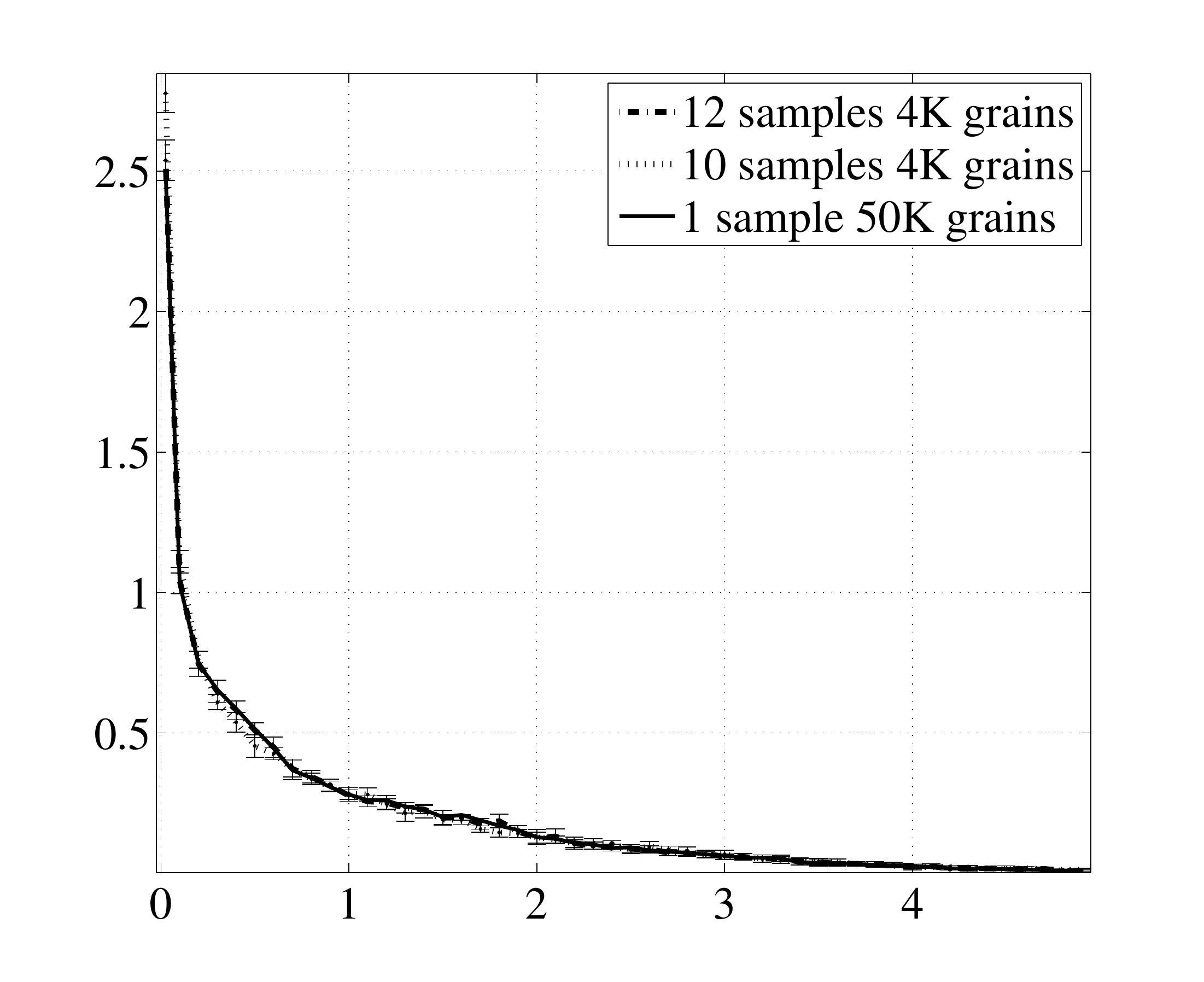}
    \caption{linear scale.}
    \label{fig:ergo_iso_01}
  \end{subfigure}
  \begin{subfigure}[b]{0.45\textwidth}
    \centering
    \includegraphics[height=5cm]{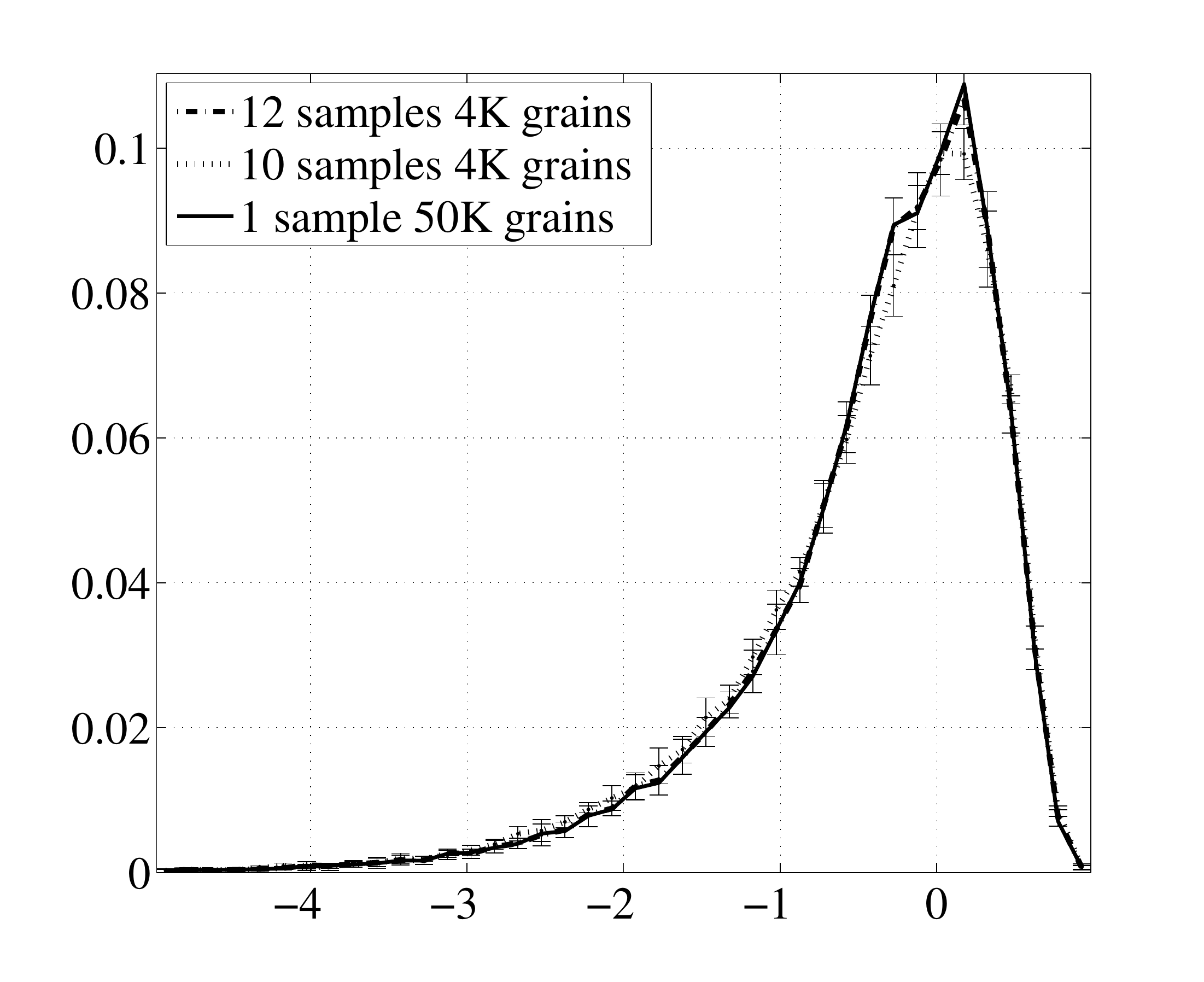}
    \caption{log scale.}
    \label{fig:ergo_iso_02}
  \end{subfigure}
   \caption{Stability of the relative grain area distribution. The average of distributions for 12 subsets of the same simulation, the average of distributions for 10 different simulations, and the distribution for one large simulation are shown. The deviation of distributions from their average for each group is indicated by error bars.}  \label{fig:ergo_area}
\end{figure}

\begin{figure}[!t]
  \begin{subfigure}[b]{0.45\textwidth}
    \centering
    \includegraphics[height=5cm]{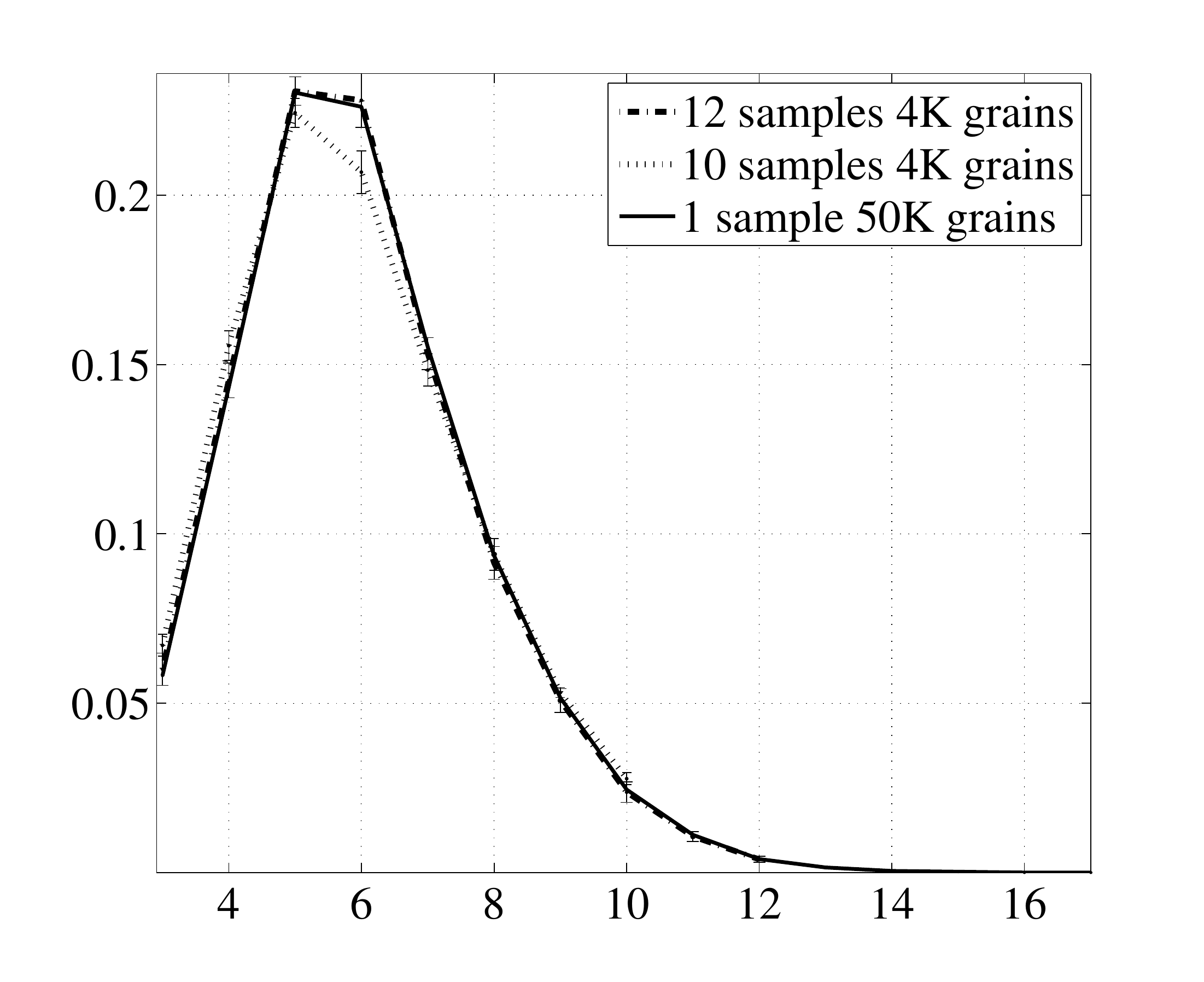}
    \caption{}
    \label{fig:ergo_iso_03}
  \end{subfigure}
  \begin{subfigure}[b]{0.45\textwidth}
    \centering
    \includegraphics[height=5cm]{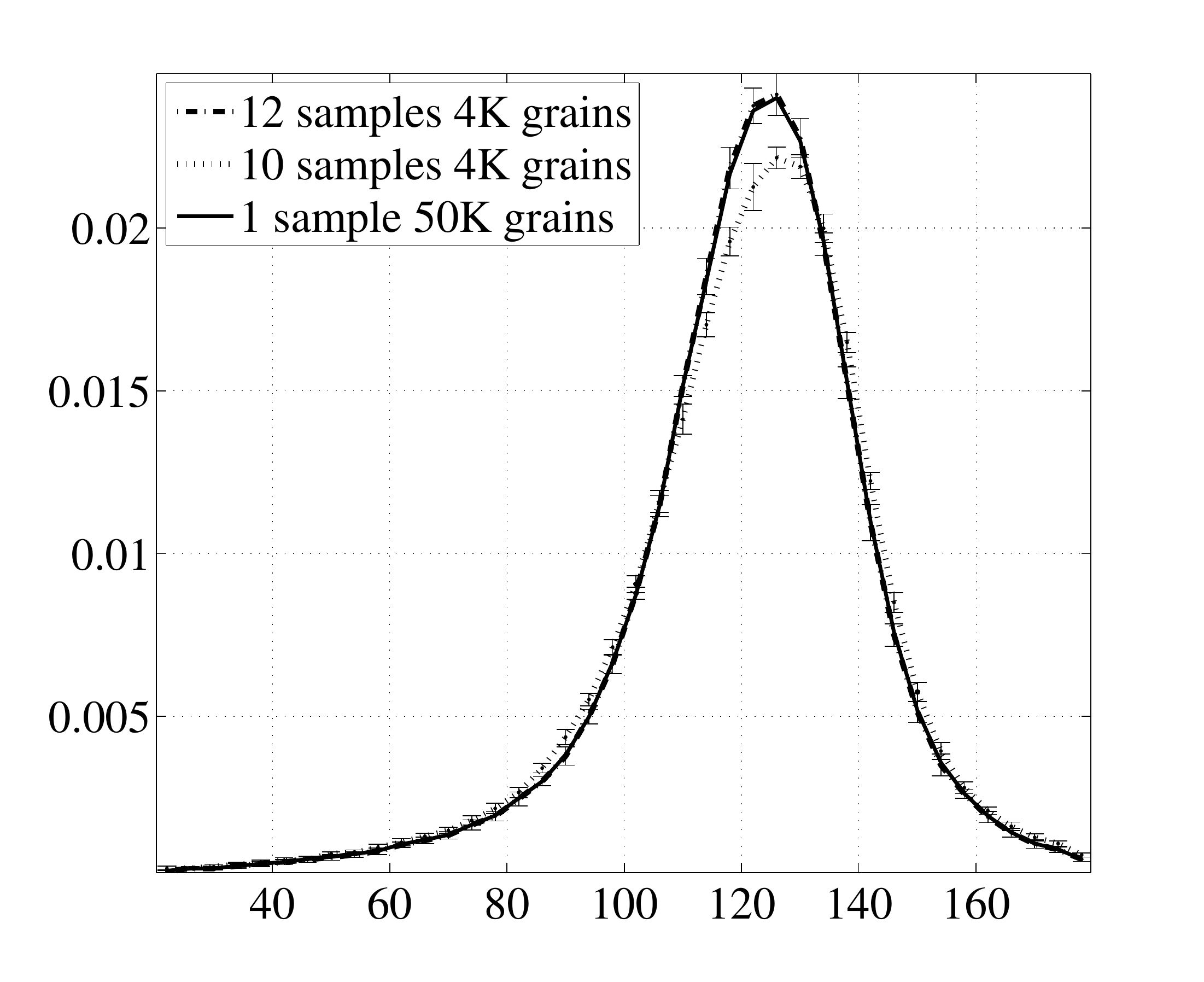}
    \caption{}
    \label{fig:ergo_iso_04}
  \end{subfigure}
  \caption{Stability of distributions for the: (a) number of sides of a grain; and (b) dihedral angle. The average of distributions for 12 subsets of the same simulation, the average of distributions for 10 different simulations, and the distribution for one large simulation are shown. The deviation of distributions from their average for each group is indicated by error bars.}
  \label{fig:ergo_num}
\end{figure}

\begin{figure}[!t]
  \begin{subfigure}[b]{0.45\textwidth}
    \centering
    \includegraphics[height=5cm]{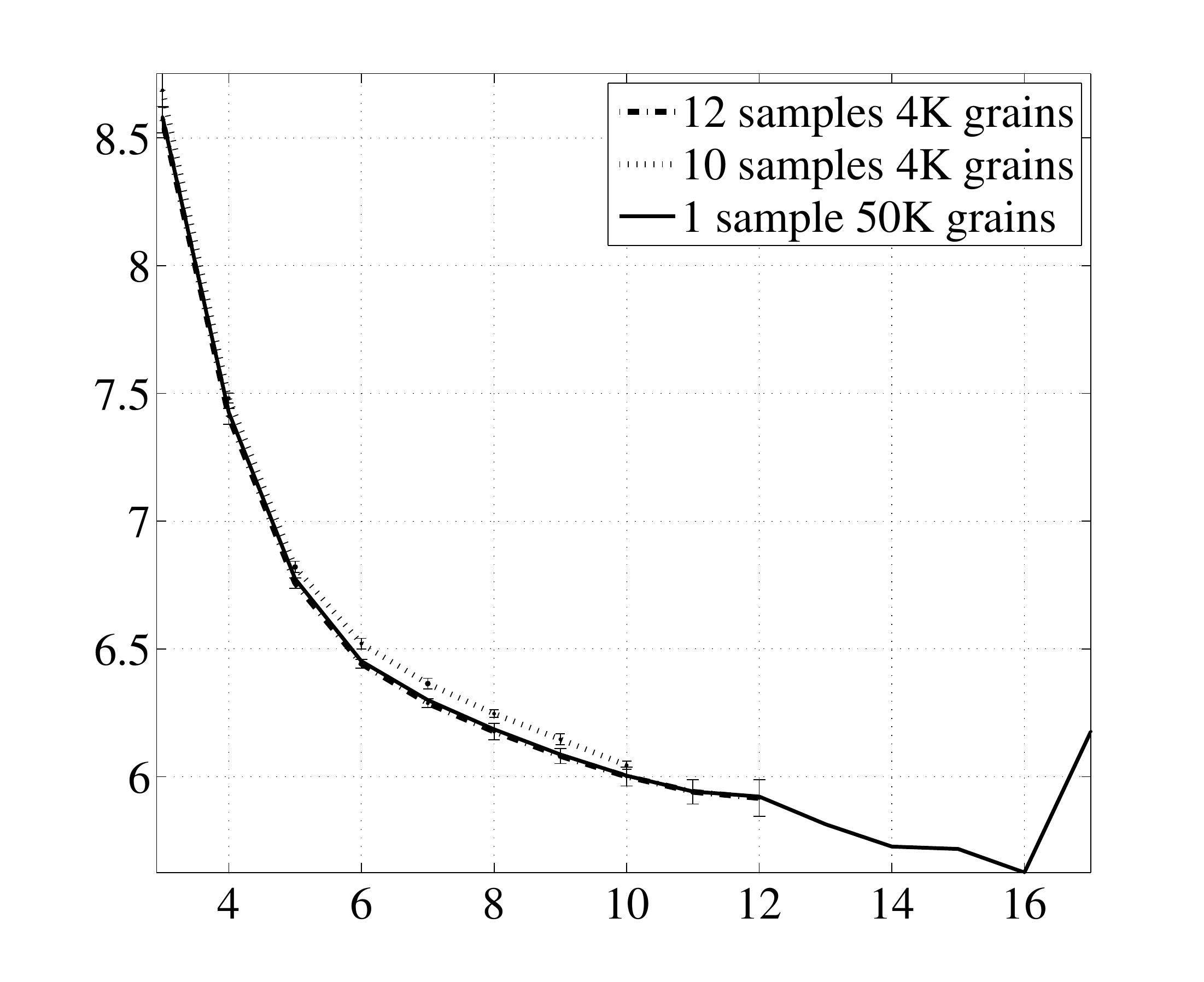}
    \caption{}
    \label{fig:ergo_iso_05}
  \end{subfigure}
  \begin{subfigure}[b]{0.45\textwidth}
    \centering
    \includegraphics[height=5cm]{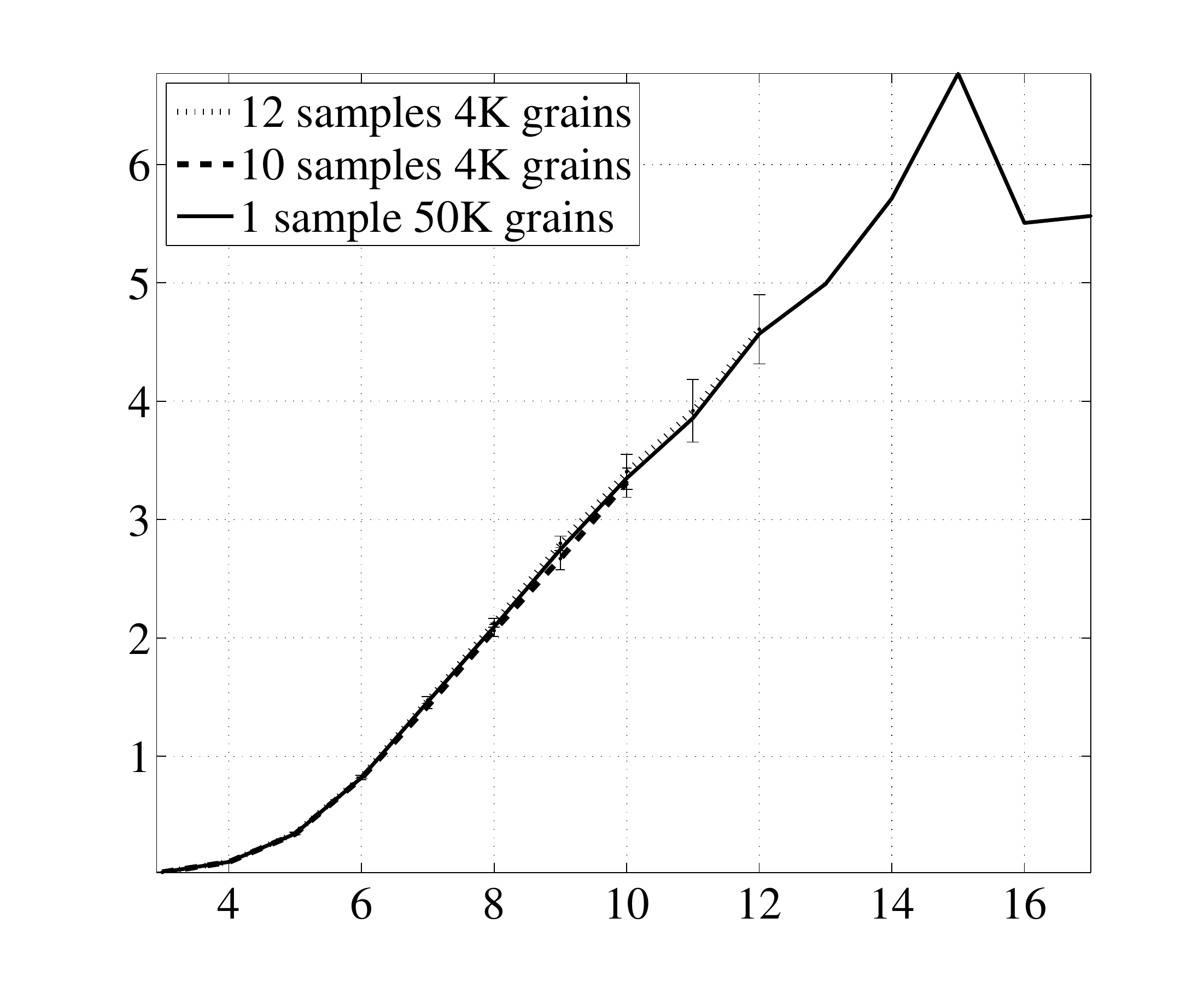}
    \caption{}
    \label{fig:ergo_iso_06}
  \end{subfigure}
 \caption{For the grains with a given number of sides: (a) Average number of sides of neighbors; (b) Average relative grain area of neighbors. The average of distributions for 12 subsets of the same simulation, the average of distributions for 10 different simulations, and the distribution for one large simulation are shown. The deviation of distributions from their average for each group is indicated by error bars.}
  \label{fig:ergo_side}
\end{figure}

Fig.~\ref{fig:ergo_area} through
Fig.~\ref{fig:ergo_side} present the outcome of this study.  This
experiment shows numerical evidence that all distributions are remarkably stable in the sense that collecting statistics over subareas of 
the network or the entire network produces the same results.  

\subsection{Quadruple junctions and their stability}
\label{quad_junction}
\begin{figure}[!t]
  \begin{subfigure}[b]{0.45\textwidth}
    \centering
    \includegraphics[height=4.5cm]{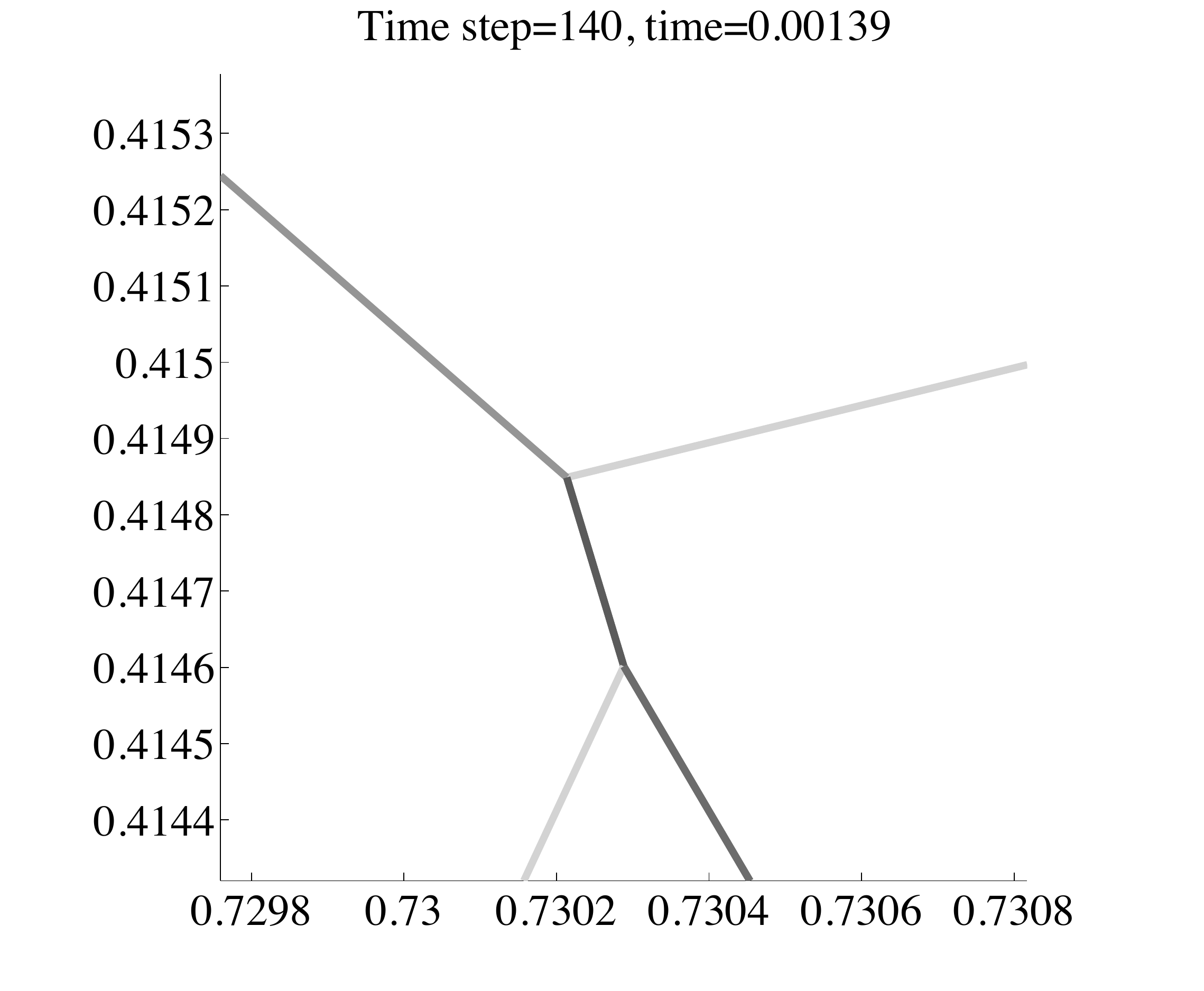}
    \caption{}
    \label{fig:quad_01}
  \end{subfigure}
  \begin{subfigure}[b]{0.45\textwidth}
    \centering
    \includegraphics[height=4.5cm]{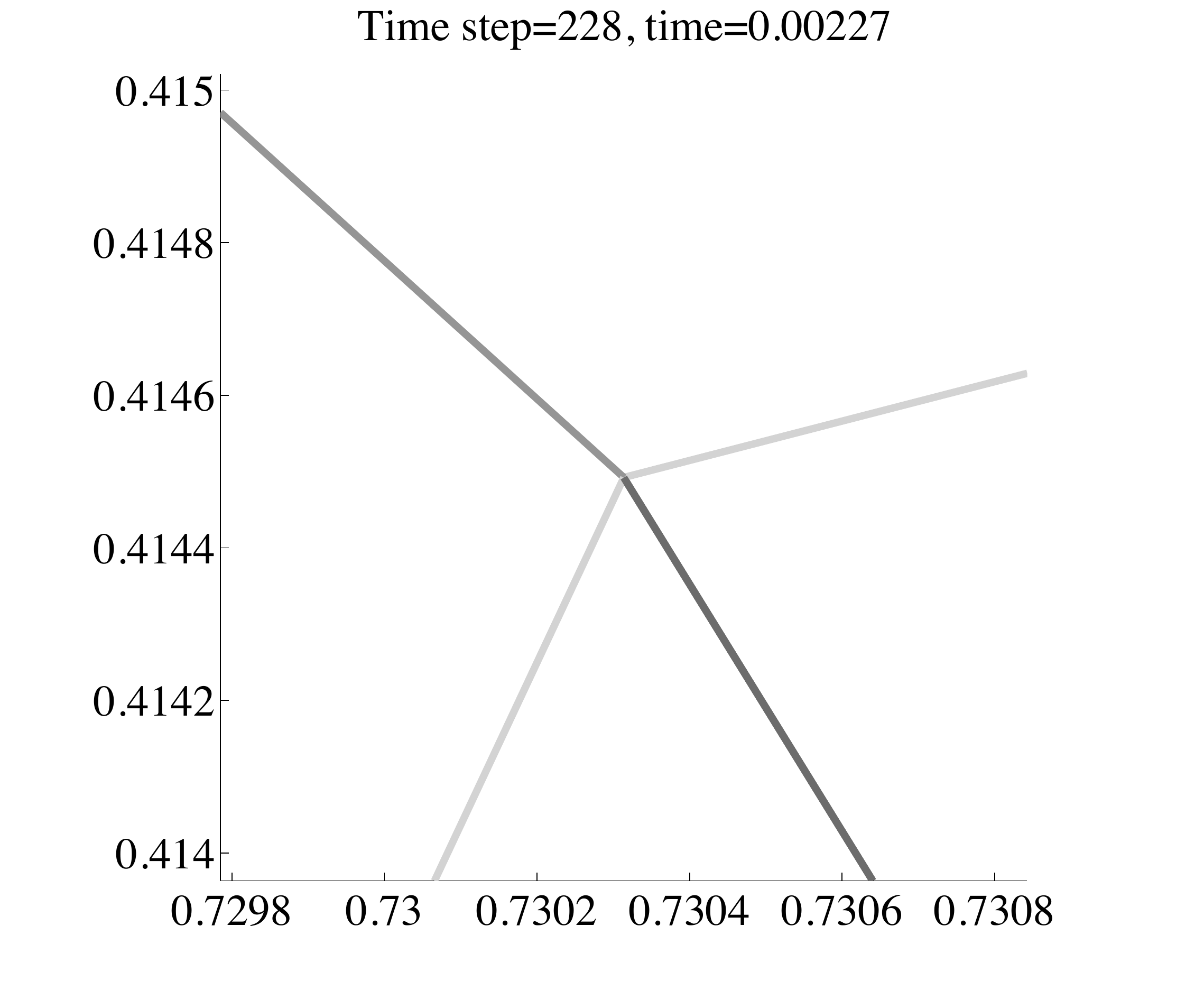}
    \caption{}
    \label{fig:quad_02}
  \end{subfigure}

\begin{subfigure}[b]{0.45\textwidth}
  \centering
  \includegraphics[height=4.5cm]{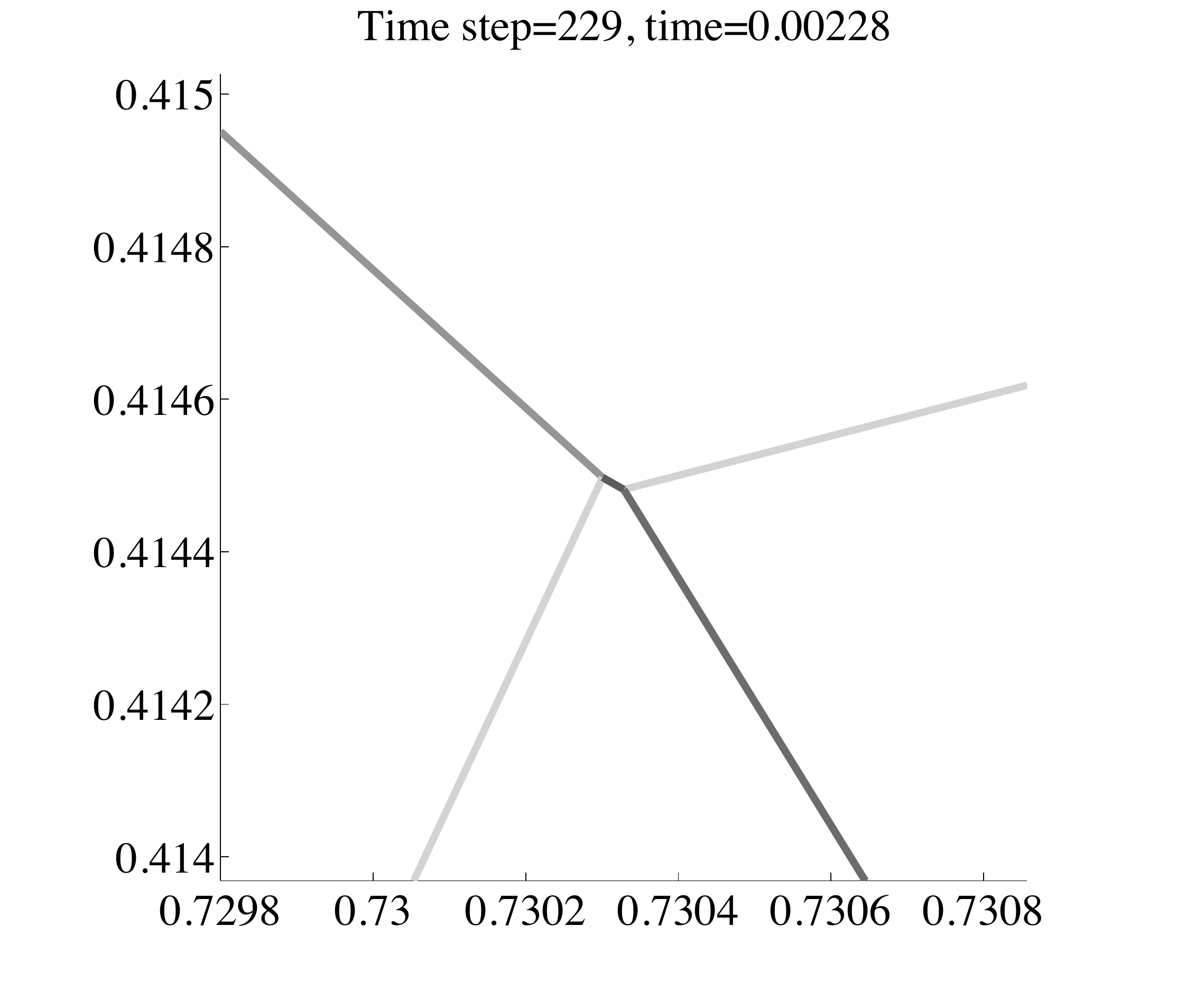}
  \caption{}
  \label{fig:quad_03}
\end{subfigure}
\begin{subfigure}[b]{0.45\textwidth}
  \centering
  \includegraphics[height=4.5cm]{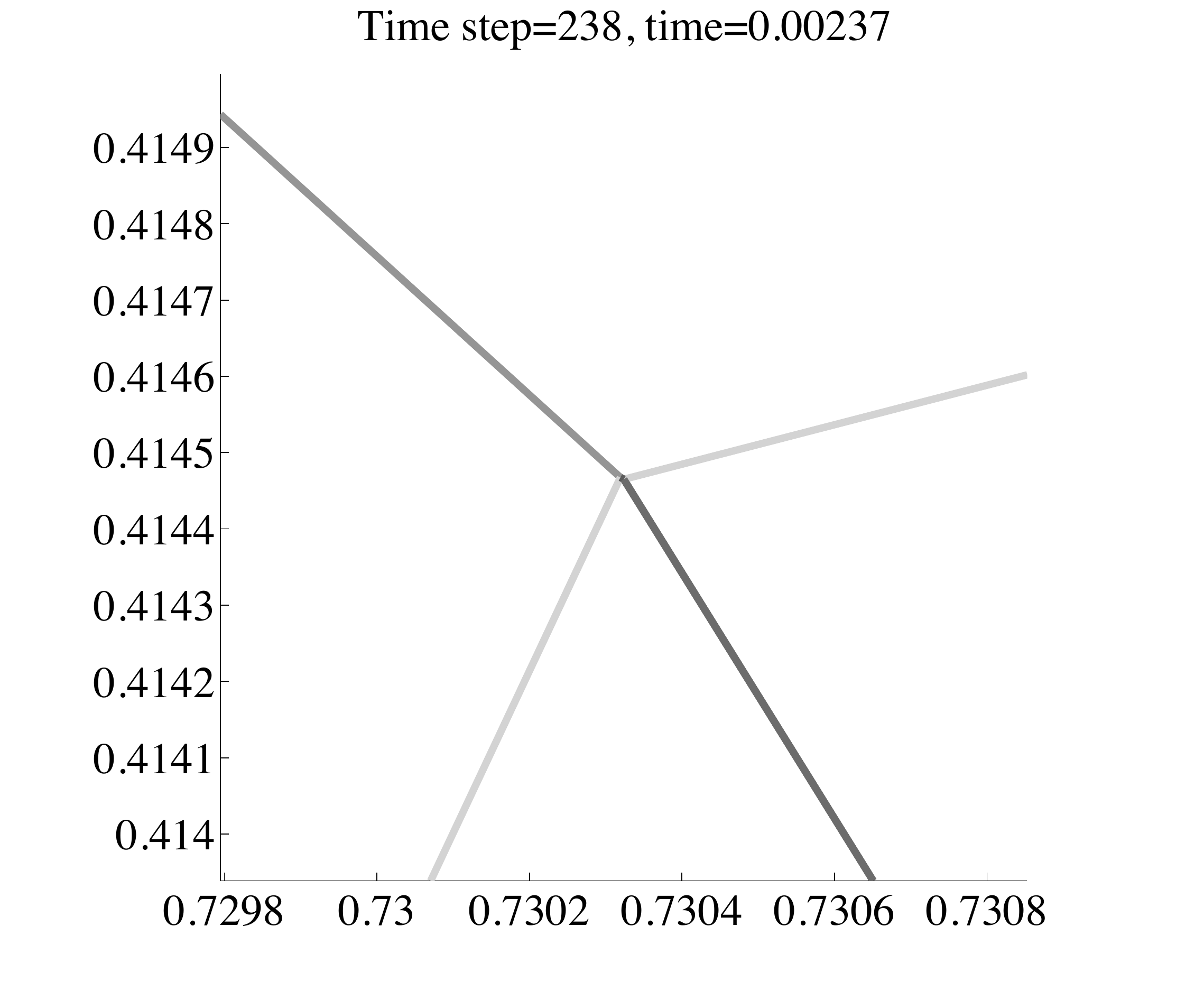}
  \caption{}
  \label{fig:quad_04}
\end{subfigure}
\caption{Evolution of two triple junction that remain together and
  emulate a quadruple junction.  Darker colors correspond to larger
  grain boundary energy.}
\label{fig:quad}
\end{figure}

In a network with a large anisotropy stable quadruple junctions may exist.  From the
implementation point of view, they are two triple junctions that almost overlap. In principle, one needs to develop a separate set of rules that 
govern neighbor switching for this type of a junction. In our algorithm, quadruple junctions are always assumed to split into triple junctions. 
However, if a quadruple junction is stable, then any new boundary created as a result of the split would shrink 
and disappear, restoring the original quadruple junction. In this way, the 
code is capable of dealing with stable quadruple junctions. The example of such junction in a grain boundary network with an anisotropic grain boundary energy $\gamma(\Delta \alpha)=0.55-0.45\,\cos^3(4\,\Delta \alpha)$ is shown in
Figs.~\ref{fig:quad}.

\section{Conclusions}

We have developed a numerical algorithm for an evolving grain boundary network described by a simple vertex model. The model can be formally derived via the assumption that the mobility of triple junctions is much lower than the mobility of grain boundaries evolving via curvature-driven motion. We have used a semi-rigorous analysis of vertex dynamics to derive the neighbor switching rules consistent with continuous evolution of vertices as well as the estimates of vertex collision times. These estimates were incorporated into in the numerical procedure to pinpoint the times corresponding to topological transitions. By simulating coarsening of the vertex model grain boundary network, we have demonstrated that the geometry of configurations that develop is described by the standard statistical measures for characterizing grain growth. These include distributions of relative areas of grains, dihedral angle, number of sides, among others. We have confirmed spatiotemporal stability of the distributions that develop in a network evolving via our numerical algorithm. We find that the distributions are essentially independent of the level of numerical resolution as the network passes through the sequence of similar states. While mesoscopic characteristics of the network appear to be robust, even with respect to changes in the rules governing topological transitions, the microscopic features of the network at a given time are sensitive to any modifications of the algorithm.

\section{Acknowledgements}

M.~E. and C.~T. were supported in part by NSF grant DMS-1056821. D.~G. was supported in part by NSF grant DMS-1009849. D.~K. was supported in part by NSF grants DMS-0806703, DMS-0635983, and OISE-0967140. S.~T. was supported in part by NSF grant DMS-1216433.

\bibliographystyle{ieeetr}
\bibliography{vertex-code-references}

\section*{Appendix 1}

Suppose that the edge connecting two vertices ${\bf x}_i$ and ${\bf
  x}_j$ disappears at the time $t_0$ and that ${\bf p}_-$, $\rho$,
$\theta$, ${\bf n}$, and $\boldsymbol{\tau}$ are as defined in Section
\ref{sec:analysis_flipping_rule}. Here we will use
\eqref{eq:right_before_flipping} to argue that, as long as
$\mathbf{p_-}\in C([t_0-\Delta t,t_0])$ for some small $\Delta t>0$
and
\[\|{\bf p}_-(t_0)\|-2\gamma_{ij}<0,\] we have that
\begin{equation}
  \label{thetalim}
  \lim_{t\to t_0^-}\mathbf{p_-}\cdot\boldsymbol\tau=0.
\end{equation}

First, since $\mathbf{p_-}\in C([t_0-\Delta t,t_0])$ and $\|{\bf
  p}_-(t_0)\|-2\gamma_{ij}<0,$ we can choose $\Delta t$ small enough
so that
$-4\gamma_{ij}\leq\mathbf{p_-}\cdot\mathbf{n}-2\,\gamma_{ij}\leq-\alpha$
on $[t_0-\Delta t,t_0]$ for some $\alpha>0.$ The system
\eqref{eq:right_before_flipping} has a continuous solution on
$[t_0-\Delta t,t_0)$ where the function $\rho$ is, in fact, continuous
on $[t_0-\Delta t,t_0]$. Integrating the equation
\eqref{eq:right_before_flipping_1} and using the condition
$\rho(t_0)=0$, we obtain
\begin{equation}
  \label{eq:rhobound}
  \alpha(t_0-t)\leq\rho(t)\leq4\gamma_{ij}(t_0-t),
\end{equation}
when $t\in[t_0-\Delta t,t_0]$.

Set ${\mathbf p}_-=\|{\mathbf p}_-\|(\cos{\theta_p},\sin{\theta_p})$,
then the equation \eqref{eq:right_before_flipping_2} takes the form
\begin{equation}
  \label{eq:a1}
  \dot\theta=\frac{\|{\mathbf p}_-\|}{\rho}\sin{(\theta_p-\theta)}.
\end{equation}
We will assume here that ${\mathbf p}_-\neq{\mathbf 0}$ on
$[t_0-\Delta t,t_0]$, then both $\|{\mathbf p}_-\|$ and $\theta_p$ are
continuous and $\|{\mathbf p}_-(t)\|>\beta$ for all $t\in[t_0-\Delta
t,t_0]$ and some constant $\beta>0$.

Suppose first that $0\leq\theta_p-\theta\leq\pi$ on $[t_0-\Delta
t,t_0)$. Then $\dot\theta\geq 0$ on $t\in[t_0-\Delta t,t_0)$, hence
$\lim_{t\to t_0^-}\theta$ exists. If this limit is finite, it
immediately follows from integrability of the right hand side of
\eqref{eq:a1} and \eqref{eq:rhobound} that the $\lim_{t\to
  t_0^-}\theta=\theta_p(t_0)$. If the $\lim_{t\to
  t_0^-}\theta=\infty$, then the assumption that $\theta_p-\theta>0$
implies that $\lim_{t\to t_0^-}\theta_p=\infty$; this violates the
continuity of $\theta_p$ on $[t_0-\Delta t,t_0]$. An analogous
argument can be used to show a symmetric result for
$-\pi\leq\theta_p-\theta\leq0$.

It remains to prove \eqref{thetalim} when there exists a sequence
$t_n\to t_0$ such that $t_n<t_{n+1}$ and
$\sin{\left(\theta_p(t_n)-\theta(t_n)\right)}=0$ for all
$n=1,2,3,\ldots$.  Since $\theta_p$ is continuous on $[t_0-\Delta
t,t_0]$, we can assume that
\begin{equation}
  \label{eq:a2}
  \left|\theta_p(t_n)-\theta_p(t_0)\right|<\pi/2
\end{equation}
for all $n=1,2,3,\ldots.$ Fix an arbitrary $n=1,2,3,\ldots$ then
$\sin{\left(\theta_p(t_n)-\theta(t_n)\right)}=0$ and
$\sin{\left(\theta_p(t_{n+1})-\theta(t_{n+1})\right)}=0$. Assume that
$\theta_p(t_{n})=\theta(t_{n})$ and $0<\theta_p(t)-\theta(t)<\pi$ on
the interval $(t_n,t_{n+1})$. Then the equation \eqref{eq:a1}
guarantees that $\dot\theta>0$ on $(t_n,t_{n+1})$ and thus $\theta$ is
increasing on $(t_n,t_{n+1})$. Further, we have that either
$\theta_p(t_{n+1})=\theta(t_{n+1})$ or
$\theta_p(t_{n+1})=\theta(t_{n+1})+\pi$. But since $\theta$ is
increasing on the interval $(t_n,t_{n+1})$, the inequality
\eqref{eq:a2} implies that
\[0\leq\theta_p(t_{n+1})-\theta(t_{n+1})=\theta_p(t_{n+1})-\theta_p(t_n)+\theta(t_n)-\theta(t_{n+1})<\theta_p(t_{n+1})-\theta_p(t_n)\leq\frac{\pi}{2}.\]
and therefore $\theta_p(t_{n+1})=\theta(t_{n+1})$. The same conclusion
holds if we assume that $-\pi<\theta_p(t)-\theta(t)<0$ on the interval
$(t_n,t_{n+1})$, except that in this case $\theta$ is monotone
decreasing on $(t_n,t_{n+1})$. It follows by an induction argument
that $\theta$ and $\theta_p$ have the same values at $t_n$ and
$\theta$ is monotone on $(t_n,t_{n+1})$ for every
$n=1,2,3,\ldots$. The fact that $\lim_{t\to
  t_0^-}\theta=\theta_p(t_0)$ is then a simple consequence of
continuity of $\theta_p$ on $[t_0-\Delta t,t_0]$. This, in particular,
implies \eqref{thetalim}.

\section*{Appendix 2}

Here we derive the equation \eqref{eq:d5}, that is, given
\begin{equation}
  \label{eq:a21}
  \Phi(\alpha,x)=\prod_{j=1}^{\lfloor x\rfloor-1}\left|1-\frac{\alpha}{1-\frac{j}{x}}\right|=\prod_{j=1}^{\lfloor x\rfloor-1}\left|\frac{x-j-\alpha x}{x-j}\right|,
\end{equation}
where $x$ is large and $0<\alpha<1$, we show that
\begin{equation}
  \label{eq:a22}
  \Phi(\alpha,x)=\frac{1}{\pi}\frac{\Gamma(1+x-\lfloor x \rfloor)\Gamma((1-\alpha)x)\Gamma(\lfloor x \rfloor-x+\alpha x)}{\Gamma(x)}\sin{\pi\left((1-\alpha)x-\lfloor(1-\alpha)x\rfloor\right)}.
\end{equation}
First, given $\lambda>0$, the following relationship
\begin{equation}
  \label{eq:a23}
  \frac{\Gamma(m+\lambda+1)}{\Gamma(l+\lambda)}=\prod_{i=l}^{m}(i+\lambda)
\end{equation}
holds for any $l,m\in\mathbb{N}$. By changing the index, $j\to\lfloor
x\rfloor-j$, we have that
\begin{equation}
  \label{eq:a24}
  \begin{split}
    \Phi(\alpha,x) & =\prod_{j=1}^{\lfloor
      x\rfloor-1}\left|\frac{x-\lfloor x\rfloor-\alpha x+j}{x-\lfloor
        x\rfloor+j}\right| \\ & =\frac{\prod_{j=1}^{\lfloor
        x\rfloor-\lfloor(1-\alpha)x\rfloor-1}\left(x-\lfloor
        x\rfloor-\alpha x+j\right)\prod_{j=\lfloor
        x\rfloor-\lfloor(1-\alpha)x\rfloor}^{\lfloor
        x\rfloor-1}\left(x-\lfloor x\rfloor-\alpha
        x+j\right)}{\prod_{j=1}^{\lfloor x\rfloor-1}\left(x-\lfloor
        x\rfloor+j\right)} \\ &
    =\frac{\Phi_1(\alpha,x)\Phi_2(\alpha,x)}{\Phi_3(\alpha,x)}.
  \end{split}
\end{equation}
We consider $\Phi_1,\ \Phi_2,$ and $\Phi_3$ separately. Using
\eqref{eq:a23}, we immediately obtain
\[\Phi_3(\alpha,x)=\prod_{j=1}^{\lfloor x\rfloor-1}\left(x-\lfloor
  x\rfloor+j\right)=\frac{\Gamma(x)}{\Gamma(1+x-\lfloor x \rfloor)}.\]
By changing the index $j\to j-\lfloor
x\rfloor+\lfloor(1-\alpha)x\rfloor$, we find
\[\Phi_1(\alpha,x)=\prod_{j=0}^{\lfloor(1-\alpha)x\rfloor-1}\left((1-\alpha)x-\lfloor
  (1-\alpha)x\rfloor+j\right)=\frac{\Gamma((1-\alpha)x)}{\Gamma((1-\alpha)x-\lfloor
  (1-\alpha)x\rfloor)},\] and by changing the index $j\to \lfloor
x\rfloor-\lfloor(1-\alpha)x\rfloor-1-j$, we determine that
\[\Phi_2(\alpha,x)=\prod_{j=0}^{\lfloor
  x\rfloor-\lceil(1-\alpha)x\rceil-1}\left(\lceil(1-\alpha)x\rceil-(1-\alpha)x+j\right)=\frac{\Gamma(\lfloor
  x\rfloor-x+\alpha
  x)}{\Gamma(\lceil(1-\alpha)x\rceil-(1-\alpha)x)}.\] It follows by
the properties of the $\Gamma$-function that
\begin{equation*}
  \begin{split}
    \Phi_1(\alpha,x)\Phi_2(\alpha,x)=\frac{\Gamma((1-\alpha)x)\Gamma(\lfloor
      x\rfloor-x+\alpha x)}{\Gamma((1-\alpha)x-\lfloor
      (1-\alpha)x\rfloor)\Gamma(\lceil(1-\alpha)x\rceil-(1-\alpha)x)}
    \\ =\frac{\Gamma((1-\alpha)x)\Gamma(\lfloor x\rfloor-x+\alpha
      x)}{\Gamma((1-\alpha)x-\lfloor
      (1-\alpha)x\rfloor)\Gamma(1-((1-\alpha)x-\lfloor
      (1-\alpha)x\rfloor))} \\
    =\frac{1}{\pi}\Gamma((1-\alpha)x)\Gamma(\lfloor x\rfloor-x+\alpha
    x)\sin{\pi\left((1-\alpha)x-\lfloor(1-\alpha)x\rfloor\right)}.
  \end{split}
\end{equation*}
Dividing this expression by $\Phi_3$, we recover \eqref{eq:a21}.

\end{document}